\documentclass[twocolumn,showpacs,superscriptaddress,preprintnumbers,prc]{revtex4-1}
\usepackage{amsmath}
\usepackage{amssymb}
\usepackage{graphicx}
\usepackage{dcolumn}
\usepackage[colorlinks,linkcolor=blue,urlcolor=blue,citecolor=blue]{hyperref}
\usepackage{orcidlink}
\usepackage{CJK}
\usepackage{newtxtext}\usepackage[varvw,bigdelims]{newtxmath}
\usepackage{bm}
\usepackage{tikz}
\newcommand{\nuc}[2][]{{}^{#1}\mathrm{#2}}

\date{October 30, 2024}
\begin{document}

\title{Activation of momentum fluctuations in wave packet molecular dynamics: Impacts on momentum distributions of projectilelike fragments}

\author{Lei Shen (\begin{CJK}{UTF8}{gbsn}沈雷\end{CJK})
\orcidlink{0000-0002-9814-8171}
}
\affiliation{School of Physical Science and Technology, ShanghaiTech University, Shanghai 201203, China}
\affiliation{Shanghai Institute of Applied Physics, Chinese Academy of Sciences, Shanghai 201800, China}
\affiliation{University of the Chinese Academy of Sciences, Beijing 100080, China}
\affiliation{Department of Physics, Tohoku University, Sendai 980-8578, Japan}

\author{Akira Ono (\begin{CJK}{UTF8}{ipxm}小野 章\end{CJK})
\orcidlink{0000-0002-0233-0252}
}
\email{akira.ono@tohoku.ac.jp} 
\affiliation{Department of Physics, Tohoku University, Sendai 980-8578, Japan}

\author{Yu-Gang Ma (\begin{CJK}{UTF8}{gbsn}马余刚\end{CJK})
\orcidlink{0000-0002-0233-9900}
}
\email{mayugang@fudan.edu.cn} 
\affiliation{Key Laboratory of Nuclear Physics and Ion-Beam Application (MOE), Institute of Modern Physics, Fudan University, Shanghai 200433, China}
\affiliation{Shanghai Research Center for Theoretical Nuclear Physics, NSFC and Fudan University, Shanghai 200438, China}

\begin{abstract}
Molecular dynamics approaches use wave packets as nucleon wave functions to simulate the time evolution of nuclear reactions. It is crucial to activate the momentum fluctuation inherent in each wave packet so that it properly affects the time evolution. In the antisymmetrized molecular dynamics (AMD) model, this has traditionally been done by splitting the wave packets, i.e., by introducing a random fluctuation to the wave packet center of each particle. The present work proposes an improved approach to activate the fluctuation in both the one-body mean-field propagation and the two-nucleon collision processes, consistently based on the gradual or sudden change of the degree of isolation, which is derived from the fragment number function used for the zero-point energy subtraction. This new method is applied to the $\nuc[12]{C}+\nuc[12]{C}$ and $\nuc[12]{C} + p$ reactions at about 100 MeV/nucleon, focusing on the momentum distribution of the $\nuc[11]{B}$ fragments produced by one-proton removal from the $\nuc[12]{C}$ projectile. The results show that, with the momentum fluctuation suitably activated, the method correctly accounts for the recoil from the removed nucleon to the residue and the $\nuc[11]{B}$ momentum distribution is significantly improved, while without activating the fluctuation the distribution is too narrow compared to the experimental data. Furthermore, the AMD results indicate that the momentum distribution consists of two components; one is the high $P_z$ component with a small shift from the beam velocity, resulting from the simple removal of a proton after an energetic collision with a target particle; the other is the low $P_z$ component with a larger peak shift resulting from the decay of an excited $\nuc[12]{C}$ nucleus in a longer time scale. The activation of momentum fluctuation mainly affects the high $P_z$ component to broaden it. The role of cluster correlations in this problem is also investigated.
\end{abstract}

\maketitle

\section{Introduction}

The nucleon momentum distribution in a nucleus contains rich information about various aspects of the nuclear structure and the interactions between nucleons. Due to the Pauli principle, the low-momentum part of the distribution is mainly determined by the uncorrelated Fermi motion, although details of the nuclear structure, such as cluster correlations, can also influence it. On the other hand, the high-momentum part is formed due to short-range correlations between nucleons in the nucleus \cite{Hen2014S,Hen2017RMP,Degli2015PR}. Quasi-free knockout reactions are used to probe the nucleon momentum distribution, which is reflected in the momentum of the residual nucleus after the removal of one or more nucleons. For single-nucleon knockout reactions, the momentum of the residual nucleus in the projectile rest frame is in the opposite direction and of approximately the same magnitude as the initial momentum of the removed nucleon, which may be in a single-particle orbit such as $s$-wave, $p$-wave, or other shells as demonstrated by experiments \cite{Hufner1981PRC,Bertulani1992PRC,Hencken1996PRC,Satou2014PLB,Holl2019PLB,Panin2019PLB}. When several or more nucleons are removed in heavy-ion collisions, the momentum of the remaining nucleus is approximately determined by the sum of the momenta of all removed nucleons, which as in the Goldhaber model should follow a Gaussian distribution \cite{Goldhaber1974PLB} and the width can be explained by the Fermi momentum of the removed nucleons \cite{Goldhaber1974PLB,Morrissey1989PRC,Ma2002PRC}. In some cases some non-trivial fragment momentum distributions are explained by more complex dynamic mechanisms \cite{Momota2017NPA,Notani2007PRC,Meier2012PRC,Mocko2006PRC,Greiner1975PRL,Caamano2004NPA,Weber1994NPA,Reinhold1998PRC,Momota2023PS,Bibber1979PRL,Kidd1988PRC}. 

Besides the nucleon removal reactions, the nucleon momentum distribution also plays an important role in various nuclear reactions such as violent heavy-ion collisions. Transport models have been widely used for the simulation of heavy-ion collisions \cite{Ono2006EPJA,Ma2018PPNP,Ono2019PPNP,Xu2019PPNP,Wolter2022PPNP,Deng2024PPNP,Sun2024NC}. There are basically two families of transport models, namely the Boltzmann-Uehling-Uhlenbeck (BUU) models \cite{Bertsch1988PR,Li2008PR,Song2023PRC,Wang2023PRC} and the Quantum Molecular Dynamics (QMD) models \cite{Aichelin1991PR,Liu2022NST,Li2022NST,Wang2022NST,Wang2023NST,Xiao2023NST,Wei2024NST}. They consist of a mean-field propagation of the one-body phase-space distribution and a two-nucleon collision term. At least, the nucleon momentum distribution in the initial state nuclei should affect the time evolution of the heavy-ion collision and may affect the particle momenta in the final state. BUU models most naturally treat the uncorrelated component of the momentum distribution and its effect on the time evolution of the one-body distribution function $f(\bm{r}, \bm{p})$ through the mean-field propagation and two-body collisions. For example, the width of the momentum distribution can lead to a broadening of the distribution in the coordinate space according to the mean-field propagation. However, the BUU models cannot properly describe the event-by-event fluctuations, and therefore it is not straightforward to describe fragment production. For example, the event-by-event momentum fluctuation of the projectile-like fragment is difficult to be described by BUU models. On the other hand, QMD models can describe event-by-event fluctuations by generating different events that evolve independently, through the stochastic processes such as nucleon-nucleon collisions. However, a problem is that the treatment of the momentum distribution is not as straightforward as in BUU. The nucleons in QMD models are described by Gaussian wave packets with finite width \cite{Aichelin1991PR}, which in principle obeys the uncertainty principle $\Delta x \Delta p \geqslant \hbar/2$, but in practice QMD models usually neglect the momentum width and set $\Delta p=0$, while $\Delta x$ is a time-independent parameter. The initial momenta in the ground state nuclei are randomly chosen in the Fermi sphere, which is also a source of different events in the QMD models. Some other models, such as the Fermionic Molecular Dynamics (FMD) model and the Extended QMD model (EQMD), treat the widths as time-dependent variables \cite{Maruyama1996PRC,Feldmeier1990NPA,Kiderlen1997NPA, Colonna1998PLB} that always satisfy $\Delta x \Delta p \geqslant \hbar/2$, and they succeeded in some low energy cases; especially $\alpha$-clustering structure as well as photonuclear reaction can be well described within EQMD \cite{He2014PRL,Huang2017PRC,Wang2022PRC,Cao2022PRC,Cao2023PRC,Ma2023NT}. 

The antisymmetrized molecular dynamics (AMD) model \cite{Ono1992PTP,Ono1992PRL,Ono2004PPNP,Ikeno2016PRC} describes the total wave function as a Slater determinant of Gaussian wave packets. The width of the Gaussian wave packet in the AMD model is constant and satisfies $\Delta x \Delta p = \hbar/2$ when the wave function is faithfully interpreted. The use of such compact wave packets is advantageous for the description of fragment formation. However, the finite momentum width $\Delta p$ causes a problem that any nucleon or the center-of-mass of any fragment has at least the zero-point kinetic energy $3\Delta p^2/2M$, and thus the threshold energies for the nucleon emission and fragmentation are not correctly treated. One solution is to subtract the zero-point kinetic energy for each fragment, as introduced by Ref. \cite{Ono1992PTP}. This corresponds to a change of the interpretation in such a way that the center-of-mass of any isolated fragment or nucleon now has a definite momentum without the momentum width. It should be noted that this modification by itself does not cause event branching, since the zero-point energy is subtracted in the deterministic equation of motion by introducing the fragment number function \cite{Ono1992PTP,Ono1993PRC,Ono1993PRC2}. On the other hand, event branching has been considered by several methods to activate momentum fluctuation, without a direct link to the zero-point energy subtraction. Ref.~\cite{Ono1996PRC} proposed to give a random momentum fluctuation when a nucleon is being emitted from a hot nucleus, and the idea was generalized in Refs.~\cite{Ono1996PRC2,Ono1999PRC,Ono2002PRC} to split wave packets based on the evolution of the phase space distribution due to the Vlasov mean-field propagation. The momentum width is also considered in the two-nucleon collision process by Lin \textit{et al.}\ \cite{Lin2016PRC} by introducing the momentum boost in determining the final state.

In the present work, we propose a new approach to activate the momentum fluctuation in both the one-body mean-field propagation and the two-nucleon collision processes, consistent with the zero-point kinetic energy subtraction for isolated nucleons and clusters. In this method, the wave packet splitting is induced by the gradual or sudden change in the degree of isolation, which is estimated from the fragment number function utilized for the zero-point energy subtraction. The formulation is given in Sec.~\ref{sec_formulation}. Although the method is applicable to general situations of nuclear reactions, the application in this paper, presented in Sec.~\ref{sec_result}, aims to confirm the validity of the new method in the simple one-nucleon removal process in $\nuc[12]{C}+\nuc[12]{C}$ and $\nuc[12]{C} + p$ reactions at about 100 MeV/nucleon. We focus on the momentum distribution of the $\nuc[11]{B}$ fragment, which we find to be quite sensitive to the treatment of the momentum fluctuation for the removed nucleon and its recoil on the residue. Detailed investigations are also given to understand the $\nuc[11]{B}$ momentum distribution, which seems to contain several components corresponding to different physical mechanisms. A summary and future perspectives are given in Sec.~\ref{sec_conclusion}.

\section{Method}\label{sec_formulation}
\subsection{AMD model}

In the AMD model, the total wave function of an $A$-nucleon system is described by a Slater determinant of Gaussian wave packets
\begin{equation}
    |\Phi(\mathbf{Z})\rangle=\hat{A}\prod^{A}_{i=1}\phi_i,
\end{equation}
where $\hat{A}$ is the full antisymmetrization operator and $\phi_i$ is the single-particle state which is a product of a Gaussian function and a spin-isospin state
\begin{equation}
    \langle\bm{r}|\phi_i\rangle=e^{-\nu(\bm{r}-\mathbf{Z}_i/\sqrt{\nu})^2}\chi_i,
\end{equation}
for $i=1,2,...,A$. Here the centroid variable $\mathbf{Z}_i$ contains the information of the position and the momentum in its real and imaginary parts, respectively. The standard choice of the width parameter is $\nu=0.16\ \mathrm{fm}^{-2}$ which corresponds to the position and the momentum uncertainties $\Delta x=1/(2\sqrt{\nu})=1.25\ \mathrm{fm}$ and $\Delta p=\hbar\sqrt{\nu}=78.9\ \mathrm{MeV}/c$. The spin-isospin state $\chi_i$ is fixed to be $p\uparrow$, $p\downarrow$, $n\uparrow$ or $n\downarrow$.

Due to the antisymmetrization, the variables {$\mathbf{Z}_i$} do not have a simple meaning. In fact, the equations of motion derived from the time-dependent variational principle show that these are not canonical variables. The Wigner transform of the one-body density for the spin-isospin state $\alpha$ is written in a complicated way as
\begin{equation}
    f_{\alpha}(\bm{r}, \bm{p})=8\sum_{i\in \alpha}\sum_{j\in \alpha}e^{-(\bm{r}-\mathbf{R}_{ij})^2/2\Delta x^2}e^{-(\bm{p}-\mathbf{P}_{ij})^2/2\Delta p^2}B_{ij}B^{-1}_{ji},
    \label{eq_dis}
\end{equation}
where $\mathbf{R}_{ij}=(\mathbf{Z}_i^*+\mathbf{Z}_j)/2\sqrt{\nu}$, $\mathbf{P}_{ij}=i\hbar\sqrt{\nu}(\mathbf{Z}_i^*-\mathbf{Z}_j)$ and the overlap matrix elements $B_{ij}=\langle\phi_i|\phi_j\rangle$ take complex value.
Depending on the purpose, it is often convenient to introduce a decomposition of the Wigner distribution function $f_{\alpha}(\bm{r}, \bm{p})$ into terms each of which may be regarded as a nucleon in some sense. For example, an approximated distribution function
\begin{equation}
    f_{\alpha}(\bm{r}, \bm{p})\approx 8\sum_{i\in\alpha}e^{-(\bm{r}-\bm{R}_{i})^2/2\Delta x^2}e^{-(\bm{p}-\bm{P}_{i})^2/2\Delta p^2}
\end{equation}
was introduced by using the so-called physical coordinates $(\bm{R}_{i}, \bm{P}_{i})$, to treat two-nucleon collisions in AMD \cite{Ono1992PTP}. A much more precise decomposition was introduced in Appendix C of Ref.~\cite{Ikeno2016PRC} when a method was formulated to sample test particles following the precise distribution function $f_{\alpha}(\bm{r}, \bm{p})$. In this method, the Wigner function of Eq.~(\ref{eq_dis}) is decomposed as 
\begin{equation}
    f_{\alpha}(\bm{r}, \bm{p})=\sum_{i\in\alpha}F_i(\bm{r}-\bm{R}_i, \bm{p}-\bm{P}_i),
    \label{eq_testp}
\end{equation}
where the function $F_i$ is implicitly defined by the method of generating test particles \cite{Ikeno2016PRC}. Each test particle $(\bm{r}^{\mathrm{tp}}, \bm{p}^{\mathrm{tp}})$ is sampled associated with one of the physical coordinates $(\bm{R}_{i}, \bm{P}_{i})$. If a wave packet is well separated in the phase space from the rest of the system, the corresponding distribution $F_i$ will be a Gaussian function with the width parameters $\Delta x$ and $\Delta p$. Generally, $F_i$ is not a Gaussian function but it is still localized around the point $(\bm{R}_{i}, \bm{P}_{i})$.
In this paper, bold italic type, such as $\bm{R}_{i}$ and $\bm{P}_{i}$, is used for the physical coordinates, while bold roman type is used for the original wave packet centroids $\mathbf{R}_i=\mathbf{R}_{ii}=\mathop{\mathrm{Re}}\mathbf{Z}_i/\sqrt{\nu}$ and $\mathbf{P}_i=\mathbf{P}_{ii}=2\hbar\sqrt{\nu}\mathop{\mathrm{Im}}\mathbf{Z}_i$ before considering antisymmetrization.

Test particles are typically used in BUU calculations to represent the phase space distribution, and their motion is tracked to solve the time evolution. In contrast, AMD follows the time evolution of the wave packet centroids $\mathbf{Z}_i$ or $(\mathbf{R}_i,\mathbf{P}_i)$. Test particles are generated randomly from the distribution function of Eq.~\eqref{eq_dis} whenever needed for some purpose.

The time evolution of the centroids $\mathbf{Z}_i$ is determined by the equation of motion derived from the time dependent variational principle. It is
\begin{equation}
    i\hbar \sum_{j\tau}C_{i\sigma,j\tau}\frac{d\mathbf{Z}_{j\tau}}{dt}=\frac{\partial\mathcal{H}}{\partial\mathbf{Z}_{i\sigma}^*}
    \label{eq_motion}
\end{equation}
with a Hermitian matrix $C_{i\sigma,j\tau}$ with $\sigma,\tau=x,y,z$ defined as
\begin{equation}
    C_{i\sigma,j\tau}=\frac{\partial^2}{\partial\mathbf{Z}_{i\sigma}^*\partial\mathbf{Z}_{j\tau}}\log\langle\Phi(\mathbf{Z})|\Phi(\mathbf{Z})\rangle.
\end{equation}
The present work uses the Skyrme-type interaction with the SLy4 parameter set \cite{Chabanat1998NPA} together with a correction of the momentum dependence of the mean field \cite{Ikeno2023PRC}. The Hamiltonian $\mathcal{H}$ in Eq.~(\ref{eq_motion}) is the expectation value
\begin{equation}
    \mathcal{H} =\frac{\langle\Phi|\hat{H}|\Phi\rangle}{\langle\Phi|\Phi\rangle}-\frac{3\hbar^2\nu}{2M}A+T_0(A-\mathcal{N}_{\mathrm{frag}}),
    \label{eq_energy}
\end{equation}
but the zero-point kinetic energies of isolated nucleons and fragments have been subtracted by introducing the number of isolated fragments $\mathcal{N}_{\mathrm{frag}}$. This function $\mathcal{N}_{\mathrm{frag}}$, described in detail in the next subsection, is very important for the present work to activate momentum fluctuation. The parameter $T_0$ is $3\Delta p^2/2M$ in principle, with $M$ being the nucleon mass, but it is adjusted to improve the overall reproduction of the binding energies of various nuclei, as in Refs.~\cite{Ono1992PTP,Ono1993PRC2}. In the present work, $T_0=8.2$ MeV is chosen for the calculation with the Skyrme SLy4 interaction and the width parameter $\nu=0.16\ \text{fm}^{-2}$. The term $-T_0\mathcal{N}_{\mathrm{frag}}$ is expected to act as a repulsive potential when a fragmentation occurs.

In the present AMD model, the two-nucleon collision process considers cluster correlation in the final state \cite{Ono2013JPCS,Ikeno2016PRC}. When two nucleons $N_1$ and $N_2$ collide, each of them may form a cluster with other spectator particles $B_1$ or $B_2$ around it. A general collision process can be described as $N_1+N_2+B_1+B_2\rightarrow C_1+C_2$. Here $B_1$ and/or $B_2$ can be empty. The collision probability is described as
\begin{equation}
  {v_{\text{i}}}\frac{d\sigma(C_1,C_2)}{d\Omega}=P(C_1,C_2,p_{\text{f}},\Omega)|M|^2\frac{p_{\text{f}}^2}{v_{\text{f}}},
    \label{eq_coll}
\end{equation}
where $v_{\text{i}}$ is the initial relative velocity of the two colliding nucleons $N_1$ and $N_2$. The relative momentum vector after the momentum transfer between them is denoted by $(p_{\text{f}},\Omega)$, and $p_{\text{f}}$ is determined to conserve the energy $\mathcal{H}$ of the system which includes the adopted effective interaction. The velocity factor $v_{\text{f}}=\partial \mathcal{H}/\partial p_{\text{f}}$ as a function of $p_{\text{f}}$ also depends on the effective interaction. The probability factor $P(C_1,C_2,p_{\text{f}},\Omega)$ for cluster formation is the overlap probability between the initial and final states which considers the non-orthogonality of the final configurations \cite{Ikeno2016PRC}. The matrix element $|M|^2$ in nuclear medium is written as 
\begin{equation}
    |M|^2=(2/m_N)^2d\sigma_{NN}/d\Omega,
\end{equation}
with
\begin{equation}
    \sigma_{NN}=\sigma_0 \mathrm{tanh}(\sigma_{NN}^{(\mathrm{free})}/\sigma_0),
\end{equation}
where $\sigma_0=0.8\rho'^{-\frac{2}{3}}$ is chosen in the present work. A kind of phase-space density $\rho'$ is averaged in some way for the initial and final momenta and is calculated by Eq.~(161) in Ref.~\cite{Wolter2022PPNP}. The free cross section $\sigma_{NN}^{(\mathrm{free})}$ is parametrized by Eq.~(1) to (4) of Ref.~\cite{Cugnon1996NIM} and is evaluated at an average energy of the initial and final states.

After a cluster is formed in the collision process, the nucleons in the cluster will tend to move together until the cluster is broken when one or some of nucleons in it is involved in another collision process. It is also possible that a cluster is broken due to the different mean-field forces acting on the nucleons in it.

In the present work, the test particles generated by the method of Ref.~\cite{Ikeno2016PRC} are utilized in the two-nucleon collision process. At every time step for the collision process, a test particle coordinate $(\bm{r}_i^{\text{tp}}, \bm{p}_i^{\text{tp}})$ is randomly generated for each nucleon $i$ following the distribution function defined by Eq.~(\ref{eq_dis}) or Eq.~(\ref{eq_testp}). The relative coordinate $\bm{r}_{N_1}^{\text{tp}} - \bm{r}_{N_2}^{\text{tp}}$ is used to judge the collision, while the relative velocity $\bm{v}_{\text{i}}=\dot{\bm{R}}_{N_1}-\dot{\bm{R}}_{N_2}$ of the wave packet centers is used as $v_\text{i}$ in Eq.~(\ref{eq_coll}). The wave packet momentum centers $\bm{P}_{N_1}$ and $\bm{P}_{N_2}$ are changed by the scattering, but the information of the test particles is taken into account in $p_{\text{f}}$ and $v_{\text{f}}$ in Eq.~(\ref{eq_coll}).

It should be noticed that although each nucleon in AMD is described by a Gaussian wave packet, the test particles do not necessarily follow the Gaussian distribution due to antisymmetrization. This is in contrast to the usual method in which momentum fluctuation is sampled from the Gaussian distribution. The test particle in the present work follows the distribution function $f_{\alpha}(\bm{r}, \bm{p})$ which contains all kinds of quantum effects from the antisymmetrization of the many-body state. For example, it can represent some typical shell model states, such as the $(0s)^4(0p)^{12}$ configuration for the $\nuc[16]{O}$ nucleus. Gaussian distribution can only represent the $0s$ state whose center may be shifted, while test particles can represent the single-particle distribution more precisely.

When the dynamical evolution is truncated at a finite time (300 fm/$c$ in this work), the produced fragments are moving outwards without strong interaction among them. Such primary fragments are usually in excited state. The decay of the excited fragments are handled by the statistical decay model \cite{Maruyama1992PTP}, to generate the final fragments in the ground state which can be compared with the experimental data. This decay model is based on the sequential binary decay model by P\"ulhofer \cite{puhlhofer1977} but allows emission of composite particles not only in their ground states but also in the excited states with the excitation energy $E^*\le 40$ MeV. The switching time from AMD to a statistical decay model can be freely chosen, provided both models describe the decay of excited fragments comparably well. In fact, Ref.~\cite{tian2018} demonstrated that the final results are largely unaffected by selecting different switching times of 300, 1000 and 3000 fm/$c$.

\subsection{Interpretation of the momentum width}\label{sec_Nf}

The momentum width of each Gaussian wave packet has a large contribution $3\Delta p^2/2M \approx 10$ MeV (per nucleon) to the kinetic energy, where $M$ is the nucleon mass. This zero-point kinetic energy is an important part of the physical energy of nucleons in a nucleus. This is, however, problematic, e.g., when a nucleon is to be emitted from a nucleus. The emitted wave packet must have an average kinetic energy of $3\Delta p^2/2M \approx 10$ MeV at least, which disallows emission of low-energy nucleon. The same problem exists for the center-of-mass motion of a nucleus. Even if one may decide to ignore the 10 MeV shift of the energy of the center-of-mass motion of the total system, this problem unphysically raises the threshold energy for a separation of a nucleus into two or more fragments. A possible solution for this problem of spurious zero-point kinetic energies was proposed in Ref.~\cite{Ono1992PTP}. As in Eq.~\eqref{eq_energy}, the expectation value of the kinetic energy for the AMD wave function was modified to
\begin{equation}
    \mathcal{T}=\sum_{i=1}^A\sum_{j=1}^A\frac{\mathbf{P}_{ij}^2}{2M}B_{ij}B_{ji}^{-1}+(A-\mathcal{N}_{\mathrm{frag}})T_0
    \label{eq_Ek}
\end{equation}
by introducing a continious function $\mathcal{N}_{\mathrm{frag}}$ of the coordinates {$\mathbf{Z}_i$} that agrees with the number of fragments when the system is clearly separated into fragments. This expression means that each of the $A$ nucleons in the system has a zero-point energy $T_0$ but the zero-point energy of the center of mass of each of the $\mathcal{N}_{\text{frag}}$ isolated fragments and isolated nucleons is interpreted as spurious. Thus, for such an isolated fragment or nucleon, its center-of-mass wave function is now regarded as having a definite momentum without momentum width.

In the present work, we define $\mathcal{N}_{\mathrm{frag}}$ in a form similar to that in Ref.~\cite{Ono1993PRC} as
\begin{equation}
    \mathcal{N}_{\mathrm{frag}}=\sum_{i=1}^A\frac{g(k_i)}{n_im_i},
\end{equation}
where
\begin{equation}
    n_i=\sum_{j=1}^A\hat{f_{ij}},\ m_i=\sum_{j=1}^A\frac{1}{n_j}f_{ij},\ k_i=\sum_{j=1}^A\bar{f}_{ij},
\end{equation}
with
\begin{align}
    \hat{f_{ij}}= F(d_{ij},v_{ij},\hat{\xi},\hat{a}),
    \\
    f_{ij}= F(d_{ij},v_{ij},\xi,a),
    \\
    \bar{f_{ij}}= F(d_{ij},v_{ij},\bar{\xi},\bar{a}),
\end{align}
where $\hat{\xi}=0.32\ \mathrm{fm}^{-2},\ \hat{a}=0.5\ \mathrm{fm},\ \xi=0.32\ \mathrm{fm}^{-2},\ a=1.5 \ \mathrm{fm},\ \bar{\xi}=0.16\ \mathrm{fm}^{-2},\ \bar{a}=1.25\ \mathrm{fm},$ are used in this work. In the above equations, $d_{ij} = |\mathbf{R}_i - \mathbf{R}|$ is the distance between the wave packet centroids in the coordinate space. The function $g(k)$ is unity usually, but it is chosen to be
\begin{equation}
    g(k)=1+g_0e^{-(k-\mathcal{M})^2/2\sigma^2}
\end{equation}
in order to remedy the underbinding problem around the $\nuc[12]{C}$ nucleus, with $g_0=1.24$, $\mathcal{M}=12$ and $\sigma=2.0$ in this work. The function $F$ is to quantify whether the two wave packets belong to the same fragment. An important point of the present work is to optionally consider a momentum dependence of $F$ by choosing a form as
\begin{equation}
    F(d_{ij},v_{ij},\xi,a)=
    \begin{cases}
         c(v_{ij},d_{ij}) \ & d_{ij}\leq a,\\
         c(v_{ij},d_{ij}) e^{-\xi(d_{ij}-a)^2} & d_{ij}>a .
    \end{cases}
\end{equation}
where $c(v_{ij},d_{ij})$ is a momentum dependent factor which is introduced as
\begin{equation}
    c(v_{ij},d_{ij})=e^{-\xi_v ((1+v_{ij}^4/v_0^4)(r_0^2+d_{ij}^2)-r_0^2)},
    \label{eq_mdnf}
\end{equation}
where $\xi_v=0.0128\ \mathrm{fm}^{-2}$, $v_0=0.1c$, $r_0=0.2\ \mathrm{fm}$ and $v_{ij}=(\mathbf{P}_i-\mathbf{P}_j)/M$ is defined for the momentum centers of the wave packets. The factor $c(v_{ij},d_{ij})$ is multiplied to make sure that the nucleons which are close in the coordinate space but far away in the momentum space will not be judged  to be in the same fragment. With such momentum dependent factor, the collision process can change the number of fragments $\mathcal{N}_{\mathrm{frag}}$, which is important for the present work. Additionally $c(v_{ij},d_{ij})$ has a $d_{ij}$ dependence which is to enhance the isolation between the particles with similar momenta. We also show some results without this momentum dependence in $\mathcal{N}_{\mathrm{frag}}$, which corresponds to setting $c(v_{ij},d_{ij})=1$.

Associated with the fragment number function $\mathcal{N}_{\mathrm{frag}}$, the present work introduces the concept of the degree of isolation $\mu_i$ of a particle $i$ defined by
\begin{equation}
    \mu_i=\mathcal{N}_{\mathrm{frag}}-\mathcal{N}_{\mathrm{frag}}(i), 
    \label{eq_doi}
\end{equation}
where $\mathcal{N}_{\mathrm{frag}}(i)$ is the number of fragments evaluated when the $i$-th wave packet is eliminated. This quantity should be $\mu_i=1$ when the $i$-th wave packet $\mathbf{Z}_i$ is well isolated, and it should be $\mu_i=0$ when there is one or more wave packets close to it. We use this quantity to estimate the momentum width that is consistent with the subtraction of the spurious zero-point kinetic energy, i.e., we interpret that the momentum width is changed from $\Delta p$ to $\sqrt{1-\mu_i}\Delta p$. According to this change, we assume that the momentum distribution of each component of Eq.~(\ref{eq_testp}) is scaled as
\begin{equation}
    F_i(\bm{r}-\bm{R}_i, \bm{p}-\bm{P}_i;\mu_i)=(1-\mu_i)^{-\frac32}F_i\Bigl(\bm{r}-\bm{R}_i,\frac{\bm{p}-\bm{P}_i}{\sqrt{1-\mu_i}}\Bigr).
    \label{eq_Fi}
\end{equation}

The above consideration on the momentum width of a nucleon should be generalized to the center-of-mass motion of a cluster composed of several nucleon wave packets. When Gaussian wave functions are assumed for the nucleons, the center-of-mass degree of freedom of a cluster $k$ consisting of $A_k$ nucleons has a phase-space distribution
\begin{equation}
    G_k(\bm{r}-\mathbf{R}_k, \bm{p}-\mathbf{P}_k)=8e^{-A_k(\bm{r}-\mathbf{R}_k)^2/2\Delta x^2}e^{-(\bm{p}-\mathbf{P}_k)^2/2A_k\Delta p^2},
\end{equation}
with
\begin{equation}
    \mathbf{R}_k=\frac{1}{A_k}\sum_{i\in k}\mathbf{R}_i\ \ \text{and}\ \ \mathbf{P}_k=\frac{1}{A_k}\sum_{i\in k}\mathbf{P}_i,
\end{equation}
and $\Delta x=1/(2\sqrt{\nu})$ and $\Delta p=\hbar\sqrt{\nu}$. Following the same idea as for a nucleon, this is now modified when the cluster is partially or fully isolated, as
\begin{equation}
    G_k(\bm{r}-\mathbf{R}_k, \bm{p}-\mathbf{P}_k;\mu_k)=(1-\mu_k)^{-\frac32}G_k\Bigl(\bm{r}-\mathbf{R}_k,\frac{\bm{p}-\mathbf{P}_k}{\sqrt{1-\mu_k}}\Bigr),
    \label{eq_Gk}
\end{equation}
where the degree of isolation $\mu_k$ for the cluster is defined by
\begin{equation}
    \mu_k=\mathcal{N}_{\mathrm{frag}}-\mathcal{N}_{\mathrm{frag}}(k).
\end{equation}
Here $\mathcal{N}_{\mathrm{frag}}(k)$ is the fragment number when all nucleons in the cluster $k$ are eliminated.

Of course, this reinterpretation of the momentum width is not faithful to the standard interpretation of the wave function in quantum mechanics. One needs to carefully avoid falling into inconsistencies. See Ref.~\cite{Ono2022PLB} for an example in which this kind of reinterpretation causes a serious confusion.

In the following, the index $i$ is used for a nucleon without specifying whether it is contained in a composite cluster or not. On the other hand, the index $k$ is used for a particle that includes both cases of the center-of-mass of a composite cluster and a non-clustered nucleon. The index $j$ may be used in statements that apply to both ways of labeling nucleons ($i$) or particles ($k$).

\subsection{Activating momentum fluctuation by wave packet splitting}

A Gaussian wave packet with the time-dependent centroid variables $(\mathbf{R}_j,\mathbf{P}_j)$ can describe the exact time evolution for a particle in a harmonic oscillator potential if the constant width of the wave packet corresponds to the curvature of the potential. Therefore, we can usually expect that the motion of a nucleon in a nucleus may be approximated by a Gaussian wave packet with a fixed width parameter. In particular, a wave packet has a momentum width $\Delta p$. However, when a nucleon, initially described by a wave packet, is emitted from a nucleus, the emission usually occurs with some probability. Namely, the high-momentum component in the wave packet should be able to go out of the nucleus even if the wave packet cannot go out as a whole in case the centroid momentum $\mathbf{P}_i$ is not sufficiently large. To describe such a situation, we should allow the wave packet to split into components based on the momentum distribution $\Delta p$, and the time evolution of each component should be independent of the other components. Such a wave packet splitting in AMD was first considered by Ref.~\cite{Ono1996PRC} in which a random fluctuation is given to the momentum $\mathbf{P}_i$ when a nucleon is being emitted from a nucleus. More general treatments were introduced in Refs.~\cite{Ono1996PRC2,Ono2002PRC} by determining the wave packet splitting based on the change of the wave packet shape in the phase space according to the Vlasov equation. See Ref.~\cite{Ono2004PPNP} for a review. The momentum boost at two-nucleon collisions introduced by Ref.~\cite{Lin2016PRC} can also be regarded as a kind of wave packet splitting. However, in these approaches, the wave packet splitting was treated without a direct relation to the consideration on the zero-point kinetic energy, while the present work is the first attempt of a consistent treatment of the wave packet splitting and the zero-point kinetic energy subtraction.

Thus the present work introduces stochastic process to treat the momentum fluctuation in a consistent way with the increase of the isolation $\mu_j$. In other words, the momentum fluctuation inherent in the original wave packet ($F_j$ or $G_j$) is activated by splitting the wave packet into pieces with different momentum centroids, being induced by the increase of $\mu_j$. Since $\mu_j$ changes by both the one-body mean-field propagation of Eq.~(\ref{eq_motion}) and the two-nucleon collision process, we consider the splitting caused by both, as explained in the following two subsections. Note that the fragment number $\mathcal{N}_{\mathrm{frag}}$ is defined to be momentum dependent by Eq.~(\ref{eq_mdnf}), and thus $\mu_i$ of nucleons change at two-nucleon collisions. The continuous or sudden change of $\mu_i$ can be thought of as a progress bar of the activation of momentum fluctuation.

When a random momentum fluctuation is added to a particle for the wave packet splitting, the total energy and momentum of the system will change. The momentum conservation should be restored by many-body correlations, which in practice is treated by giving a momentum recoil to surrounding particles. We will see later in the application that the freedom of the energy source for the energy conservation can significantly affect the results.

\subsection{One-body momentum fluctuation}\label{sec_one}

\begin{figure}
    \centering
\includegraphics{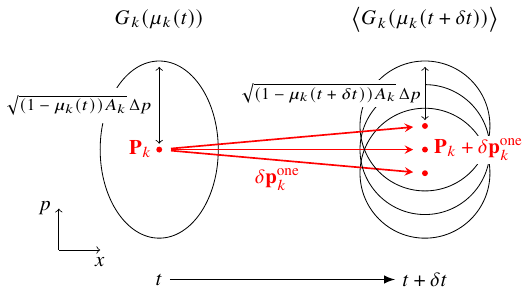}
    \caption{One-body momentum fluctuation for the $k$-th particle in the time step from $t$ to $t+\delta t$ in the phase space. According to the increase of the degree of isolation $\mu_k$, the momentum width of the Gaussian phase space distribution $G_k$ decreases and a random fluctuation $\delta\mathbf{p}_k^{\text{one}}$ is added to the centroid $\mathbf{P}_k$.}
    \label{fig_onebody}
\end{figure}

Let us first formulate the wave packet splitting induced by the deterministic motion of the particles in the mean field according to Eq.~(\ref{eq_motion}). As explained in Sec.~\ref{sec_Nf}, each particle $k$, which may be partially isolated, is interpreted as having a Gaussian phase space distribution $G_k(\bm{r}-\mathbf{R}_k, \bm{p}-\mathbf{P}_k;\mu_k)$ defined by Eq.~(\ref{eq_Gk}). According to the increase of the degree of isolation $\mu_k$ from time $t$ to $t+\delta t$ due to the equation of motion, a random fluctuation $\delta\mathbf{p}^{\text{one}}_k$ is added to the centroid $\mathbf{P}_k$ as illustrated in Fig.~\ref{fig_onebody}. This fluctuation should be introduced to satisfy
\begin{equation}
\begin{aligned}
    G_k&(\bm{r}-\mathbf{R}_k, \bm{p}-\mathbf{P}_k;\mu_k(t))\\
    &=\Bigl<G_k(\bm{r}-\mathbf{R}_k, \bm{p}-\mathbf{P}_k-\delta\mathbf{p}^{\text{one}}_k;\mu_k(t+\delta t))\Bigr>,
\end{aligned}
\end{equation}
where the brackets $\langle ... \rangle$ stand for the average over the events of $\delta\mathbf{p}^{\text{one}}_k$. Considering up to the second moment, the mean and variance of $\delta\mathbf{p}^{\text{one}}_k$ are determined as
\begin{align}
    \langle \delta\mathbf{p}^{\text{one}}_k \rangle&=0,\\
    \langle (\delta\mathbf{p}^{\text{one}}_k)^2 \rangle&=3A_k\Delta p^2(\mu_k(t+\delta t)-\mu_k(t)),
\end{align}
and the direction of $\delta\mathbf{p}^{\text{one}}$ is assumed to be isotropic. Corresponding to the adjustment of the $T_0$ parameter, the factor $3A_k\Delta p^2$ above can be replaced by $2M_k T_0$, where $M_k$ is the mass of the particle $k$. This one-body fluctuation $\delta\mathbf{p}^{\text{one}}_k$ is considered only when $\mu_k(t+\delta t)>\mu_k(t)$.

To conserve the momentum, an environment Env$_{k}^P$ is defined for the particle $k$. With the intention to find Env$_{k}^P$ that is close enough and which the particle $k$ is moving away from, a priority value $\mathcal{P}(l,k)$ is defined for each particle $l$ ($l\neq k$) as
\begin{equation}
    \mathcal{P}(l,k)=(\mathbf{R}_l-\mathbf{R}_k)^2\frac{1.2-\cos\theta_{l,k}}{A_l},
\end{equation}
where $\theta_{l,k}$ is the angle of $\mathbf{R}_l-\mathbf{R}_k$ and $\mathbf{P}_l-\mathbf{P}_k$, and $A_l$ is the number of the nucleons within 4 fm of the center-of-mass of particle $l$. The particle $l_{\text{min}}$ has the lowest $\mathcal{P}(l,k)$, then the particle $l_{\text{min}}$ and the 6--11 neighbor nucleons that are closest to the particle $l_{\text{min}}$ are selected as Env$_{k}^P$. Then the recoil momentum $-\delta\mathbf{p}^{\text{one}}_k$ is added to the center-of-mass of Env$_{k}^P$. For the energy conservation, two particles in Env$_{k}^P$ are selected; one is the particle $l_{\text{min}}$, and another particle $m_{\text{min}}$ is selected to minimize the priority value $\mathcal{P}'(m,l_{\text{min}})$ which is defined as
\begin{equation}
\begin{aligned}
    &\mathcal{P}'(m,l_{\text{min}})=\frac{\sqrt{1+(\mathbf{R}_m-\mathbf{R}_{l_{\text{min}}})^2/(2\ \text{fm})^2}}{1-\exp(-(\mathbf{P}_m/M_m-\mathbf{P}_{l_{\text{min}}}/M_l)^2/({0.1c})^2)},
\end{aligned}
\end{equation}
with an intention to choose smaller relative distance and larger relative velocity. The relative momentum between the two particles $l_{\text{min}}$ and $m_{\text{min}}$ is scaled to conserve the energy, to achieve a numerical precision of the order of 0.1 MeV in the total energy. It is still possible that the energy conservation is not possible even if the relative momentum is eliminated. In such a case, the fluctuation $\delta\mathbf{p}^{\text{one}}_k$ is set to be 0.

\subsection{Momentum fluctuation at two-nucleon collisions}\label{sec_coll}
Here we consider a general two-nucleon collision $N_1+N_2+B_1+B_2\rightarrow C_1+C_2$, where the scattered two nucleons $N_1$ and $N_2$ may form clusters $C_1$ and $C_2$ in the final state with particles around them. Let us label the nucleons $N_1$ and $N_2$ by the indices $i=1$ and $i=2$, respectively. In the standard treatment without activation of momentum fluctuations, the momentum centroids $\bm{P}_1$ and $\bm{P}_2$ are scattered to intermediate momenta as
\begin{equation}
    \bm{P}_i'=\frac{1}{2}(\bm{P}_1+\bm{P}_2)\pm p_{\text{rel}}\hat{\bm\Omega}.
\end{equation}
where the unit vector $\hat{\bm\Omega}$ represents the chosen scattering angle, while the relative momentum $p_{\text{rel}}$ is to be determined below. Then each nucleon $N_i$ ($i=1$ and 2) may form a cluster with other particle(s) $B_i$. This is achieved by moving the nucleon wave packets to the same phase space point $(\mathbf{R}^{\text{fin}}_i,\mathbf{P}^{\text{fin}}_i)$, which is the center-of-mass of the $N_i+B_i$ subsystem. The total energy $\mathcal{H}$ of the final state after forming clusters is conserved by adjusting the value of $p_{\text{rel}}$. With this standard procedure, the momentum $\bm{q}=\frac{1}{2}(\bm{P}_2-\bm{P}_1)+p_{\text{rel}}\hat{\bm\Omega}$ is transferred from the $N_2+B_2$ subsystem to the $N_1+B_1$ subsystem, while all other nucleons in the system remain unaffected by the collision.

In the present model, the degree of isolation $\mu_i$ of a nucleon is measured in the phase space with the momentum dependence of $\mathcal{N}_{\mathrm{frag}}$, aiming to activate fluctuations at two-nucleon collisions. Due to the scattering of the nucleon momenta from $(\bm{P}_1, \bm{P}_2)$ to $(\bm{P}'_1, \bm{P}'_2)$, the fragment number changes from $\mathcal{N}_{\mathrm{frag}}$ to $\mathcal{N}'_{\mathrm{frag}}$, and thus the total change of the degree of isolation is
\begin{equation}
    \sum_i(\mu'_i-\mu_i)=\mathcal{N}'_{\mathrm{frag}}-\mathcal{N}_{\mathrm{frag}}.
\end{equation}
To simplify the computation, we do not calculate the changes of isolation $\mu'_i-\mu_i$ for each nucleon $i$ but the total change is decomposed with some weights as
\begin{equation}
    \mu'_i-\mu_i=\frac{1-\mu_i}{(1-\mu_1)+(1-\mu_2)}(\mathcal{N}'_{\mathrm{frag}}-\mathcal{N}_{\mathrm{frag}})\ \ \mathrm{for}\ i=1,2,
\end{equation}
under the assumption that the less isolated nucleon can change its isolation more. Based on this change of isolation, the momentum fluctuation is activated below.

\begin{figure}[htb]
    \centering
    \includegraphics[width=\columnwidth]{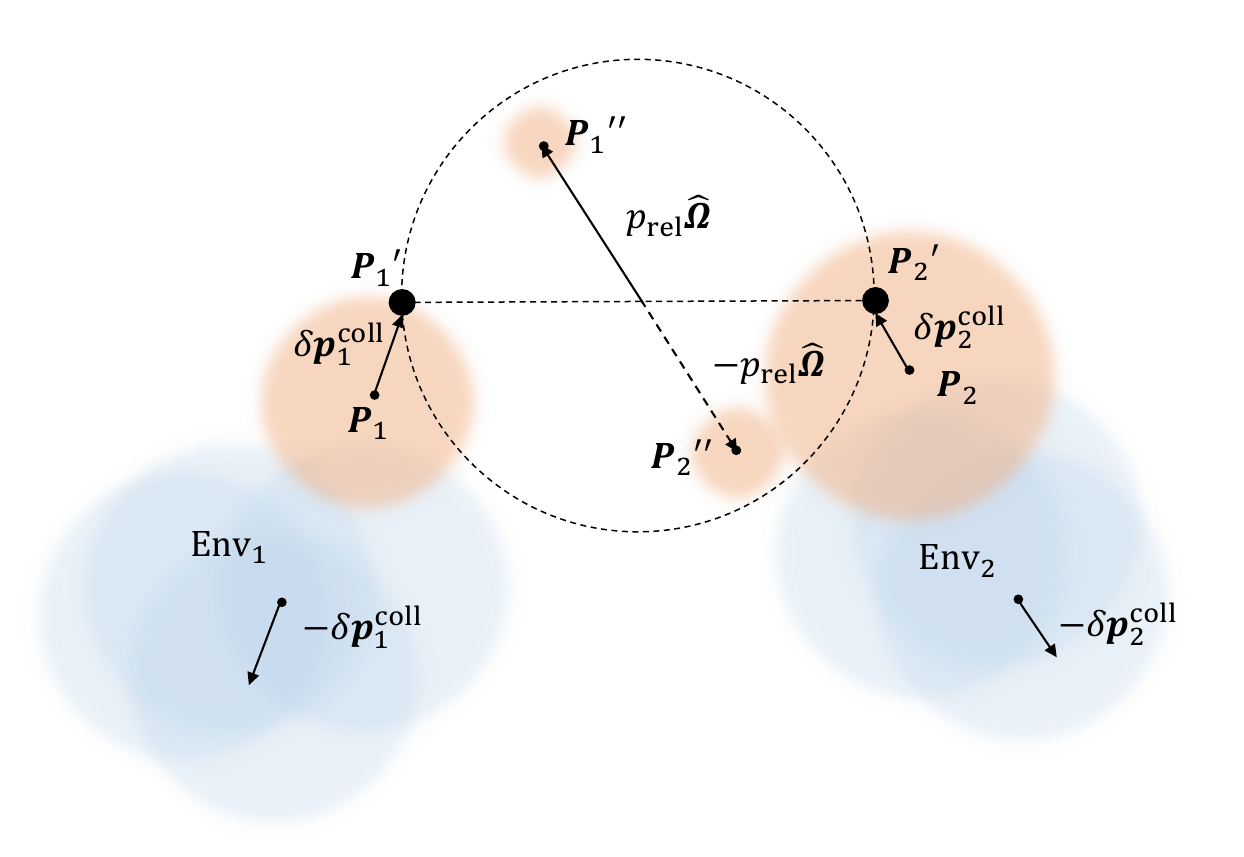}
    \caption{Activation of momentum fluctuation in a two-nucleon collision $N_1+N_2\to N_1+N_2$ in the momentum space form the initial momenta $\bm{P}_1$ and $\bm{P}_2$ to the final momenta $\bm{P}_1''$ and $\bm{P}_2''$. The orange circles refer to the two colliding nucleons, and the blue circles refer to the surrounding nucleons. This figure explains a case that the two colliding nucleons become more isolated after the two-nucleon collision. According to the increase of the degree of isolation $\mu_i$, the momentum width of both colliding nucleons (the radius of the orange circles) decrease.}
    \label{fig_coll}
\end{figure}

In the collision procedure, test particles that follow the Wigner distribution function of Eq.~(\ref{eq_dis}) or Eq.~(\ref{eq_testp}) are utilized for several purposes. In this case, the nucleon $i$, before the scattering, is associated with a distribution function $F_i(\bm{r}-\bm{R}_i, \bm{p}-\bm{P}_i)$ which can take into account the Pauli principle more appropriately than a Gaussian distribution. When the nucleon is partially or fully isolated, the distribution is reinterpreted to be $F_i(\bm{r}-\bm{R}_i, \bm{p}-\bm{P}_i;\mu_i)$ as defined by Eq.~(\ref{eq_Fi}). According to the change of isolation from $\mu_i$ to $\mu'_i$ by the scattering, the momentum fluctuation $\delta \bm{p}^{\text{coll}}_i$ should be added to $\bm{P}_i$ to satisfy
\begin{equation}
    F_i(\bm{r}-\bm{R}_i, \bm{p}-\bm{P}_i;\mu_i)=\Bigl< F_i(\bm{r}-\bm{R}_i, \bm{p}-\bm{P}_i-\delta \bm{p}^{\text{coll}}_i;\mu'_i)\Bigr>,
\end{equation}
where the brackets $\langle...\rangle$ stand for the average over the events of $\delta \bm{p}^{\text{coll}}_i$. By using the test particle $(\bm{r}^{\text{tp}}_i, \bm{p}^{\text{tp}}_i)$ that has been sampled following the distribution function $F_i(\bm{r}-\bm{R}_i, \bm{p}-\bm{P}_i)$, the random fluctuation can then be selected as
\begin{equation}
    \delta \bm{p}^{\text{coll}}_i=(\mu'_i-\mu_i)(\bm{p}_i^{\text{tp}}-\bm{P}_i)\ \ \mathrm{for}\ i=1,2.
\end{equation}
As shown in Fig.~\ref{fig_coll}, these momentum fluctuations are given to the nucleons before the scattering, and the intermediate momenta of the scattered nucleons are now written as
\begin{equation}
    \bm{P}_i''=\frac{1}{2}(\bm{P}_1+\bm{P}_2+\delta\bm{p}_1^{\text{coll}}+\delta\bm{p}_2^{\text{coll}})\pm p_{\text{rel}}\hat{\bm\Omega}.
\end{equation}
The cluster formation is then considered as explained above.

An important choice of the model is the degrees of freedom for the conservation of energy and momentum, which would otherwise be violated by the fluctuation $\delta \bm{p}^{\text{coll}}_i$. The present model chooses to adjust the relative momentum $p_{\mathrm{rel}}$ for the energy conservation, to achieve a numerical precision of the order of 0.1 MeV in the total energy. For energetic collisions, we expect that the relative motion is a sufficient source of energy to successfully achieve the energy conservation. On the other hand, for the momentum conservation, an environment $\text{Env}^P_i$ is defined for each scattered nucleon $i$ by selecting the particles within relative distance less than 5 fm and the relative velocity less than $0.2c$, and the recoil momentum $-\delta \bm{p}^{\text{coll}}_i$ is added to the center-of-mass of $\text{Env}^P_i$. It should be emphasized that the recoil momenta are given to subsystems that are not directly involved in the process of $N_1+N_2+B_1+B_2\rightarrow C_1+C_2$.

Ref.~\cite{Lin2016PRC} by Lin \textit{et al.}\ introduced the Fermi boost as a method to activate the momentum fluctuation in two-nucleon collisions. One difference between the present work and Ref.~\cite{Lin2016PRC} is that Ref.~\cite{Lin2016PRC} generates the fluctuated momentum by the Gaussian distribution, while the present work generates the $\delta \bm{p}^{\text{coll}}$ by the test particle method in order to consider a more reasonable distribution with antisymmetrization. Another difference is that the present work keeps the consistency between the degree of isolation $\mu$ and the activation of momentum fluctuations, which avoids double counting of fluctuation. Furthermore the idea to deal with the energy conservation is also different. In the present work, the energy for the fluctuation is supplied from the relative motion between the scattered two nucleons, while Ref.~\cite{Lin2016PRC} defines a temporary cluster in the coordinate space and its internal state is adjusted for the energy conservation.

In the present work, to activate the fluctuation $\delta \bm{p}_i^{\text{coll}}$ at two-nucleon collisions, it is essential to introduce the momentum dependence of the fragment number $\mathcal{N}_{\mathrm{frag}}$ with the momentum dependent factor $c(v_{ij},d_{ij})$ as defined by Eq.~(\ref{eq_mdnf}). On the other hand, in the calculations with $\delta \bm{p}_i^{\text{coll}}$ turned off, we choose $\mathcal{N}_{\mathrm{frag}}$  to be momentum independent, by setting $c(v_{ij},d_{ij})=1$.

\section{Results}\label{sec_result}

\begin{figure}[htp]
    \centering
    \includegraphics[width=\columnwidth]{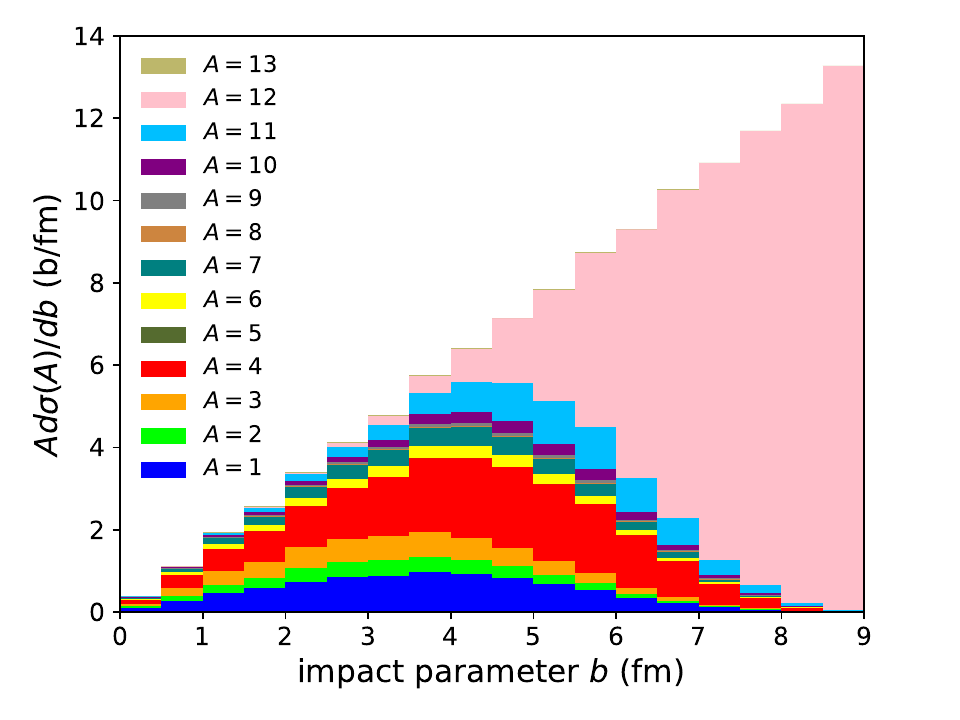}
    \caption{The mass weighted fragment production cross section $Ad\sigma(A)/db$ in the form of the impact parameter distribution, from the AMD calculation with $\delta {\mathbf p}^{\text{one}}$ and $\delta \bm{p}^{\text{coll}}$ for the reaction of $\nuc[12]{C}+\nuc[12]{C}$ at 95 MeV/nucleon.}
    \label{fig_frag}
\end{figure}

Reactions of $\nuc[12]{C}+\nuc[12]{C}$ at the beam energy of around $100$ MeV/nucleon are simulated by the AMD model with different methods of momentum fluctuations. To show the global character of fragmentation in the $\nuc[12]{C}+\nuc[12]{C}$ reaction at 95 MeV/nucleon, Fig.~\ref{fig_frag} displays the mass-weighted fragment production cross sections $Ad\sigma(A)/db$ as a function of the impact parameter $b$. Compositions of different fragment masses $A$ are accumulated in the figure. It can be seen that $\alpha$ particle production ($A=4$) is a dominant fragmentation channel at all impact parameters, while one-nucleon removal reaction ($A=11$) occurs with a considerable probability at $b\gtrsim 4$ fm. This figure shows the calculated result with both $\delta\mathbf{p}^{\text{one}}$ and $\delta\bm{p}^{\text{coll}}$ activated. For the same reaction system, in Ref.~\cite{tian2018}, the AMD calculation without activation of momentum fluctuations has been compared with the experimental data, with a reasonable success of overall reproduction of various clusters and fragments.

In the following, we focus on the single-nucleon removal channel which produces an $A=11$ residual nucleus from the projectile. The momentum of the residual nucleus should in principle reflect the single-nucleon momentum distribution in the initial nucleus, and this pure reaction channel is a suitable probe of the improvement of momentum fluctuation activation. 

\subsection{Effect of activation of momentum fluctuations}\label{sec_result1}
\begin{figure}
    \includegraphics[width=\columnwidth]{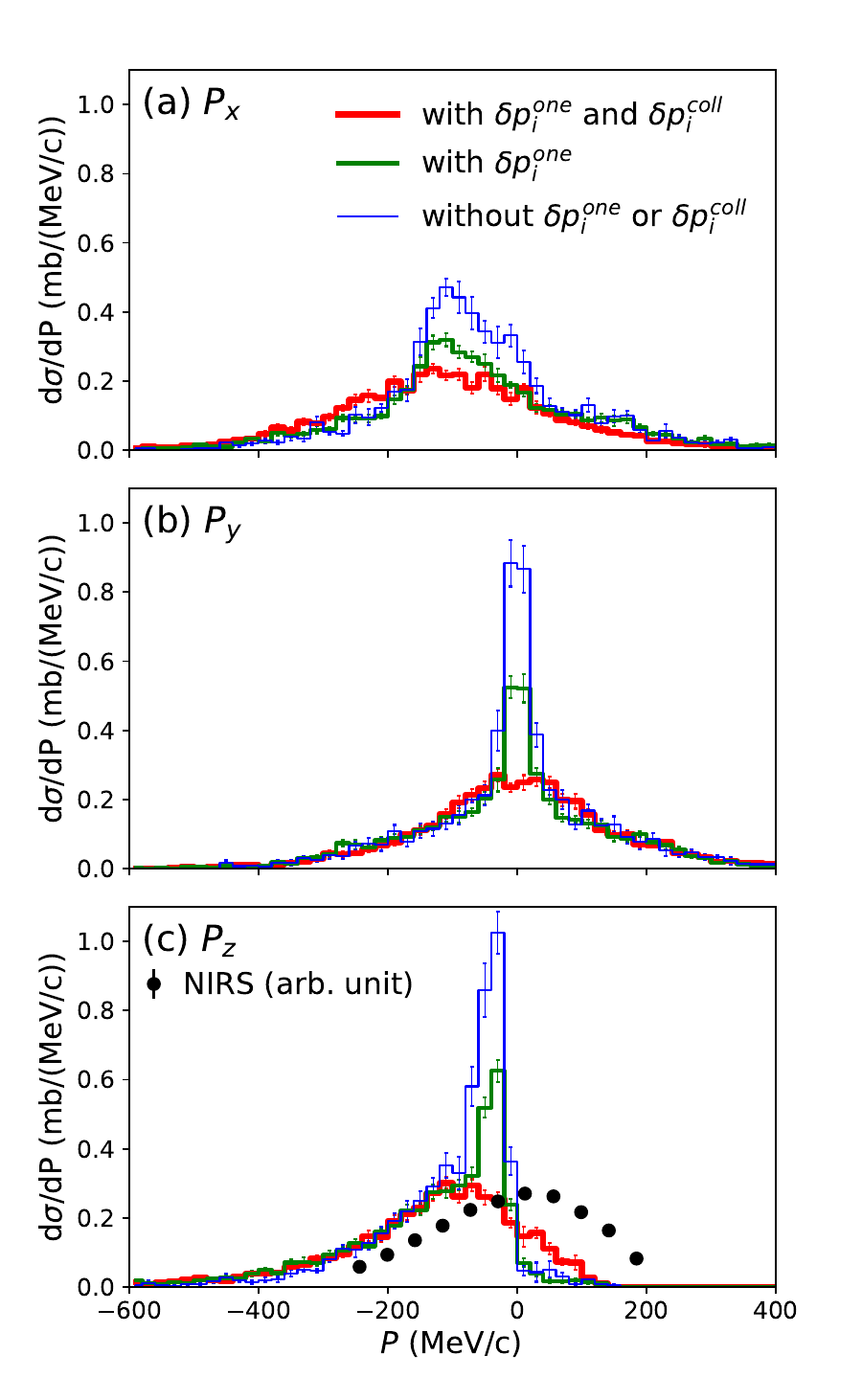}
    \caption{The distribution of the momentum components $(P_x, P_y, P_z)$ of $\nuc[11]{B}$ fragments in the projectile rest frame, for the reaction of $\nuc[12]{C}+\nuc[12]{C}$ at 100 MeV/nucleon. The blue thin histogram is the AMD result without either $\delta\mathbf{p}^{\text{one}}$ or $\delta\bm{p}^{\text{coll}}$, the green histogram is with only $\delta\mathbf{p}^{\text{one}}$ activated, and the red thick histogram is with both $\delta\mathbf{p}^{\text{one}}$ and $\delta\bm{p}^{\text{coll}}$ activated. The black points in panel (c) show the experimental data from Ref.~\cite{Momota2023PS} which have been normalized by the total cross section of the red thick histogram.}
    \label{fig_11B_C}
\end{figure}

For $\nuc[11]{B}$ fragments produced by removing one proton from the $\nuc[12]{C}$ projectile, the panels of Fig.~\ref{fig_11B_C} show the distribution of the momentum components $P_x$, $P_y$ and $P_z$, respectively, in the projectile rest frame. We take the beam direction as the $z$ axis, and the $x$ axis is the transverse direction in the reaction plane. The $y$ direction is perpendicular to the reaction plane.

\begin{figure}
    \centering
    \includegraphics[width=\columnwidth]{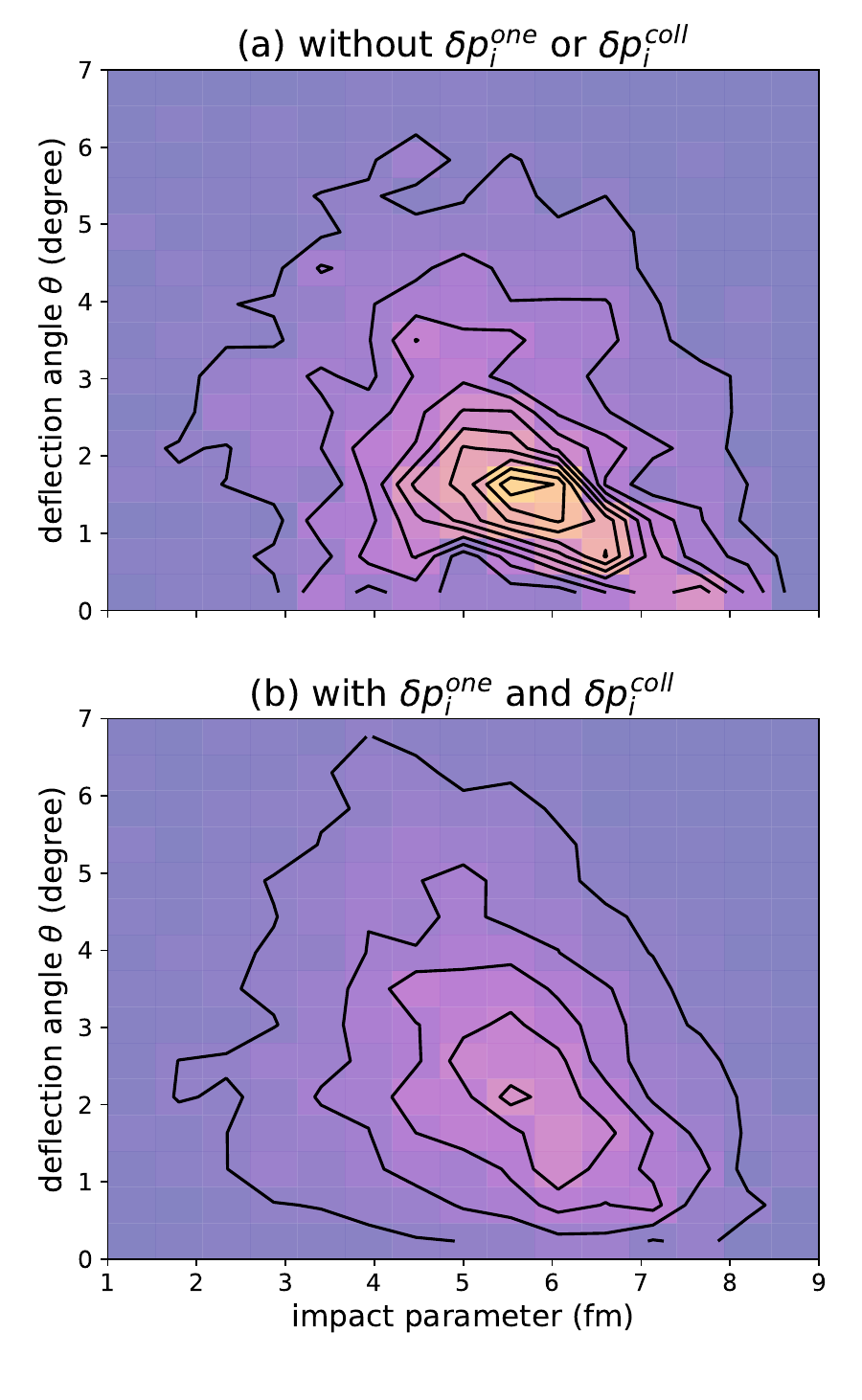}
    \caption{The distribution of the deflection angle and the impact parameter of $\nuc[11]{B}$ fragments from the $\nuc[12]{C}+\nuc[12]{C}$ reaction at 95 MeV/nucleon. Panel (a) is without either $\delta {\mathbf p}^{\text{one}}$ or $\delta \bm{p}^{\text{coll}}$, while panel (b) is with both $\delta {\mathbf p}^{\text{one}}$ and $\delta \bm{p}^{\text{coll}}$ activated.}
    \label{fig_theta_b}
\end{figure}

\begin{figure}
    \centering
    \includegraphics[width=\columnwidth]{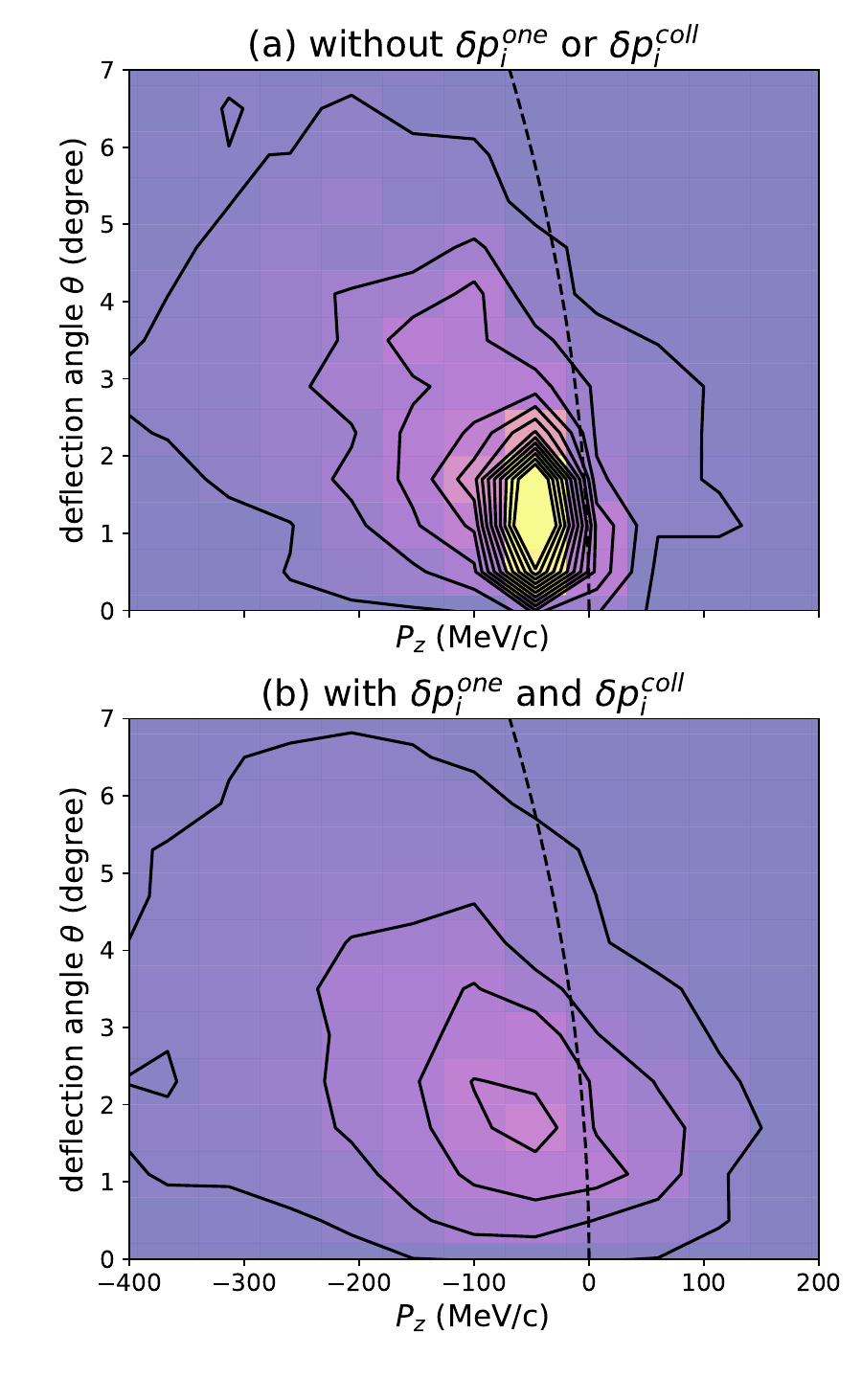}
    \caption{The distribution of $P_z$ and the deflection angle of $\nuc[11]{B}$ fragments from the $\nuc[12]{C}+\nuc[12]{C}$ reaction at 95 MeV/nucleon. Panel (a) is without either $\delta {\mathbf p}^{\text{one}}$ or $\delta \bm{p}^{\text{coll}}$, while panel (b) is with both $\delta {\mathbf p}^{\text{one}}$ and $\delta \bm{p}^{\text{coll}}$ activated. The dashed line refers to $P_z=-\sqrt{2ME_{\text{lab}}}\sin^2 \theta$, which represents the elastic scattering.}
    \label{fig_pz_theta}
\end{figure}

First, let us focus on the blue thin histograms in the panels of Fig.~\ref{fig_11B_C}, which represent the result without activation of momentum fluctuations. It can be seen that the $P_z$ distribution in panel (c) is very sharp compared to the experimental data taken at National Institute of Radiological Sciences (NIRS) \cite{Momota2023PS} shown by black points in arbitrary normalization. In particular, the calculation hardly produces the residual $\nuc[11]{B}$ nucleus at velocities faster than the beam ($P_z>0$) even though the distribution extends to the slower side $(P_z<0)$. This is in contrast to the experimental data which extend broadly to the side of $P_z>0$ as well as to the side of $P_z<0$. The Gaussian wave packet of the removed proton has a momentum width of $\Delta p=\hbar\sqrt{\nu}=78.9$ MeV/$c$, but the calculated result indicates that this width is not properly taken into account in the recoil momentum of the residue nucleus, in this calculation without activation of momentum fluctuations. The $P_y$ distribution in panel (b) also shows a sharp peak at $P_y=0$. The $P_x$ distribution in panel (a) is broader, which is due to the deflection angle that depends on the impact parameter. The peak position shifted to the negative side of $P_x$ indicates an attractive interaction. Fig.~\ref{fig_theta_b}(a) shows a strong correlation between the deflection angle and the impact parameter in the 2D distribution of these quantities. Fig.~\ref{fig_pz_theta}(a) shows the 2D distribution of $P_z$ and the deflection angle, which is very concentrated around a point, together with a broader distribution with some correlation, in this calculation without activation of momentum fluctuations.

Next, the green histograms in the panels in Fig.~\ref{fig_11B_C} are obtained by the improved AMD model with only $\delta {\mathbf p}^{\text{one}}$ activated (Sec.~\ref{sec_one}), where the changes of degree of isolation $\mu$ are taken care by only the one-body momentum fluctuation $\delta {\mathbf p}^{\text{one}}$. The peak is still not wide enough in comparison to the experimental data and to the momentum width $\Delta p=78.9\ \mathrm{MeV}/c$ of the wave packet, indicating that the momentum width is not sufficiently activated by the one-body momentum fluctuation $\delta {\mathbf p}^{\text{one}}$.

\begin{figure*}
\centering
\includegraphics[width=\textwidth]{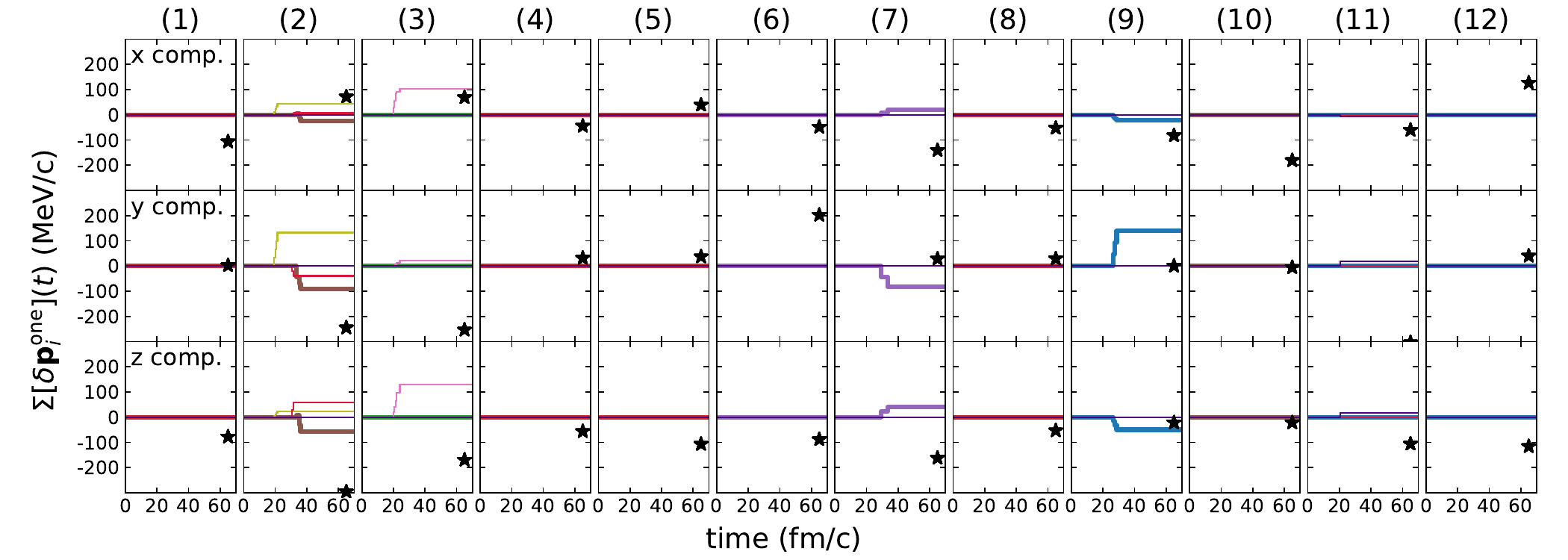}
\caption{\label{fig:cumpflone}
The time evolution of the cumulative fluctuation $\Sigma[\delta\mathbf{p}_i^{\text{one}}]$ for nucleons originating from the projectile in the $\nuc[12]{C}+\nuc[12]{C}$ reaction at 95 MeV/nucleon, for the AMD calculation with only $\delta\mathbf{p}^{\text{one}}$ activated. The figure presents 12 events in which a stable projectile-like $\nuc[11]{B}$ is produced before $t=300$ fm/$c$, with each event displayed in a separate column. The $x$, $y$ and $z$ components of the cumulative fluctuation are shown in the three rows, respectively. The thick line represents the proton emitted from the projectile, and the black star indicates the final momentum of $\nuc[11]{B}$.
}
\end{figure*}

Figure \ref{fig:cumpflone} illustrates the effect of one-body fluctuations on individual nucleons during time evolution. The cumulative fluctuation for each nucleon $i$ up to time $t$ is defined as
\begin{equation}
\Sigma[\delta\mathbf{p}_i^{\text{one}}](t)=\int_0^t
\frac{M}{M_k}\frac{\delta\mathbf{p}_k^{\text{one}}(t')}{\delta t'}dt'
,
\end{equation}
where $k$ denotes the index of the particle (which may be a cluster) that includes the nucleon $i$ at time $t'$, and $M_k$ is the mass of that particle. The figure shows the time evolution of $\Sigma[\delta\mathbf{p}_i^{\text{one}}]$ for the nucleons originating from the projectile. Twelve events, in which a stable projectile-like $\nuc[11]{B}$ is produced before $t=300$ fm/$c$, are selected and displayed in separate columns. As expected, most of the nucleons that remain in the projectile experience no fluctuations, resulting in their lines overlapping at zero in the figure. For all observed fluctuations in these events, we confirmed that the sum of the mass number $A_k$ of the fluctuated particle and the mass number of the environment $\text{Env}_k^P$ lies between 9 and 12, which suggests that the residual part of the projectile is suitably selected as $\text{Env}_k^P$. In the figure, the proton that is eventually emitted is indicated by a thick line, for which we observe that in the majority of events the proton is emitted without experiencing any fluctuations, and even in events with non-zero fluctuations the $z$ component is relatively small compared to the wave packet momentum width $\Delta p=78.9$ MeV/$c$. This may be because the fluctuation $\delta\mathbf{p}_k^{\text{one}}$ is often canceled when the residual nucleus is close to the ground state, since energy conservation cannot be restored by reducing the internal energy of the residual nucleus. This problem suggests the need for another source of energy for momentum fluctuation.

Finally, the red thick histograms in the panels of Fig.~\ref{fig_11B_C} are obtained with both momentum fluctuations $\delta {\mathbf p}^{\text{one}}$ and $\delta \bm{p}^{\text{coll}}$ activated (Sec.~\ref{sec_one} and \ref{sec_coll}). The $P_z$ distribution in panel (c) shows an encouraging result that the activation of the momentum fluctuations has greatly improved the momentum distribution of $\nuc[11]{B}$, resulting in a broader shape that is similar to the experimental data, which mainly benefits from $\delta \bm{p}^{\text{coll}}$. In particular, the distribution more reasonably extends to the high momentum side, and the region faster than the beam ($P_z>0$) now has a considerable probability. The broadening of the peak is observed also in the $P_y$ and $P_x$ distributions in panels (b) and (a).

\begin{figure*}
\centering
\includegraphics[width=\textwidth]{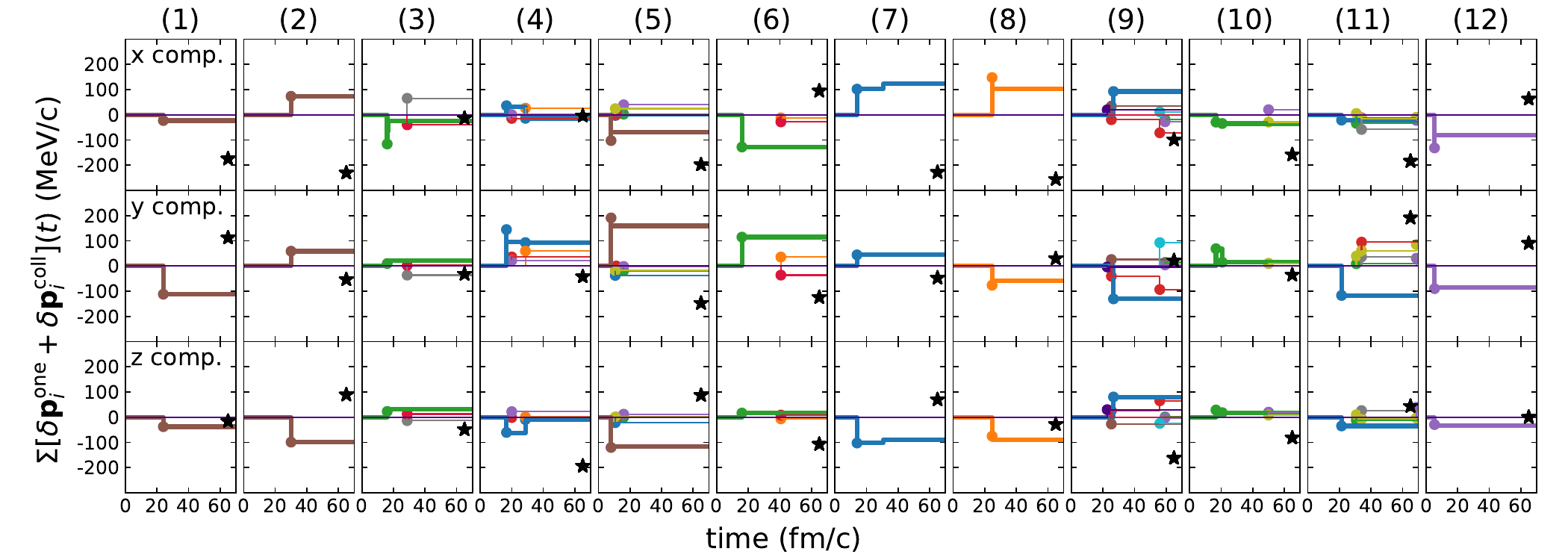}
\caption{\label{fig:cumpfl}
The time evolution of the cumulative fluctuation $\Sigma[\delta\mathbf{p}_i^{\text{one}}+\delta\bm{p}_i^{\text{coll}}]$ for nucleons originating from the projectile in the $\nuc[12]{C}+\nuc[12]{C}$ reaction at 95 MeV/nucleon, for the AMD calculation with both $\delta\mathbf{p}^{\text{one}}$ and $\delta\bm{p}^{\text{one}}$ activated. The figure presents 12 events in which a stable projectile-like $\nuc[11]{B}$ is produced before $t=300$ fm/$c$, with each event displayed in a separate column. The $x$, $y$ and $z$ components of the cumulative fluctuation are shown in the three rows, respectively. The filled circles denote changes in the cumulative fluctuation due to $\delta\bm{p}^{\text{coll}}$, while the other changes are due to $\delta\mathbf{p}^{\text{one}}$. The thick line represents the proton emitted from the projectile, and the black star indicates the final momentum of $\nuc[11]{B}$.
}
\end{figure*}

The time evolution of the cumulative fluctuation, now incorporating both $\delta\mathbf{p}^{\text{one}}$ and $\delta\bm{p}^{\text{coll}}$, is shown in Fig.~\ref{fig:cumpfl} for events where a stable $\nuc[11]{B}$ is produced before $t=300$ fm/$c$. The cumulative fluctuation is defined as
\begin{equation}
\Sigma[\delta\mathbf{p}_i^{\text{one}}+\delta\bm{p}_i^{\text{coll}}](t)
=\Sigma[\delta\mathbf{p}_i^{\text{one}}](t)
+\sum_{\text{coll}}^{(t_{\text{coll}}<t)}\delta\bm{p}_i^{\text{coll}}
,
\end{equation}
where the summation in the second term is for all collisions involving the nucleon $i$ (as $N_1$ or $N_2$) that occurred before time $t$. In the figure, the changes in the cumulative fluctuation due to $\delta\bm{p}^{\text{coll}}$ at two-nucleon collisions are indicated by solid circles. In all these 12 events, the emitted proton, indicated by a thick line, has experienced a significant fluctuation $\delta\bm{p}^{\text{coll}}$ at a two-nucleon collision at an early time, and the fluctuation is inversely correlated with the final momentum of $\nuc[11]{B}$, represented by a solid star in the figure. This observation confirms that the nucleon momentum fluctuation is appropriately reflected in the momentum of the residue in the AMD calculation when $\delta\bm{p}^{\text{coll}}$ is activated.

\begin{figure}
\centering
\includegraphics[width=\columnwidth]{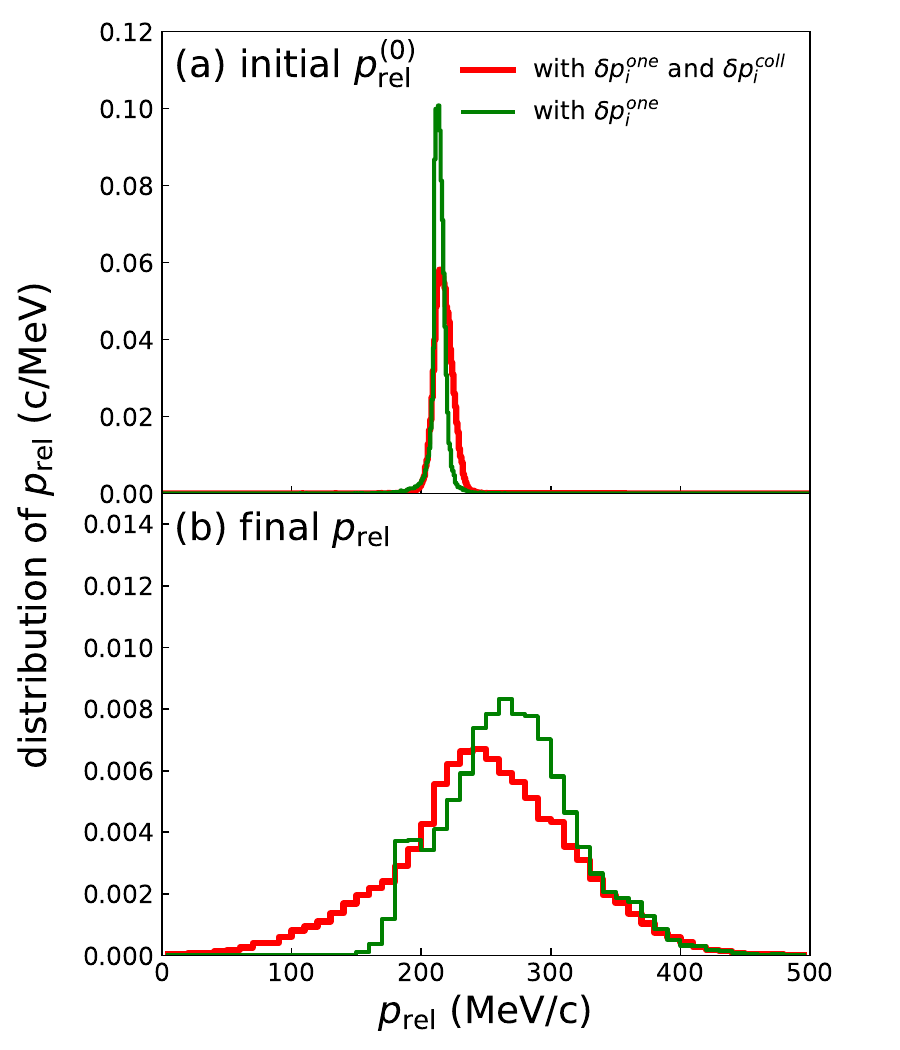}
\caption{\label{fig:preldist}
The distribution of the relative momentum between the two nucleons scattered by the first energetic two-nucleon collision in the $\nuc[12]{C}+\nuc[12]{C}$ reaction at 95 MeV/nucleon. The upper panel represents the initial relative momentum  $p_{\text{rel}}^{(0)}=\frac12|\bm{P}_1-\bm{P}_2|$ before the scattering, while the lower panel shows the final relative momentum $p_{\text{rel}}$ determined by the energy conservation. The green histogram is for the AMD calculation with only $\delta\mathbf{p}^{\text{one}}$ activated, while the red thick histogram is for the calculation with both $\delta\mathbf{p}^{\text{one}}$ and $\delta\bm{p}^{\text{coll}}$ activated.
}
\end{figure}

To illustrate how the two-nucleon collision process works with the fluctuation $\delta\bm{p}^{\text{coll}}$ and the energy conservation, Fig.~\ref{fig:preldist} shows the distribution of the relative momentum between scattered nucleons. The upper panel displays the distribution for the initial state before the scattering, while the lower panel shows it for the final state after scattering. Only the first energetic collisions are analyzed here, selecting the earliest two-nucleon collision with a collision energy $E_{NN}>47.9$ MeV from each event. In the upper panel, the distribution of the initial value, defined as $p_{\text{rel}}^{(0)}=\frac12|\bm{P}_1-\bm{P}_2|$ for the momentum centroids $\bm{P}_1$ and $\bm{P}_2$, is sharply peaked near the momentum corresponding to the beam energy, as the momentum width is not taken into account in this quantity. In contrast, the relative momentum $p_{\text{rel}}$ after the scattering, displayed in the lower panel, has a much broader distribution. This occurs because the condition for the energy conservation is not simply $p_{\text{rel}}=p_{\text{rel}}^{(0)}$, due to various effects such as cluster formation, antisymmetrization, and the momentum dependence of the mean field. Furthermore, the difference between the case with $\delta\bm{p}^{\text{coll}}$ activated (red thick line) and the case without $\delta\bm{p}^{\text{coll}}$ (green line) reveals that introducing the fluctuation shifts the distribution to the lower side, suggesting that the energy for the fluctuation is supplied by reducing $p_{\text{rel}}$ on average. Notably, the distribution extends to very low $p_{\text{rel}}$ values, corresponding to $\bm{P}_1''\approx\bm{P}_2''$, when the fluctuation is activated.

\begin{figure}
    \centering
    \includegraphics[width=\columnwidth]{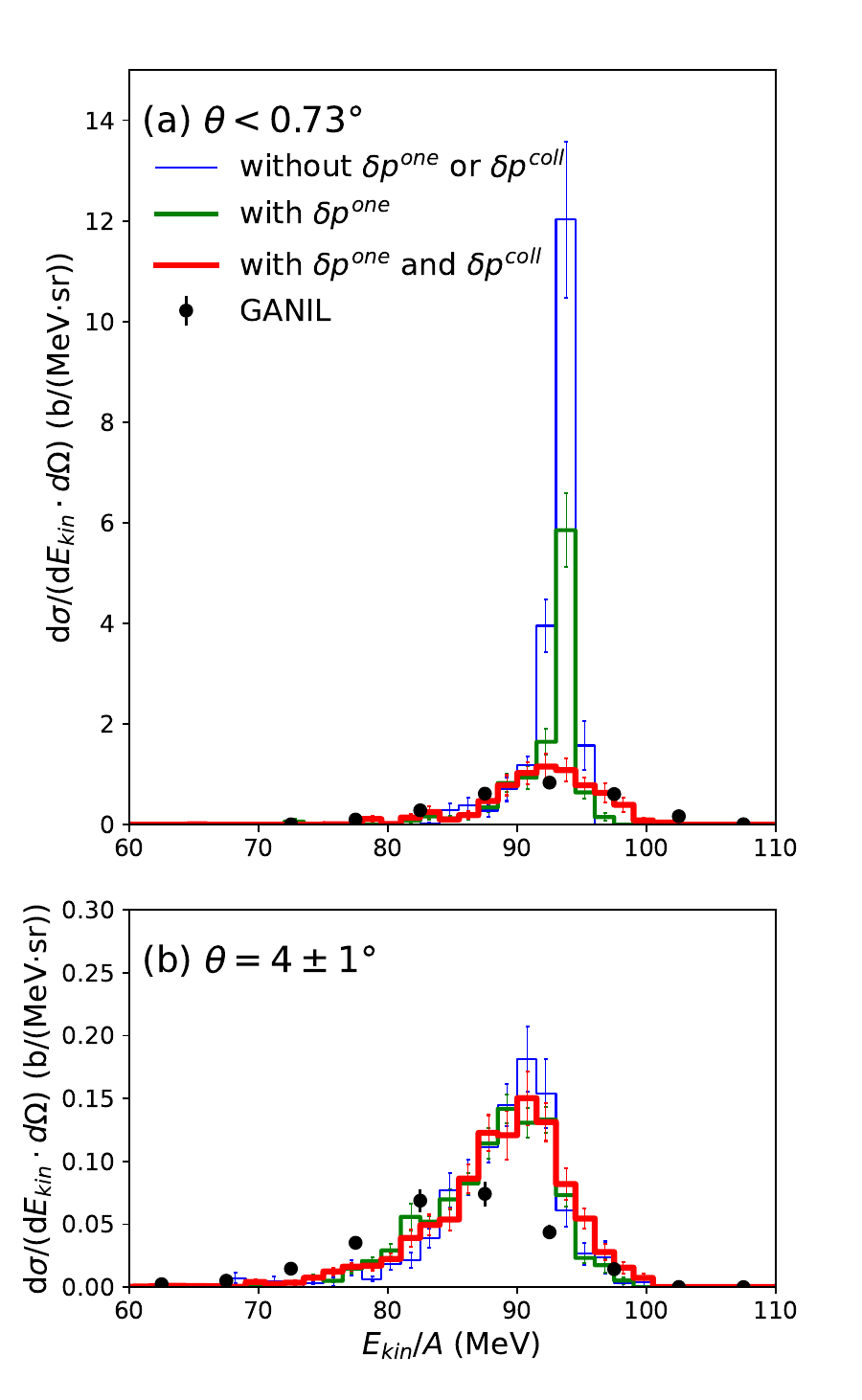}
    \caption{The kinetic energy distribution of $\nuc[11]{B}$ fragments emitted at (a) $\theta < 0.73^\circ$ and (b) $\theta =4\pm1 ^\circ$ in the reaction of $\nuc[12]{C}+\nuc[12]{C}$ at 95 MeV/nucleon. The blue thin histogram is the AMD result without either $\delta\mathbf{p}^{\text{one}}$ or $\delta\bm{p}^{\text{coll}}$, the green histogram is with only $\delta\mathbf{p}^{\text{one}}$ activated, and the red thick histogram is with both $\delta\mathbf{p}^{\text{one}}$ and $\delta\bm{p}^{\text{coll}}$ activated. The black points are the experimental data from Ref.~\cite{GANIL}.}
    \label{fig_Ek0+4}
\end{figure}

\begin{figure}
  \centering
  \includegraphics[width=\columnwidth]{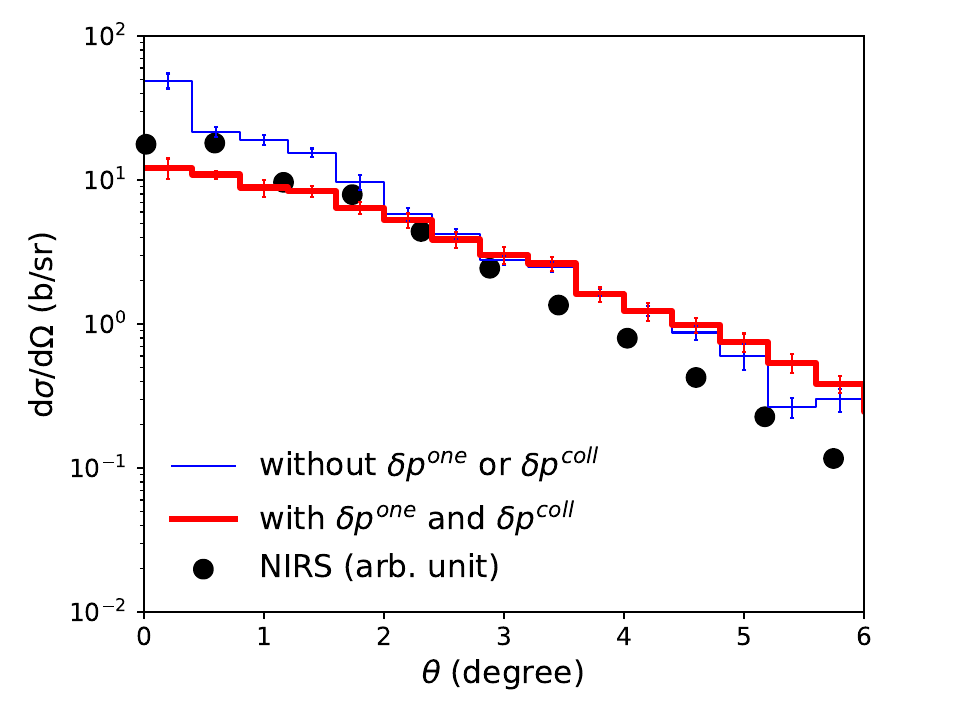}
  \caption{\label{fig:omega_NIRS}
    The angular distribution of $\nuc[11]{B}$ fragments from the $\nuc[12]{C}+\nuc[12]{C}$ reaction at 100 MeV/nucleon. The blue thin histogram represents the AMD result without either $\delta\mathbf{p}^{\text{one}}$ or $\delta\bm{p}^{\text{coll}}$, and the red thick histogram shows the result with both $\delta\mathbf{p}^{\text{one}}$ and $\delta\bm{p}^{\text{coll}}$ activated. The black points represent the experimental data from Ref.~\cite{Momota2023PS}, displayed in an arbitrary unit.
  }
\end{figure}

The absolute yield in Fig.~\ref{fig_11B_C} cannot be compared because the NIRS experimental results are arbitrarily normalized. Instead, Fig.~\ref{fig_Ek0+4} allows the comparison in the absolute scale for the kinetic energy distribution of $\nuc[11]{B}$ at (a) $\theta<0.73^\circ$ and (b) $\theta =4^\circ\pm1 ^\circ$ in the reaction of $\nuc[12]{C}+\nuc[12]{C}$ at 95 MeV/nucleon in the laboratory frame, with the experimental data taken in Grand Acc$\mathrm{\acute{e}}$l$\mathrm{\acute{e}}$rateur National d'Ions Lourds (GANIL) \cite{Dudouet2014PRC,GANIL}. Fig.~\ref{fig_Ek0+4}(a) shows a result similar to Fig.~\ref{fig_11B_C}(c) that the broad peak extends to the region faster than the beam ($E_{\text{kin}}/A>95$ MeV) when both $\delta {\mathbf p}^{\text{one}}$ and $\delta \bm{p}^{\text{coll}}$ are activated (red thick histogram). Fig.~\ref{fig_Ek0+4}(a) also shows a good agreement of the absolute value of the cross section between the experimental data and the AMD result with both $\delta {\mathbf p}^{\text{one}}$ and $\delta \bm{p}^{\text{coll}}$ activated. The peak of the $P_z$ distribution in Fig.~\ref{fig_11B_C}(c) is shifted to the negative side by about 100 MeV/$c$ which corresponds to about 4 MeV shift from 95 MeV in the energy per nucleon, and thus the shift is not seen very clearly in Fig.~\ref{fig_Ek0+4}(a) in the comparison with the GANIL data. When the momentum fluctuations are not activated (blue thin histogram), the spectrum of $\nuc[11]{B}$ is sharply peaked with a small shift from the beam energy of 95 MeV/nucleon, at this forward angle $\theta<0.73^\circ$. From Figs.~\ref{fig_theta_b}(a) and \ref{fig_pz_theta}(a), it is evident that the emission of $\nuc[11]{B}$ at this forward angle occurs in peripheral collisions ($b\gtrsim 6$ fm) with a small shift of $P_z$, which is expected for quasi-free knockout reaction where a scattered proton is simply removed. In this case, if the residual $\nuc[11]{B}$ fragment is not affected at all, it maintains its initial velocity and moves forward. On the other hand, when a fluctuation $\delta\bm{p}^{\text{coll}}$ is given to the scattered proton by the activation of momentum fluctuations, the residual $\nuc[11]{B}$ is usually chosen as the environment $\text{Env}^P$ for the momentum conservation, and is given the recoil momentum $-\delta\bm{p}^{\text{coll}}$, which will make the $\nuc[11]{B}$ momentum deviate from its initial motion. Thus, as shown by the red thick histogram in Fig.~\ref{fig_Ek0+4}(a), the yield at $\theta<0.73^\circ$ is reduced a lot by the activation of momentum fluctuations. In contrast, the peak in panel (b) for $\theta\approx 4^\circ$ is already broad before activating momentum fluctuations and it is little affected by $\delta\mathbf{p}^{\text{(one)}}$ and $\delta\bm{p}^{\text{(coll)}}$, which indicates that the emission at this angle is not simply caused by quasi-free knockout reaction. However, compared to the experimental data, the yield at $\theta\approx 4^\circ$ is larger and the peak energy is higher. This peak in the spectrum corresponds to the ridge in both panels of Fig.~\ref{fig_pz_theta} that extends from the peak in this 2D distribution ($\theta\approx1\text{--}2^{\circ}$ and $P_z\approx-(50\text{--}100)$ MeV/$c$) to larger angles, reaching $\theta\approx 4^\circ$ with $P_z\approx -125$ MeV/$c$, which is a relatively small shift of $P_z$ from the dashed line for elastic scattering. The deviation of the calculated result at $\theta\approx 4^\circ$ from the experimental data may indicate that the deflection angle in the calculation is too large, which can depend on the choice of the mean field interaction and/or some mechanisms for transverse momentum transfer due to the cluster formation in the two-nucleon collision process. The comparison in Fig.~\ref{fig:omega_NIRS} with the NIRS data with an arbitrary normalization also indicates that the present AMD calculations show broader angular distribution than the experimental data at $\theta\gtrsim 2^\circ$, already before activating momentum fluctuations.

\subsection{Decomposition of the momentum distribution}\label{sec_result2}

\begin{figure}
    \centering
    \includegraphics[width=\columnwidth]{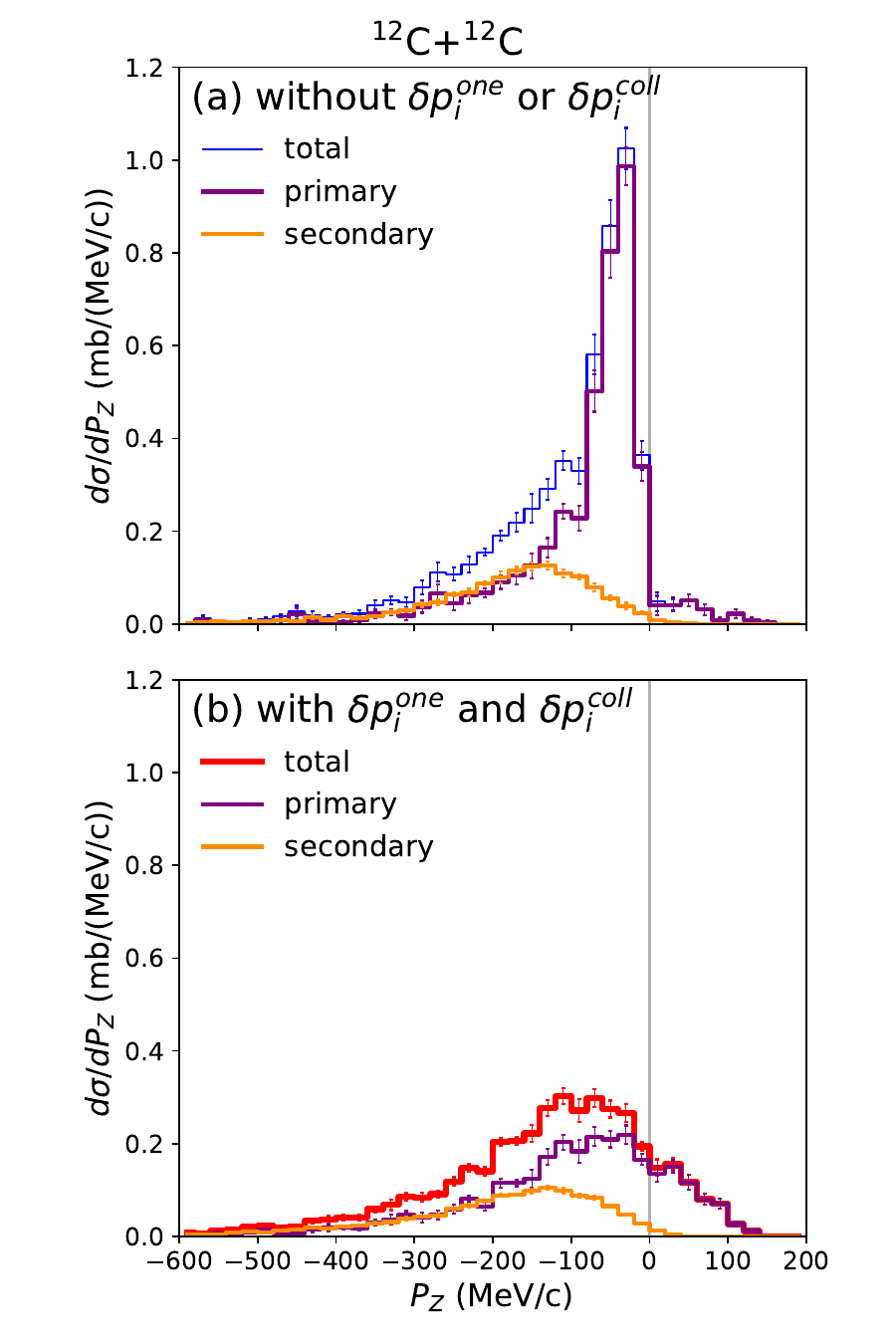}
    \caption{The primary (purple) and the secondary (orange) contributions to the momentum $P_z$ distribution of $\nuc[11]{B}$ in the projectile rest frame for the reaction of $\nuc[12]{C}+\nuc[12]{C}$ at 95 MeV/nucleon. Panel (a) is for the AMD calculation without either $\delta\mathbf{p}^{\text{one}}$ or $\delta\bm{p}^{\text{coll}}$, and panel (b) is with both $\delta\mathbf{p}^{\text{one}}$ and $\delta\bm{p}^{\text{coll}}$ activated. The total distribution is shown by the blue thin and red thick histograms.}
    \label{fig_decay_C}
\end{figure}

In Fig.~\ref{fig_decay_C}, we decompose the $P_z$ distribution into two contributions. The purple histogram is the primary products, which are the stable $\nuc[11]{B}$ fragments produced directly by the AMD calculation which is continued to $t=300$ fm/$c$, and the orange histogram is the secondary products resulting from the decay of the excited $\nuc[12]{C}$ in the statistical decay calculation. In panel (a), without momentum fluctuation activation, the primary contribution (purple histogram) has a main peak at around $-50$ MeV/$c$ and shows an asymmetric distribution that is extending more broadly to the low momentum side than to the high momentum side. The secondary contribution (orange histogram) shows a broad peak centered around $-150$ MeV/$c$. We can consider that the total distribution consists of two components, i.e., a low $P_z$ component which is broadly distributed around $-150$ MeV/$c$, and a high $P_z$ component which is sharply peaked around $-50$ MeV/$c$. The secondary contribution purely includes only the low $P_z$ component, while we may interpret that the primary contribution contains both components, i.e., its broad extension to the low momentum side is assumed to have the same physical origin as the statistical decay of excited $\nuc[12]{C}$ nuclei. This is reasonable because in some events the evaporation of a proton from an excited $\nuc[12]{C}$ nucleus can occur before $t=300$ fm/$c$, which is described in the AMD calculation. Thus the low $P_z$ component can be understood as due to the decay of excited $\nuc[12]{C}$ nuclei, which occurs in a long time scale after the $\nuc[12]{C}$ projectile is excited in the AMD simulation, while the high $P_z$ component can be interpreted as originating from a simple one-nucleon knockout process by an energetic $NN$ collision at an early time. Therefore, the high $P_z$ component is strongly affected by $\delta \bm{p}^{\text{coll}}$, which can be seen in the change from panel (b) to (a) in Fig.~\ref{fig_decay_C} and also in Fig.~\ref{fig_11B_C}. On the other hand, the low $P_z$ component seems to be little affected by the momentum fluctuation activation.

In Ref.~\cite{Momota2023PS}, Momota \textit{et al.}\ also pointed out that there are two components in the longitudinal momentum distribution. Their explanation was that the two components correspond to different reaction mechanisms. One is the pure abrasion channel which is similar to the quasi-free knockout. For the other component, as one probable reaction process, they considered a model by Souliotis \textit{et al.}~\cite{Souliotis1992} for a two-step reaction channel of two-nucleon removal and one-nucleon pickup. Unlike this model of Ref.~\cite{Souliotis1992}, the AMD results provide a different explanation for the latter component as argued above. A common point of the explanation of the AMD model and that of Ref.~\cite{Souliotis1992} is that the low $P_z$ component indicates a more complicated reaction mechanism than a one-step process.

\begin{figure}[htb]
    \centering
    \includegraphics[width=\columnwidth]{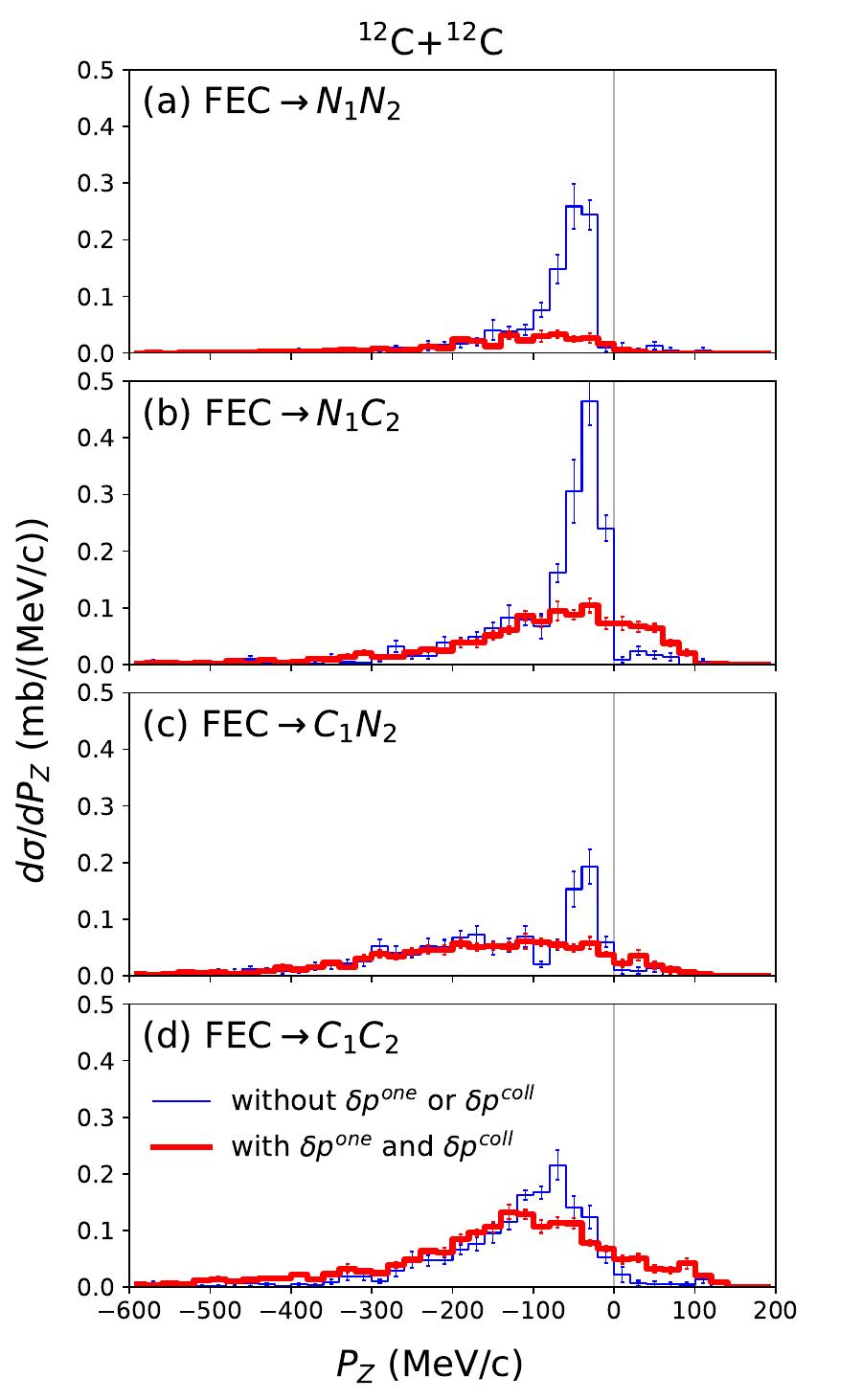}
    \caption{The $P_z$ distribution of $\nuc[11]{B}$ decomposed to panels (a)--(d) according to different cluster formation channels in the first energetic collision (FEC) for the reaction of $\nuc[12]{C}+\nuc[12]{C}$ at 95 MeV/nucleon. The blue thin histogram shows the result without either $\delta\mathbf{p}^{\text{one}}$ or $\delta\bm{p}^{\text{coll}}$, and the red thick histogram is with both $\delta\mathbf{p}^{\text{one}}$ and $\delta\bm{p}^{\text{coll}}$ activated.}
    \label{fig_fec_C}
\end{figure}

We also attempt another type of decomposition by expecting that the production of $\nuc[11]{B}$ is determined by the energetic collision at an early time, i.e., we decompose the $P_z$ distribution by classifying the events according to the cluster formation channels in the first energetic collision (FEC) as shown in Fig.~\ref{fig_fec_C}. The FEC, with the $NN$ collision energy $E_{NN}>47.9$ MeV, only occurs between a nucleon from the projectile and a nucleon from the target. Each colliding nucleon may form a cluster in the final state of the FEC, so there are four possible cluster formation channels, i.e., $N_1N_2$, $N_1C_2$, $C_1N_2$ and $C_1C_2$, where the index 1 (or 2) refers to the colliding nucleon from projectile (or target). For example, $N_1C_2$ stands for an event in which the nucleon form the projectile did not form a cluster and the the nucleon from the target formed a cluster with $A\ge2$ in the final state of the FEC. In panel (a) for the events without any cluster formation ($N_1N_2$) in the FEC, the result without any momentum fluctuation (blue thin histogram) is sharply peaked near $P_z=0$, which is similar to the high $P_z$ component discussed above. This supports the interpretation that the high $P_z$ component refers to a simple energetic two-nucleon collision which directly produces $\nuc[11]{B}$ with small energy dissipation. In panel (b) for the events ($N_1C_2$) of the FEC cluster formation only by the nucleon from the target, the result without any momentum fluctuation (blue thin histogram) is similar to that in panel (a) for $N_1N_2$, which is reasonably understood because in both cases the projectile nucleon can be simply knocked out without forming a cluster. On closer look, the peak shift from $P_z=0$ is smaller when the cluster is formed by the nucleon from the target ($N_1C_2$), which can be explained by the larger momentum transfer to the projectile nucleon when a cluster is formed in the target. In panel (d), the contribution from the events with cluster formations on both sides ($C_1C_2$) is distributed more broadly, which is similar to the low $P_z$ component discussed above for Fig.~\ref{fig_decay_C}. In most of such $C_1C_2$ events, one of the clusters must be formed by involving some nucleons in the projectile, and to produce $\nuc[11]{B}$ the formed cluster must not be directly emitted from the projectile. Instead, the cluster should stay in the projectile producing an excited $\nuc[12]{C}$ nucleus, and a proton should be emitted later in the following evolution within AMD or in the statistical decay process. In panel (c) for the events of $C_1N_2$ in the FEC, the distribution (blue thin histogram) seems to contain both the high $P_z$ and the low $P_z$ components. This mixing is due to the existence of two possibilities, either the scattered nucleon in the projectile forms a cluster $C_1$ with some other nucleons in the projectile, or with some nucleons in the target. In the former case, the $\nuc[11]{B}$ production mechanism is the same as in the $C_1C_2$ case in panel (d), and the latter case is effectively similar to the $N_1C_2$ case in panel (b). Note that the latter case can occur when the two-nucleon scattering angle is large to exchange the roles of the two nucleons.

When the momentum fluctuation is turned on in Figs.~\ref{fig_decay_C} and \ref{fig_fec_C} (red thick histogram), it is clear that the peak of the high $P_z$ becomes much broader, and also the yield of the high $P_z$ component is reduced, especially in the $N_1N_2$ case (Fig.~\ref{fig_fec_C}(a)). We consider that this is partly because the factor $p_{\text{f}}$ in Eq.~(\ref{eq_coll}) is reduced on average to supply energy for the fluctuation $\delta\bm{p}^{\text{coll}}$, resulting in the decrease of the production cross section of $\nuc[11]{B}$. Another reason may be that the momentum fluctuation activation enhances cluster formation in the FEC. In fact, the high momentum $\nuc[11]{B}$ faster than the beam ($P_z>0$) increases in the channels of cluster formation at FEC, as seen in panels (b), (c) and (d).

\begin{figure}
    \centering
    \includegraphics[width=\columnwidth]{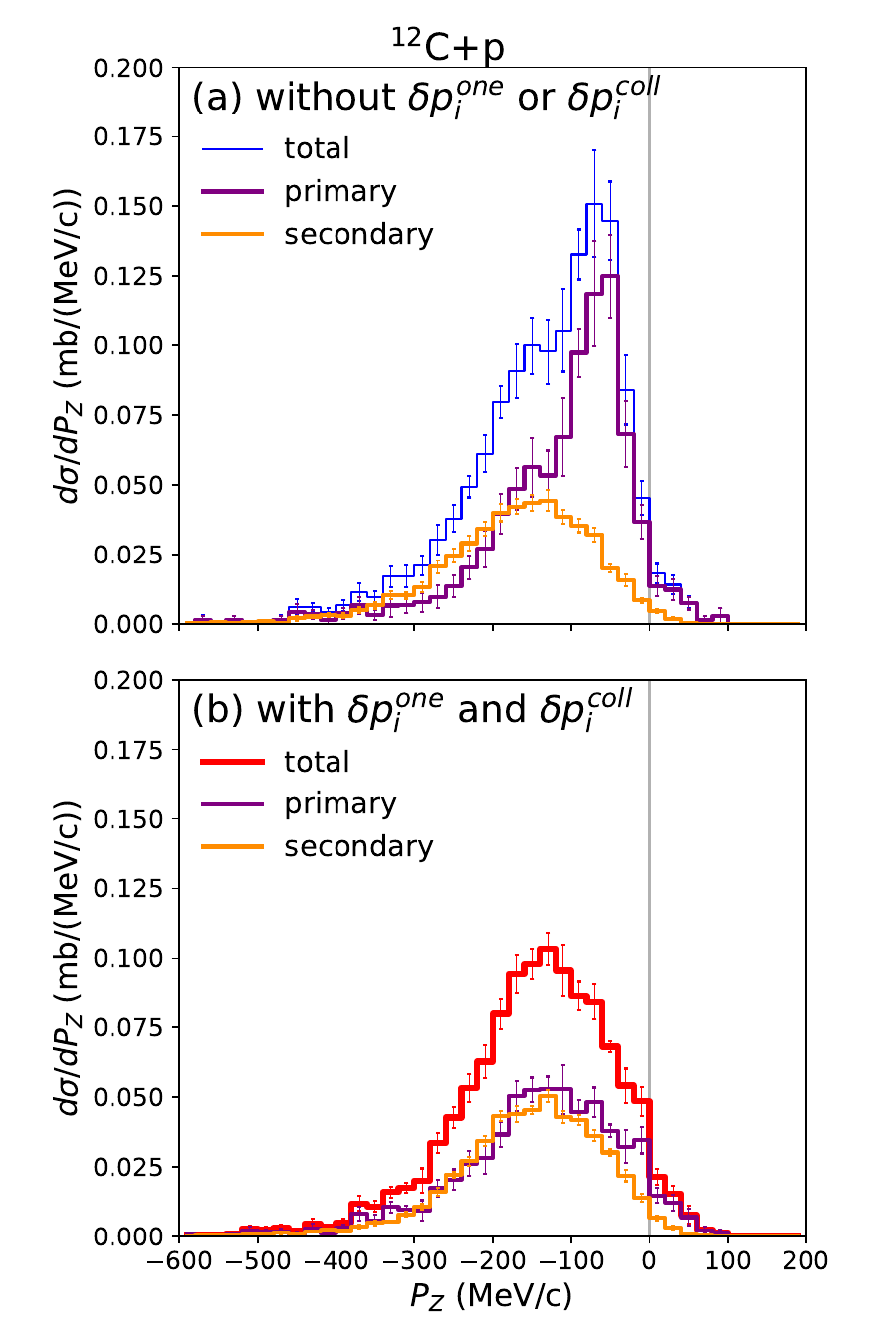}
    \caption{The primary (purple) and the secondary (orange) contributions to the momentum $P_z$ distribution of $\nuc[11]{B}$ in the projectile rest frame for the reaction of $\nuc[12]{C}+p$ at 95 MeV/nucleon. Panel (a) is for the AMD calculation without either $\delta\mathbf{p}^{\text{one}}$ or $\delta\bm{p}^{\text{coll}}$, and panel (b) is with both $\delta\mathbf{p}^{\text{one}}$ and $\delta\bm{p}^{\text{coll}}$ activated. The total distribution is shown by the blue thin and red thick histograms.}
    \label{fig_decay_p}
\end{figure}

\begin{figure}[htb]
    \centering
    \includegraphics[width=\columnwidth]{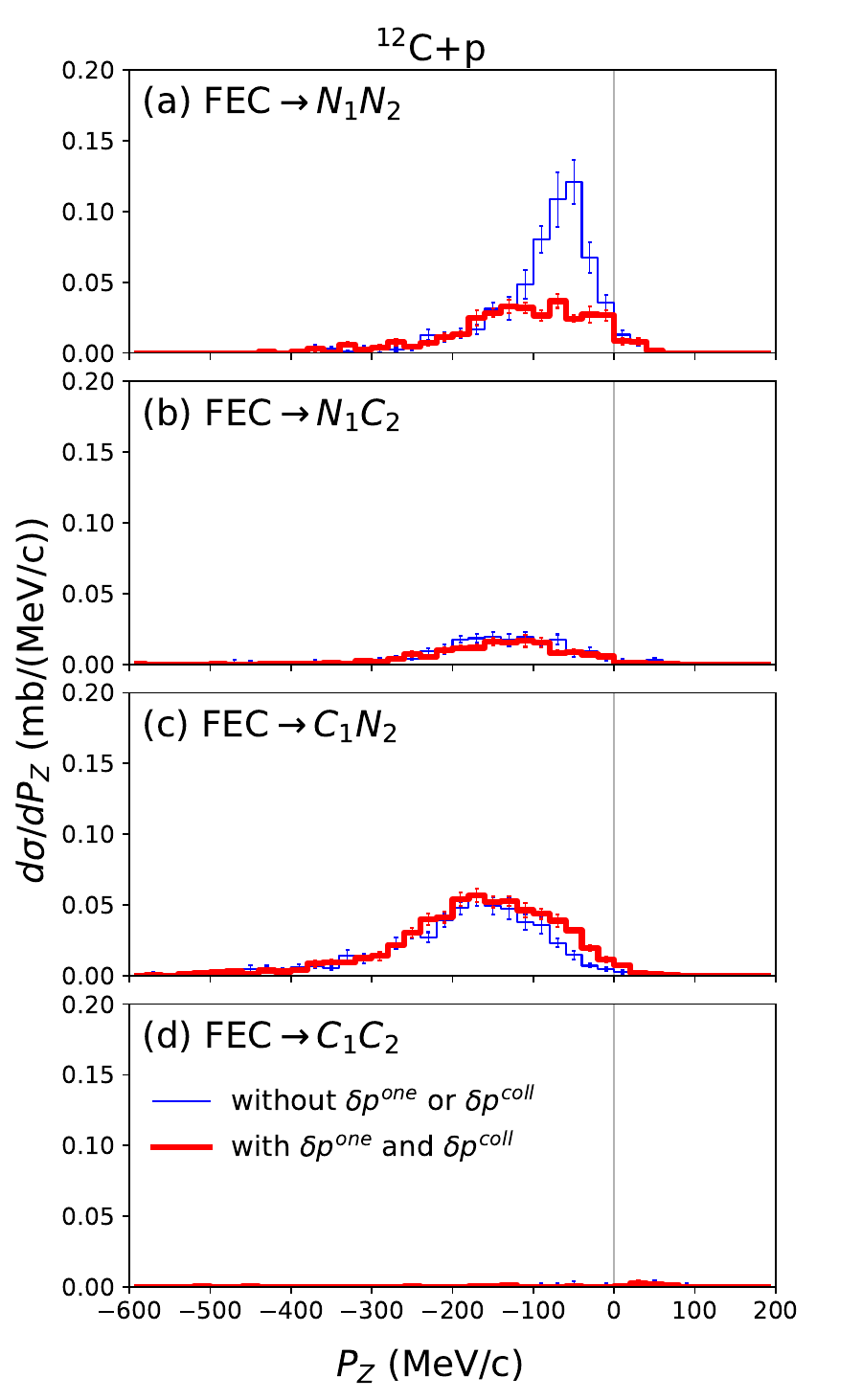}
    \caption{The $P_z$ distribution of $\nuc[11]{B}$ decomposed to panels (a)--(d) according to different cluster formation channels in the first energetic collision (FEC) for the reaction of $\nuc[12]{C}+p$ at 95 MeV/nucleon. The blue thin histogram shows the result without either $\delta\mathbf{p}^{\text{one}}$ or $\delta\bm{p}^{\text{coll}}$, and the red thick histogram is with both $\delta\mathbf{p}^{\text{one}}$ and $\delta\bm{p}^{\text{coll}}$ activated.}
    \label{fig_fec_p}
\end{figure}

The reaction of $\nuc[12]{C}+p$ at 95 MeV/nucleon was also simulated by the AMD model. In this case, because the degree of isolation $\mu$ of the target proton is already 1 in the initial state, the target proton does not have momentum width and the fluctuation is activated only for the nucleons in the projectile in the calculation with both $\delta\mathbf{p}^{\text{one}}$ and $\delta\bm{p}^{\text{coll}}$ activated. This is one of the essential differences compared to the case of the $\nuc[12]{C}$ target. The $\nuc[11]{B}$ momentum distribution is decomposed in the same way as for the $\nuc[12]{C}$ target case, i.e., Fig.~\ref{fig_decay_p} shows the decomposition to the primary and secondary contributions, and Fig.~\ref{fig_fec_p} shows the decomposition by the FEC cluster formation channels. We notice that the strength of the high $P_z$ component relative to that of the low $P_z$ component is now smaller than in the $\nuc[12]{C}$ target case. The results in these figures are understandable in the same line as our explanation above for the $\nuc[12]{C}$ target case. However, in the reaction of $\nuc[12]{C}+p$, a cluster can be formed in the FEC by always employing some nucleons in the projectile, and therefore the panels (b) and (c) of Fig.~\ref{fig_fec_p} for the $N_1C_2$ and $C_1N_2$ channels, respectively, do not contain the high $P_z$ component. For the proton target case, it is quite rare that two clusters ($C_1C_2$) are formed in the FEC as noticed in panel (d), but it is worth mentioning that the events with the highest $P_z$ ($>0$) belong to this $C_1C_2$ channel.

\begin{figure}[htb]
    \centering
    \includegraphics[width=\columnwidth]{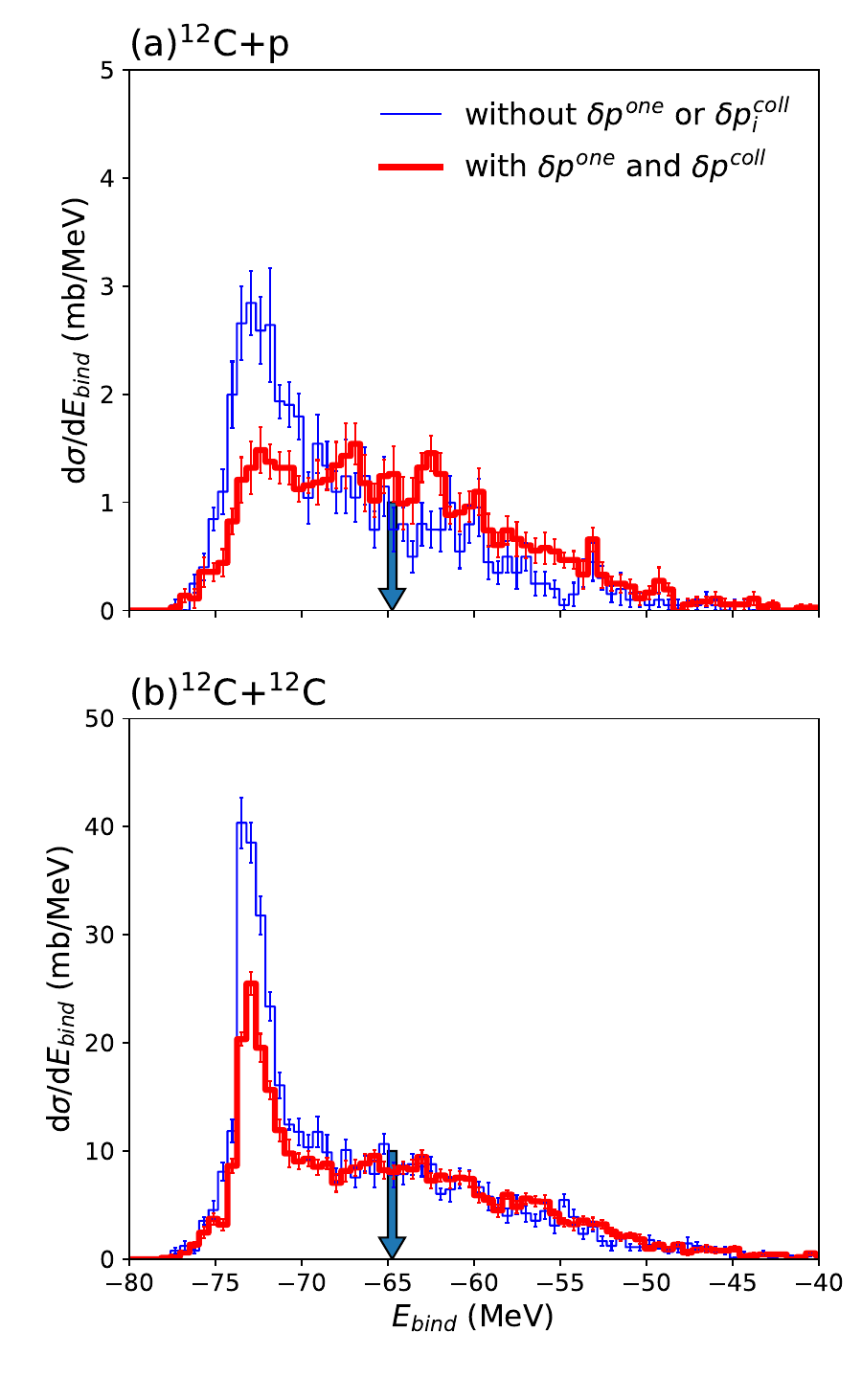}
    \caption{The binding energy distribution of primary $\nuc[11]{B}$ fragments in the final state of the AMD calculation at $t=300$ fm/$c$, in the reaction of (a)$\nuc[12]{C}+p$ and (b)$\nuc[12]{C}+\nuc[12]{C}$ at 95 MeV/nucleon. The arrow indicates the binding energy of $\nuc[10]{Be}$, which is the decay threshold.}
    \label{fig_11B_Eb}
\end{figure}

Fig.~\ref{fig_11B_Eb} shows the binding energy distribution of the excited $\nuc[11]{B}$ fragments before the statistical decay process, from the AMD calculations with the different treatments of momentum fluctuations. The decay threshold, which is the binding energy of $\nuc[10]{Be}$, is represented by the arrow. The part below the decay threshold will remain as $\nuc[11]{B}$ after the statistical decay process, while the part above the threshold is likely to decay into another nucleus. In panel (b) the distribution in the $\nuc[12]{C}+\nuc[12]{C}$ case without momentum fluctuation (blue thin histogram) has a sharp peak at about $E_{\text{bind}}=-73$ MeV. Similarly, in panel (a), the distribution in the $\nuc[12]{C}+p$ case also shows a peak at the same position but it is now lower and/or broader than in the $\nuc[12]{C}+\nuc[12]{C}$ case. This peak is likely related to the $\nuc[11]{B}$ in the high $P_z$ component which are produced by a simple knockout reaction after the FEC without cluster formation in the projectile, and this is more prominent in the $\nuc[12]{C}$ target case than in the proton target case.

As well as in the $\nuc[12]{C}$ target case, the $P_z$ peak shift is observed in the reaction of $\nuc[12]{C}+p$ in Fig~\ref{fig_decay_p}. In Ref.~\cite{Momota2018PRC}, the relationship between the peak shift of the longitudinal momentum and the fragment mass was investigated. The conclusion suggests that the peak position of $\nuc[11]{B}$ should be close to $P_z=0$. In the AMD calculation, the part of $P_z>0$ corresponding to the $\nuc[11]{B}$ fragments accelerated to a velocity higher than the beam is always underestimated. At the time of the first energetic collision (FEC), the fluctuation is symmetric with respect to $\delta \bm{p}^{\text{coll}}_i=0$, and thus the momentum distribution of the $\nuc[11]{B}$ prefragment should also be symmetric with respect to $P_z=0$ immediately after the recoil from the FEC. The peak shift is probably due to some interaction after the FEC. In fact, the peak of the $P_z$ distribution is located at a similar position in both cases of $\nuc[12]{C}+\nuc[12]{C}$ and $\nuc[12]{C}+p$, which suggests that the interaction that slows down the $\nuc[11]{B}$ residue does not strongly depend on the target mass number. This implies that the residual interaction is with the knocked out proton rather than the target nucleus.

\section{Conclusion}\label{sec_conclusion}
The present work proposes a new approach to activate the momentum fluctuation in both the one-body propagation (Sec.~\ref{sec_one}) and the two-nucleon collision process (Sec.~\ref{sec_coll}) in the AMD model. The degree of isolation $\mu$ is introduced using the fragment number function $\mathcal{N}_{\mathrm{frag}}$ which has been improved by considering the momentum dependence. Compared to some similar studies in Refs.~\cite{Ono1996PRC,Ono1996PRC2,Ono2002PRC,Lin2016PRC}, this work aims at a unified treatment of the fluctuation activated by the different mechanisms based on the change of the degree of isolation $\mu$.

The reaction of $\nuc[12]{C}+\nuc[12]{C}$ at about 100 MeV/nucleon was simulated by the AMD model. We focused on the single-nucleon knockout reaction channel which is considered as a suitable probe of the improvement. The results show that the introduced momentum fluctuation can well improve the momentum distribution of the produced $\nuc[11]{B}$ (Sec.~\ref{sec_result1}). In particular, the momentum fluctuation of the knocked out nucleon is suitably reflected to the $\nuc[11]{B}$ residue as the recoil, and consequently the peak of the longitudinal momentum distribution becomes sufficiently broad with a considerable probability of production of $\nuc[11]{B}$ faster than the beam velocity.

Then the components of the $\nuc[11]{B}$ momentum distribution were discussed (Sec.~\ref{sec_result2}). The momentum distribution of the residue consists of two components, as also observed in the experimental data \cite{Momota2023PS}. Our analysis of the AMD results indicates that the high $P_z$ component corresponds to stable $\nuc[11]{B}$ fragments produced directly by the first energetic collision without cluster formation in the projectile side, while the low $P_z$ component arises from more complicated processes, such as cluster formation in the projectile, resulting in an excited $\nuc[12]{C}$ nucleus that decays later in a long time scale.

In comparison with the experimental results, a peak shift of about 50--100 MeV/$c$ is observed in both $\nuc[12]{C}+\nuc[12]{C}$ case and $\nuc[12]{C}+p$ case. We consider that the interaction between the residue nucleus and the knocked out proton is the source of this peak shift in the calculation. The study should be extended by investigating the dependence on the beam energy and also on the combination of the projectile and the target.

Based on the present studies on simple reaction channels, the activation of the momentum fluctuations is proved to be a successful improvement to the AMD model. We expect that this can also improve the simulations in the future for various observables in more general reactions, including violent heavy-ion collisions.

\begin{acknowledgments}
The authors thank Sadao Momota for valuable discussions and comments. This work was supported in part by the National Natural Science Foundation of China under contract Nos.\ 12147101, 11890710 and 11890714, the Guangdong Major Project of Basic and Applied Basic Research No.\ 2020B0301030008, the Natural Science Foundation of Shanghai under Grant No.\ 23JC1400200, the STCSM under Grant No.\ 23590780100, the China Scholarship Council No.\ 202208310181. This work was also supported by JSPS KAKENHI Grants No.\ JP17K05432 and No.\ JP21K03528.
\end{acknowledgments}

\bibliography{AMD}

\begin{thebibliography}{74}%
\makeatletter
\providecommand \@ifxundefined [1]{%
 \@ifx{#1\undefined}
}%
\providecommand \@ifnum [1]{%
 \ifnum #1\expandafter \@firstoftwo
 \else \expandafter \@secondoftwo
 \fi
}%
\providecommand \@ifx [1]{%
 \ifx #1\expandafter \@firstoftwo
 \else \expandafter \@secondoftwo
 \fi
}%
\providecommand \natexlab [1]{#1}%
\providecommand \enquote  [1]{``#1''}%
\providecommand \bibnamefont  [1]{#1}%
\providecommand \bibfnamefont [1]{#1}%
\providecommand \citenamefont [1]{#1}%
\providecommand \href@noop [0]{\@secondoftwo}%
\providecommand \href [0]{\begingroup \@sanitize@url \@href}%
\providecommand \@href[1]{\@@startlink{#1}\@@href}%
\providecommand \@@href[1]{\endgroup#1\@@endlink}%
\providecommand \@sanitize@url [0]{\catcode `\\12\catcode `\$12\catcode
  `\&12\catcode `\#12\catcode `\^12\catcode `\_12\catcode `\%12\relax}%
\providecommand \@@startlink[1]{}%
\providecommand \@@endlink[0]{}%
\providecommand \url  [0]{\begingroup\@sanitize@url \@url }%
\providecommand \@url [1]{\endgroup\@href {#1}{\urlprefix }}%
\providecommand \urlprefix  [0]{URL }%
\providecommand \Eprint [0]{\href }%
\providecommand \doibase [0]{http://dx.doi.org/}%
\providecommand \selectlanguage [0]{\@gobble}%
\providecommand \bibinfo  [0]{\@secondoftwo}%
\providecommand \bibfield  [0]{\@secondoftwo}%
\providecommand \translation [1]{[#1]}%
\providecommand \BibitemOpen [0]{}%
\providecommand \bibitemStop [0]{}%
\providecommand \bibitemNoStop [0]{.\EOS\space}%
\providecommand \EOS [0]{\spacefactor3000\relax}%
\providecommand \BibitemShut  [1]{\csname bibitem#1\endcsname}%
\let\auto@bib@innerbib\@empty
\bibitem [{\citenamefont {Hen}\ \emph {et~al.}(2014)\citenamefont {Hen},
  \citenamefont {Sargsian}, \citenamefont {Weinstein} \emph
  {et~al.}}]{Hen2014S}%
  \BibitemOpen
  \bibfield  {author} {\bibinfo {author} {\bibfnamefont {O.}~\bibnamefont
  {Hen}}, \bibinfo {author} {\bibfnamefont {M.}~\bibnamefont {Sargsian}},
  \bibinfo {author} {\bibnamefont {Weinstein}},  \emph {et~al.},\ }\href
  {https://www.science.org/doi/abs/10.1126/science.1256785} {\bibfield
  {journal} {\bibinfo  {journal} {Science}\ }\textbf {\bibinfo {volume}
  {346}},\ \bibinfo {pages} {614} (\bibinfo {year} {2014})}\BibitemShut
  {NoStop}%
\bibitem [{\citenamefont {Hen}\ \emph {et~al.}(2017)\citenamefont {Hen},
  \citenamefont {Miller}, \citenamefont {Piasetzky},\ and\ \citenamefont
  {Weinstein}}]{Hen2017RMP}%
  \BibitemOpen
  \bibfield  {author} {\bibinfo {author} {\bibfnamefont {O.}~\bibnamefont
  {Hen}}, \bibinfo {author} {\bibfnamefont {G.~A.}\ \bibnamefont {Miller}},
  \bibinfo {author} {\bibfnamefont {E.}~\bibnamefont {Piasetzky}}, \ and\
  \bibinfo {author} {\bibfnamefont {L.~B.}\ \bibnamefont {Weinstein}},\ }\href
  {\doibase 10.1103/RevModPhys.89.045002} {\bibfield  {journal} {\bibinfo
  {journal} {Rev. Mod. Phys.}\ }\textbf {\bibinfo {volume} {89}},\ \bibinfo
  {pages} {045002} (\bibinfo {year} {2017})}\BibitemShut {NoStop}%
\bibitem [{\citenamefont {degli Atti}(2015)}]{Degli2015PR}%
  \BibitemOpen
  \bibfield  {author} {\bibinfo {author} {\bibfnamefont {C.~C.}\ \bibnamefont
  {degli Atti}},\ }\href {https://doi.org/10.1016/j.physrep.2015.06.002}
  {\bibfield  {journal} {\bibinfo  {journal} {Phys. Rep.}\ }\textbf {\bibinfo
  {volume} {590}},\ \bibinfo {pages} {1} (\bibinfo {year} {2015})}\BibitemShut
  {NoStop}%
\bibitem [{\citenamefont {H\"ufner}\ and\ \citenamefont
  {Nemes}(1981)}]{Hufner1981PRC}%
  \BibitemOpen
  \bibfield  {author} {\bibinfo {author} {\bibfnamefont {J.}~\bibnamefont
  {H\"ufner}}\ and\ \bibinfo {author} {\bibfnamefont {M.~C.}\ \bibnamefont
  {Nemes}},\ }\href {\doibase 10.1103/PhysRevC.23.2538} {\bibfield  {journal}
  {\bibinfo  {journal} {Phys. Rev. C}\ }\textbf {\bibinfo {volume} {23}},\
  \bibinfo {pages} {2538} (\bibinfo {year} {1981})}\BibitemShut {NoStop}%
\bibitem [{\citenamefont {Bertulani}\ and\ \citenamefont
  {McVoy}(1992)}]{Bertulani1992PRC}%
  \BibitemOpen
  \bibfield  {author} {\bibinfo {author} {\bibfnamefont {C.~A.}\ \bibnamefont
  {Bertulani}}\ and\ \bibinfo {author} {\bibfnamefont {K.~W.}\ \bibnamefont
  {McVoy}},\ }\href {\doibase 10.1103/PhysRevC.46.2638} {\bibfield  {journal}
  {\bibinfo  {journal} {Phys. Rev. C}\ }\textbf {\bibinfo {volume} {46}},\
  \bibinfo {pages} {2638} (\bibinfo {year} {1992})}\BibitemShut {NoStop}%
\bibitem [{\citenamefont {Hencken}\ \emph {et~al.}(1996)\citenamefont
  {Hencken}, \citenamefont {Bertsch},\ and\ \citenamefont
  {Esbensen}}]{Hencken1996PRC}%
  \BibitemOpen
  \bibfield  {author} {\bibinfo {author} {\bibfnamefont {K.}~\bibnamefont
  {Hencken}}, \bibinfo {author} {\bibfnamefont {G.}~\bibnamefont {Bertsch}}, \
  and\ \bibinfo {author} {\bibfnamefont {H.}~\bibnamefont {Esbensen}},\ }\href
  {\doibase 10.1103/PhysRevC.54.3043} {\bibfield  {journal} {\bibinfo
  {journal} {Phys. Rev. C}\ }\textbf {\bibinfo {volume} {54}},\ \bibinfo
  {pages} {3043} (\bibinfo {year} {1996})}\BibitemShut {NoStop}%
\bibitem [{\citenamefont {Satou}\ \emph {et~al.}(2014)\citenamefont {Satou},
  \citenamefont {Hwang}, \citenamefont {Kim} \emph {et~al.}}]{Satou2014PLB}%
  \BibitemOpen
  \bibfield  {author} {\bibinfo {author} {\bibfnamefont {Y.}~\bibnamefont
  {Satou}}, \bibinfo {author} {\bibfnamefont {J.}~\bibnamefont {Hwang}},
  \bibinfo {author} {\bibfnamefont {S.}~\bibnamefont {Kim}},  \emph {et~al.},\
  }\href {\doibase https://doi.org/10.1016/j.physletb.2013.12.024} {\bibfield
  {journal} {\bibinfo  {journal} {Phys. Lett. B}\ }\textbf {\bibinfo {volume}
  {728}},\ \bibinfo {pages} {462} (\bibinfo {year} {2014})}\BibitemShut
  {NoStop}%
\bibitem [{\citenamefont {Holl}\ \emph {et~al.}(2019)\citenamefont {Holl},
  \citenamefont {Panin}, \citenamefont {{\'A}lvarez-Pol} \emph
  {et~al.}}]{Holl2019PLB}%
  \BibitemOpen
  \bibfield  {author} {\bibinfo {author} {\bibfnamefont {M.}~\bibnamefont
  {Holl}}, \bibinfo {author} {\bibfnamefont {V.}~\bibnamefont {Panin}},
  \bibinfo {author} {\bibfnamefont {H.}~\bibnamefont {{\'A}lvarez-Pol}},  \emph
  {et~al.},\ }\href {https://doi.org/10.1016/j.physletb.2019.06.069} {\bibfield
   {journal} {\bibinfo  {journal} {Phys. Lett. B}\ }\textbf {\bibinfo {volume}
  {795}},\ \bibinfo {pages} {682} (\bibinfo {year} {2019})}\BibitemShut
  {NoStop}%
\bibitem [{\citenamefont {Panin}\ \emph {et~al.}(2019)\citenamefont {Panin},
  \citenamefont {Holl}, \citenamefont {Taylor} \emph {et~al.}}]{Panin2019PLB}%
  \BibitemOpen
  \bibfield  {author} {\bibinfo {author} {\bibfnamefont {V.}~\bibnamefont
  {Panin}}, \bibinfo {author} {\bibfnamefont {M.}~\bibnamefont {Holl}},
  \bibinfo {author} {\bibfnamefont {J.~T.}\ \bibnamefont {Taylor}},  \emph
  {et~al.},\ }\href {https://doi.org/10.1016/j.physletb.2019.134802} {\bibfield
   {journal} {\bibinfo  {journal} {Phys. Lett. B}\ }\textbf {\bibinfo {volume}
  {797}},\ \bibinfo {pages} {134802} (\bibinfo {year} {2019})}\BibitemShut
  {NoStop}%
\bibitem [{\citenamefont {Goldhaber}(1974)}]{Goldhaber1974PLB}%
  \BibitemOpen
  \bibfield  {author} {\bibinfo {author} {\bibfnamefont {A.~S.}\ \bibnamefont
  {Goldhaber}},\ }\href {https://doi.org/10.1016/0370-2693(74)90388-8}
  {\bibfield  {journal} {\bibinfo  {journal} {Phys. Lett. B}\ }\textbf
  {\bibinfo {volume} {53}},\ \bibinfo {pages} {306} (\bibinfo {year}
  {1974})}\BibitemShut {NoStop}%
\bibitem [{\citenamefont {Morrissey}(1989)}]{Morrissey1989PRC}%
  \BibitemOpen
  \bibfield  {author} {\bibinfo {author} {\bibfnamefont {D.~J.}\ \bibnamefont
  {Morrissey}},\ }\href {\doibase 10.1103/PhysRevC.39.460} {\bibfield
  {journal} {\bibinfo  {journal} {Phys. Rev. C}\ }\textbf {\bibinfo {volume}
  {39}},\ \bibinfo {pages} {460} (\bibinfo {year} {1989})}\BibitemShut
  {NoStop}%
\bibitem [{\citenamefont {Ma}\ \emph {et~al.}(2002)\citenamefont {Ma},
  \citenamefont {Wada}, \citenamefont {Hagel}, \citenamefont {Murray},
  \citenamefont {Wang}, \citenamefont {Qin}, \citenamefont {Makeev},
  \citenamefont {Smith}, \citenamefont {Natowitz},\ and\ \citenamefont
  {Ono}}]{Ma2002PRC}%
  \BibitemOpen
  \bibfield  {author} {\bibinfo {author} {\bibfnamefont {Y.~G.}\ \bibnamefont
  {Ma}}, \bibinfo {author} {\bibfnamefont {R.}~\bibnamefont {Wada}}, \bibinfo
  {author} {\bibfnamefont {K.}~\bibnamefont {Hagel}}, \bibinfo {author}
  {\bibfnamefont {M.}~\bibnamefont {Murray}}, \bibinfo {author} {\bibfnamefont
  {J.~S.}\ \bibnamefont {Wang}}, \bibinfo {author} {\bibfnamefont {L.~J.}\
  \bibnamefont {Qin}}, \bibinfo {author} {\bibfnamefont {A.}~\bibnamefont
  {Makeev}}, \bibinfo {author} {\bibfnamefont {P.}~\bibnamefont {Smith}},
  \bibinfo {author} {\bibfnamefont {J.~B.}\ \bibnamefont {Natowitz}}, \ and\
  \bibinfo {author} {\bibfnamefont {A.}~\bibnamefont {Ono}},\ }\href {\doibase
  10.1103/PhysRevC.65.051602} {\bibfield  {journal} {\bibinfo  {journal} {Phys.
  Rev. C}\ }\textbf {\bibinfo {volume} {65}},\ \bibinfo {pages} {051602(R)}
  (\bibinfo {year} {2002})}\BibitemShut {NoStop}%
\bibitem [{\citenamefont {Momota}\ \emph {et~al.}(2017)\citenamefont {Momota},
  \citenamefont {Kanazawa}, \citenamefont {Kitagawa},\ and\ \citenamefont
  {Sato}}]{Momota2017NPA}%
  \BibitemOpen
  \bibfield  {author} {\bibinfo {author} {\bibfnamefont {S.}~\bibnamefont
  {Momota}}, \bibinfo {author} {\bibfnamefont {M.}~\bibnamefont {Kanazawa}},
  \bibinfo {author} {\bibfnamefont {A.}~\bibnamefont {Kitagawa}}, \ and\
  \bibinfo {author} {\bibfnamefont {S.}~\bibnamefont {Sato}},\ }\href {\doibase
  https://doi.org/10.1016/j.nuclphysa.2016.12.004} {\bibfield  {journal}
  {\bibinfo  {journal} {Nucl. Phys. A}\ }\textbf {\bibinfo {volume} {958}},\
  \bibinfo {pages} {219} (\bibinfo {year} {2017})}\BibitemShut {NoStop}%
\bibitem [{\citenamefont {Notani}\ \emph {et~al.}(2007)\citenamefont {Notani},
  \citenamefont {Sakurai}, \citenamefont {Aoi}, \citenamefont {Iwasaki},
  \citenamefont {Fukuda}, \citenamefont {Liu}, \citenamefont {Yoneda},
  \citenamefont {Ogawa}, \citenamefont {Teranishi}, \citenamefont {Nakamura},
  \citenamefont {Okuno}, \citenamefont {Yoshida}, \citenamefont {Watanabe},
  \citenamefont {Momota}, \citenamefont {Inabe}, \citenamefont {Kubo},
  \citenamefont {Ito}, \citenamefont {Ozawa}, \citenamefont {Suzuki},
  \citenamefont {Tanihata},\ and\ \citenamefont {Ishihara}}]{Notani2007PRC}%
  \BibitemOpen
  \bibfield  {author} {\bibinfo {author} {\bibfnamefont {M.}~\bibnamefont
  {Notani}}, \bibinfo {author} {\bibfnamefont {H.}~\bibnamefont {Sakurai}},
  \bibinfo {author} {\bibfnamefont {N.}~\bibnamefont {Aoi}}, \bibinfo {author}
  {\bibfnamefont {H.}~\bibnamefont {Iwasaki}}, \bibinfo {author} {\bibfnamefont
  {N.}~\bibnamefont {Fukuda}}, \bibinfo {author} {\bibfnamefont
  {Z.}~\bibnamefont {Liu}}, \bibinfo {author} {\bibfnamefont {K.}~\bibnamefont
  {Yoneda}}, \bibinfo {author} {\bibfnamefont {H.}~\bibnamefont {Ogawa}},
  \bibinfo {author} {\bibfnamefont {T.}~\bibnamefont {Teranishi}}, \bibinfo
  {author} {\bibfnamefont {T.}~\bibnamefont {Nakamura}}, \bibinfo {author}
  {\bibfnamefont {H.}~\bibnamefont {Okuno}}, \bibinfo {author} {\bibfnamefont
  {A.}~\bibnamefont {Yoshida}}, \bibinfo {author} {\bibfnamefont {Y.~X.}\
  \bibnamefont {Watanabe}}, \bibinfo {author} {\bibfnamefont {S.}~\bibnamefont
  {Momota}}, \bibinfo {author} {\bibfnamefont {N.}~\bibnamefont {Inabe}},
  \bibinfo {author} {\bibfnamefont {T.}~\bibnamefont {Kubo}}, \bibinfo {author}
  {\bibfnamefont {S.}~\bibnamefont {Ito}}, \bibinfo {author} {\bibfnamefont
  {A.}~\bibnamefont {Ozawa}}, \bibinfo {author} {\bibfnamefont
  {T.}~\bibnamefont {Suzuki}}, \bibinfo {author} {\bibfnamefont
  {I.}~\bibnamefont {Tanihata}}, \ and\ \bibinfo {author} {\bibfnamefont
  {M.}~\bibnamefont {Ishihara}},\ }\href {\doibase 10.1103/PhysRevC.76.044605}
  {\bibfield  {journal} {\bibinfo  {journal} {Phys. Rev. C}\ }\textbf {\bibinfo
  {volume} {76}},\ \bibinfo {pages} {044605} (\bibinfo {year}
  {2007})}\BibitemShut {NoStop}%
\bibitem [{\citenamefont {Meierbachtol}\ \emph {et~al.}(2012)\citenamefont
  {Meierbachtol}, \citenamefont {Morrissey}, \citenamefont {Mosby},\ and\
  \citenamefont {Bazin}}]{Meier2012PRC}%
  \BibitemOpen
  \bibfield  {author} {\bibinfo {author} {\bibfnamefont {K.}~\bibnamefont
  {Meierbachtol}}, \bibinfo {author} {\bibfnamefont {D.~J.}\ \bibnamefont
  {Morrissey}}, \bibinfo {author} {\bibfnamefont {M.}~\bibnamefont {Mosby}}, \
  and\ \bibinfo {author} {\bibfnamefont {D.}~\bibnamefont {Bazin}},\ }\href
  {\doibase 10.1103/PhysRevC.85.034608} {\bibfield  {journal} {\bibinfo
  {journal} {Phys. Rev. C}\ }\textbf {\bibinfo {volume} {85}},\ \bibinfo
  {pages} {034608} (\bibinfo {year} {2012})}\BibitemShut {NoStop}%
\bibitem [{\citenamefont {Mocko}\ \emph {et~al.}(2006)\citenamefont {Mocko},
  \citenamefont {Tsang}, \citenamefont {Andronenko}, \citenamefont
  {Andronenko}, \citenamefont {Delaunay}, \citenamefont {Famiano},
  \citenamefont {Ginter}, \citenamefont {Henzl}, \citenamefont {Henzlov\'a},
  \citenamefont {Hua}, \citenamefont {Lukyanov}, \citenamefont {Lynch},
  \citenamefont {Rogers}, \citenamefont {Steiner}, \citenamefont {Stolz},
  \citenamefont {Tarasov}, \citenamefont {Goethem}, \citenamefont {Verde},
  \citenamefont {Wallace},\ and\ \citenamefont {Zalessov}}]{Mocko2006PRC}%
  \BibitemOpen
  \bibfield  {author} {\bibinfo {author} {\bibfnamefont {M.}~\bibnamefont
  {Mocko}}, \bibinfo {author} {\bibfnamefont {M.~B.}\ \bibnamefont {Tsang}},
  \bibinfo {author} {\bibfnamefont {L.}~\bibnamefont {Andronenko}}, \bibinfo
  {author} {\bibfnamefont {M.}~\bibnamefont {Andronenko}}, \bibinfo {author}
  {\bibfnamefont {F.}~\bibnamefont {Delaunay}}, \bibinfo {author}
  {\bibfnamefont {M.}~\bibnamefont {Famiano}}, \bibinfo {author} {\bibfnamefont
  {T.}~\bibnamefont {Ginter}}, \bibinfo {author} {\bibfnamefont
  {V.}~\bibnamefont {Henzl}}, \bibinfo {author} {\bibfnamefont
  {D.}~\bibnamefont {Henzlov\'a}}, \bibinfo {author} {\bibfnamefont
  {H.}~\bibnamefont {Hua}}, \bibinfo {author} {\bibfnamefont {S.}~\bibnamefont
  {Lukyanov}}, \bibinfo {author} {\bibfnamefont {W.~G.}\ \bibnamefont {Lynch}},
  \bibinfo {author} {\bibfnamefont {A.~M.}\ \bibnamefont {Rogers}}, \bibinfo
  {author} {\bibfnamefont {M.}~\bibnamefont {Steiner}}, \bibinfo {author}
  {\bibfnamefont {A.}~\bibnamefont {Stolz}}, \bibinfo {author} {\bibfnamefont
  {O.}~\bibnamefont {Tarasov}}, \bibinfo {author} {\bibfnamefont {M.-J.~v.}\
  \bibnamefont {Goethem}}, \bibinfo {author} {\bibfnamefont {G.}~\bibnamefont
  {Verde}}, \bibinfo {author} {\bibfnamefont {W.~S.}\ \bibnamefont {Wallace}},
  \ and\ \bibinfo {author} {\bibfnamefont {A.}~\bibnamefont {Zalessov}},\
  }\href {\doibase 10.1103/PhysRevC.74.054612} {\bibfield  {journal} {\bibinfo
  {journal} {Phys. Rev. C}\ }\textbf {\bibinfo {volume} {74}},\ \bibinfo
  {pages} {054612} (\bibinfo {year} {2006})}\BibitemShut {NoStop}%
\bibitem [{\citenamefont {Greiner}\ \emph {et~al.}(1975)\citenamefont
  {Greiner}, \citenamefont {Lindstrom}, \citenamefont {Heckman}, \citenamefont
  {Cork},\ and\ \citenamefont {Bieser}}]{Greiner1975PRL}%
  \BibitemOpen
  \bibfield  {author} {\bibinfo {author} {\bibfnamefont {D.~E.}\ \bibnamefont
  {Greiner}}, \bibinfo {author} {\bibfnamefont {P.~J.}\ \bibnamefont
  {Lindstrom}}, \bibinfo {author} {\bibfnamefont {H.~H.}\ \bibnamefont
  {Heckman}}, \bibinfo {author} {\bibfnamefont {B.}~\bibnamefont {Cork}}, \
  and\ \bibinfo {author} {\bibfnamefont {F.~S.}\ \bibnamefont {Bieser}},\
  }\href {\doibase 10.1103/PhysRevLett.35.152} {\bibfield  {journal} {\bibinfo
  {journal} {Phys. Rev. Lett.}\ }\textbf {\bibinfo {volume} {35}},\ \bibinfo
  {pages} {152} (\bibinfo {year} {1975})}\BibitemShut {NoStop}%
\bibitem [{\citenamefont {Caamano}\ \emph {et~al.}(2004)\citenamefont
  {Caamano}, \citenamefont {Cortina-Gil}, \citenamefont {S{\"u}mmerer},
  \citenamefont {Benlliure}, \citenamefont {Casarejos}, \citenamefont
  {Geissel}, \citenamefont {M{\"u}nzenberg},\ and\ \citenamefont
  {Pereira}}]{Caamano2004NPA}%
  \BibitemOpen
  \bibfield  {author} {\bibinfo {author} {\bibfnamefont {M.}~\bibnamefont
  {Caamano}}, \bibinfo {author} {\bibfnamefont {D.}~\bibnamefont
  {Cortina-Gil}}, \bibinfo {author} {\bibfnamefont {K.}~\bibnamefont
  {S{\"u}mmerer}}, \bibinfo {author} {\bibfnamefont {J.}~\bibnamefont
  {Benlliure}}, \bibinfo {author} {\bibfnamefont {E.}~\bibnamefont
  {Casarejos}}, \bibinfo {author} {\bibfnamefont {H.}~\bibnamefont {Geissel}},
  \bibinfo {author} {\bibfnamefont {G.}~\bibnamefont {M{\"u}nzenberg}}, \ and\
  \bibinfo {author} {\bibfnamefont {J.}~\bibnamefont {Pereira}},\ }\href
  {https://doi.org/10.1016/j.nuclphysa.2004.01.070} {\bibfield  {journal}
  {\bibinfo  {journal} {Nucl. Phys. A}\ }\textbf {\bibinfo {volume} {733}},\
  \bibinfo {pages} {187} (\bibinfo {year} {2004})}\BibitemShut {NoStop}%
\bibitem [{\citenamefont {Weber}\ \emph {et~al.}(1994)\citenamefont {Weber},
  \citenamefont {Donzaud}, \citenamefont {Dufour} \emph
  {et~al.}}]{Weber1994NPA}%
  \BibitemOpen
  \bibfield  {author} {\bibinfo {author} {\bibfnamefont {M.}~\bibnamefont
  {Weber}}, \bibinfo {author} {\bibfnamefont {C.}~\bibnamefont {Donzaud}},
  \bibinfo {author} {\bibfnamefont {J.~P.}\ \bibnamefont {Dufour}},  \emph
  {et~al.},\ }\href {https://doi.org/10.1016/0375-9474(94)90766-8} {\bibfield
  {journal} {\bibinfo  {journal} {Nucl. Phys. A}\ }\textbf {\bibinfo {volume}
  {578}},\ \bibinfo {pages} {659} (\bibinfo {year} {1994})}\BibitemShut
  {NoStop}%
\bibitem [{\citenamefont {Reinhold}\ \emph {et~al.}(1998)\citenamefont
  {Reinhold}, \citenamefont {Friese}, \citenamefont {K\"orner}, \citenamefont
  {Schneider}, \citenamefont {Zeitelhack}, \citenamefont {Geissel},
  \citenamefont {Magel}, \citenamefont {M\"unzenberg},\ and\ \citenamefont
  {S\"ummerer}}]{Reinhold1998PRC}%
  \BibitemOpen
  \bibfield  {author} {\bibinfo {author} {\bibfnamefont {J.}~\bibnamefont
  {Reinhold}}, \bibinfo {author} {\bibfnamefont {J.}~\bibnamefont {Friese}},
  \bibinfo {author} {\bibfnamefont {H.-J.}\ \bibnamefont {K\"orner}}, \bibinfo
  {author} {\bibfnamefont {R.}~\bibnamefont {Schneider}}, \bibinfo {author}
  {\bibfnamefont {K.}~\bibnamefont {Zeitelhack}}, \bibinfo {author}
  {\bibfnamefont {H.}~\bibnamefont {Geissel}}, \bibinfo {author} {\bibfnamefont
  {A.}~\bibnamefont {Magel}}, \bibinfo {author} {\bibfnamefont
  {G.}~\bibnamefont {M\"unzenberg}}, \ and\ \bibinfo {author} {\bibfnamefont
  {K.}~\bibnamefont {S\"ummerer}},\ }\href {\doibase 10.1103/PhysRevC.58.247}
  {\bibfield  {journal} {\bibinfo  {journal} {Phys. Rev. C}\ }\textbf {\bibinfo
  {volume} {58}},\ \bibinfo {pages} {247} (\bibinfo {year} {1998})}\BibitemShut
  {NoStop}%
\bibitem [{\citenamefont {Momota}\ \emph {et~al.}(2023)\citenamefont {Momota},
  \citenamefont {Ohtsubo}, \citenamefont {Honma}, \citenamefont {Kitagawa},\
  and\ \citenamefont {Sato}}]{Momota2023PS}%
  \BibitemOpen
  \bibfield  {author} {\bibinfo {author} {\bibfnamefont {S.}~\bibnamefont
  {Momota}}, \bibinfo {author} {\bibfnamefont {T.}~\bibnamefont {Ohtsubo}},
  \bibinfo {author} {\bibfnamefont {A.}~\bibnamefont {Honma}}, \bibinfo
  {author} {\bibfnamefont {A.}~\bibnamefont {Kitagawa}}, \ and\ \bibinfo
  {author} {\bibfnamefont {S.}~\bibnamefont {Sato}},\ }\href {\doibase
  10.1088/1402-4896/acdf93} {\bibfield  {journal} {\bibinfo  {journal} {Phys.
  Scri.}\ }\textbf {\bibinfo {volume} {98}},\ \bibinfo {pages} {085301}
  (\bibinfo {year} {2023})}\BibitemShut {NoStop}%
\bibitem [{\citenamefont {Van~Bibber}\ \emph {et~al.}(1979)\citenamefont
  {Van~Bibber}, \citenamefont {Hendrie}, \citenamefont {Scott}, \citenamefont
  {Weiman}, \citenamefont {Schroeder}, \citenamefont {Geaga}, \citenamefont
  {Cessin}, \citenamefont {Treuhaft}, \citenamefont {Grossiord}, \citenamefont
  {Rasmussen},\ and\ \citenamefont {Wong}}]{Bibber1979PRL}%
  \BibitemOpen
  \bibfield  {author} {\bibinfo {author} {\bibfnamefont {K.}~\bibnamefont
  {Van~Bibber}}, \bibinfo {author} {\bibfnamefont {D.~L.}\ \bibnamefont
  {Hendrie}}, \bibinfo {author} {\bibfnamefont {D.~K.}\ \bibnamefont {Scott}},
  \bibinfo {author} {\bibfnamefont {H.~H.}\ \bibnamefont {Weiman}}, \bibinfo
  {author} {\bibfnamefont {L.~S.}\ \bibnamefont {Schroeder}}, \bibinfo {author}
  {\bibfnamefont {J.~V.}\ \bibnamefont {Geaga}}, \bibinfo {author}
  {\bibfnamefont {S.~A.}\ \bibnamefont {Cessin}}, \bibinfo {author}
  {\bibfnamefont {R.}~\bibnamefont {Treuhaft}}, \bibinfo {author}
  {\bibfnamefont {Y.~J.}\ \bibnamefont {Grossiord}}, \bibinfo {author}
  {\bibfnamefont {J.~O.}\ \bibnamefont {Rasmussen}}, \ and\ \bibinfo {author}
  {\bibfnamefont {C.~Y.}\ \bibnamefont {Wong}},\ }\href {\doibase
  10.1103/PhysRevLett.43.840} {\bibfield  {journal} {\bibinfo  {journal} {Phys.
  Rev. Lett.}\ }\textbf {\bibinfo {volume} {43}},\ \bibinfo {pages} {840}
  (\bibinfo {year} {1979})}\BibitemShut {NoStop}%
\bibitem [{\citenamefont {Kidd}\ \emph {et~al.}(1988)\citenamefont {Kidd},
  \citenamefont {Lindstrom}, \citenamefont {Crawford},\ and\ \citenamefont
  {Woods}}]{Kidd1988PRC}%
  \BibitemOpen
  \bibfield  {author} {\bibinfo {author} {\bibfnamefont {J.~M.}\ \bibnamefont
  {Kidd}}, \bibinfo {author} {\bibfnamefont {P.~J.}\ \bibnamefont {Lindstrom}},
  \bibinfo {author} {\bibfnamefont {H.~J.}\ \bibnamefont {Crawford}}, \ and\
  \bibinfo {author} {\bibfnamefont {G.}~\bibnamefont {Woods}},\ }\href
  {\doibase 10.1103/PhysRevC.37.2613} {\bibfield  {journal} {\bibinfo
  {journal} {Phys. Rev. C}\ }\textbf {\bibinfo {volume} {37}},\ \bibinfo
  {pages} {2613} (\bibinfo {year} {1988})}\BibitemShut {NoStop}%
\bibitem [{\citenamefont {Ono}\ and\ \citenamefont
  {Randrup}(2006)}]{Ono2006EPJA}%
  \BibitemOpen
  \bibfield  {author} {\bibinfo {author} {\bibfnamefont {A.}~\bibnamefont
  {Ono}}\ and\ \bibinfo {author} {\bibfnamefont {J.}~\bibnamefont {Randrup}},\
  }\href {https://doi.org/10.1140/epja/i2006-10110-1} {\bibfield  {journal}
  {\bibinfo  {journal} {Eur. Phys. J. A}\ ,\ \bibinfo {pages} {109}} (\bibinfo
  {year} {2006})}\BibitemShut {NoStop}%
\bibitem [{\citenamefont {Ma}\ and\ \citenamefont {Ma}(2018)}]{Ma2018PPNP}%
  \BibitemOpen
  \bibfield  {author} {\bibinfo {author} {\bibfnamefont {C.~W.}\ \bibnamefont
  {Ma}}\ and\ \bibinfo {author} {\bibfnamefont {Y.~G.}\ \bibnamefont {Ma}},\
  }\href {\doibase 10.1016/j.ppnp.2018.01.002} {\bibfield  {journal} {\bibinfo
  {journal} {Prog. Part. Nucl. Phys.}\ }\textbf {\bibinfo {volume} {99}},\
  \bibinfo {pages} {120} (\bibinfo {year} {2018})}\BibitemShut {NoStop}%
\bibitem [{\citenamefont {Ono}(2019)}]{Ono2019PPNP}%
  \BibitemOpen
  \bibfield  {author} {\bibinfo {author} {\bibfnamefont {A.}~\bibnamefont
  {Ono}},\ }\href {\doibase 10.1016/j.ppnp.2018.11.001} {\bibfield  {journal}
  {\bibinfo  {journal} {Prog. Part. Nucl. Phys.}\ }\textbf {\bibinfo {volume}
  {105}},\ \bibinfo {pages} {139} (\bibinfo {year} {2019})}\BibitemShut
  {NoStop}%
\bibitem [{\citenamefont {Xu}(2019)}]{Xu2019PPNP}%
  \BibitemOpen
  \bibfield  {author} {\bibinfo {author} {\bibfnamefont {J.}~\bibnamefont
  {Xu}},\ }\href {\doibase 10.1016/j.ppnp.2019.02.009} {\bibfield  {journal}
  {\bibinfo  {journal} {Prog. Part. Nucl. Phys.}\ }\textbf {\bibinfo {volume}
  {106}},\ \bibinfo {pages} {312} (\bibinfo {year} {2019})}\BibitemShut
  {NoStop}%
\bibitem [{\citenamefont {Wolter}\ \emph {et~al.}(2022)\citenamefont {Wolter},
  \citenamefont {Colonna}, \citenamefont {Cozma} \emph
  {et~al.}}]{Wolter2022PPNP}%
  \BibitemOpen
  \bibfield  {author} {\bibinfo {author} {\bibfnamefont {H.}~\bibnamefont
  {Wolter}}, \bibinfo {author} {\bibfnamefont {M.}~\bibnamefont {Colonna}},
  \bibinfo {author} {\bibfnamefont {D.}~\bibnamefont {Cozma}},  \emph
  {et~al.},\ }\href {\doibase 10.1016/j.ppnp.2022.103962} {\bibfield  {journal}
  {\bibinfo  {journal} {Prog. Part. Nucl. Phys.}\ }\textbf {\bibinfo {volume}
  {125}},\ \bibinfo {pages} {103962} (\bibinfo {year} {2022})}\BibitemShut
  {NoStop}%
\bibitem [{\citenamefont {Deng}\ \emph {et~al.}(2024)\citenamefont {Deng},
  \citenamefont {Fang},\ and\ \citenamefont {Ma}}]{Deng2024PPNP}%
  \BibitemOpen
  \bibfield  {author} {\bibinfo {author} {\bibfnamefont {X.~G.}\ \bibnamefont
  {Deng}}, \bibinfo {author} {\bibfnamefont {D.~Q.}\ \bibnamefont {Fang}}, \
  and\ \bibinfo {author} {\bibfnamefont {Y.~G.}\ \bibnamefont {Ma}},\ }\href
  {\doibase https://doi.org/10.1016/j.ppnp.2023.104095} {\bibfield  {journal}
  {\bibinfo  {journal} {Prog. Part. Nucl. Phys.}\ }\textbf {\bibinfo {volume}
  {136}},\ \bibinfo {pages} {104095} (\bibinfo {year} {2024})}\BibitemShut
  {NoStop}%
\bibitem [{\citenamefont {Sun}\ \emph {et~al.}(2024)\citenamefont {Sun},
  \citenamefont {Wang}, \citenamefont {Ko}, \citenamefont {Ma},\ and\
  \citenamefont {Shen}}]{Sun2024NC}%
  \BibitemOpen
  \bibfield  {author} {\bibinfo {author} {\bibfnamefont {K.~J.}\ \bibnamefont
  {Sun}}, \bibinfo {author} {\bibfnamefont {R.}~\bibnamefont {Wang}}, \bibinfo
  {author} {\bibfnamefont {C.~M.}\ \bibnamefont {Ko}}, \bibinfo {author}
  {\bibfnamefont {Y.~G.}\ \bibnamefont {Ma}}, \ and\ \bibinfo {author}
  {\bibfnamefont {C.}~\bibnamefont {Shen}},\ }\href
  {https://doi.org/10.1038/s41467-024-45474-x} {\bibfield  {journal} {\bibinfo
  {journal} {Nat. Commun}\ }\textbf {\bibinfo {volume} {15}},\ \bibinfo {pages}
  {1074} (\bibinfo {year} {2024})}\BibitemShut {NoStop}%
\bibitem [{\citenamefont {Bertsch}\ and\ \citenamefont {{Das
  Gupta}}(1988)}]{Bertsch1988PR}%
  \BibitemOpen
  \bibfield  {author} {\bibinfo {author} {\bibfnamefont {G.}~\bibnamefont
  {Bertsch}}\ and\ \bibinfo {author} {\bibfnamefont {S.}~\bibnamefont {{Das
  Gupta}}},\ }\href {\doibase https://doi.org/10.1016/0370-1573(88)90170-6}
  {\bibfield  {journal} {\bibinfo  {journal} {Phys. Rep.}\ }\textbf {\bibinfo
  {volume} {160}},\ \bibinfo {pages} {189} (\bibinfo {year}
  {1988})}\BibitemShut {NoStop}%
\bibitem [{\citenamefont {Li}\ \emph {et~al.}(2008)\citenamefont {Li},
  \citenamefont {Chen},\ and\ \citenamefont {Ko}}]{Li2008PR}%
  \BibitemOpen
  \bibfield  {author} {\bibinfo {author} {\bibfnamefont {B.-A.}\ \bibnamefont
  {Li}}, \bibinfo {author} {\bibfnamefont {L.-W.}\ \bibnamefont {Chen}}, \ and\
  \bibinfo {author} {\bibfnamefont {C.~M.}\ \bibnamefont {Ko}},\ }\href
  {\doibase https://doi.org/10.1016/j.physrep.2008.04.005} {\bibfield
  {journal} {\bibinfo  {journal} {Phys. Rep.}\ }\textbf {\bibinfo {volume}
  {464}},\ \bibinfo {pages} {113} (\bibinfo {year} {2008})}\BibitemShut
  {NoStop}%
\bibitem [{\citenamefont {Song}\ \emph {et~al.}(2023)\citenamefont {Song},
  \citenamefont {Wang}, \citenamefont {Zhang},\ and\ \citenamefont
  {Ma}}]{Song2023PRC}%
  \BibitemOpen
  \bibfield  {author} {\bibinfo {author} {\bibfnamefont {Y.-D.}\ \bibnamefont
  {Song}}, \bibinfo {author} {\bibfnamefont {R.}~\bibnamefont {Wang}}, \bibinfo
  {author} {\bibfnamefont {Z.}~\bibnamefont {Zhang}}, \ and\ \bibinfo {author}
  {\bibfnamefont {Y.-G.}\ \bibnamefont {Ma}},\ }\href {\doibase
  10.1103/PhysRevC.108.064603} {\bibfield  {journal} {\bibinfo  {journal}
  {Phys. Rev. C}\ }\textbf {\bibinfo {volume} {108}},\ \bibinfo {pages}
  {064603} (\bibinfo {year} {2023})}\BibitemShut {NoStop}%
\bibitem [{\citenamefont {Wang}\ \emph
  {et~al.}(2023{\natexlab{a}})\citenamefont {Wang}, \citenamefont {Ma},
  \citenamefont {Chen}, \citenamefont {Ko}, \citenamefont {Sun},\ and\
  \citenamefont {Zhang}}]{Wang2023PRC}%
  \BibitemOpen
  \bibfield  {author} {\bibinfo {author} {\bibfnamefont {R.}~\bibnamefont
  {Wang}}, \bibinfo {author} {\bibfnamefont {Y.-G.}\ \bibnamefont {Ma}},
  \bibinfo {author} {\bibfnamefont {L.-W.}\ \bibnamefont {Chen}}, \bibinfo
  {author} {\bibfnamefont {C.~M.}\ \bibnamefont {Ko}}, \bibinfo {author}
  {\bibfnamefont {K.-J.}\ \bibnamefont {Sun}}, \ and\ \bibinfo {author}
  {\bibfnamefont {Z.}~\bibnamefont {Zhang}},\ }\href {\doibase
  10.1103/PhysRevC.108.L031601} {\bibfield  {journal} {\bibinfo  {journal}
  {Phys. Rev. C}\ }\textbf {\bibinfo {volume} {108}},\ \bibinfo {pages}
  {L031601} (\bibinfo {year} {2023}{\natexlab{a}})}\BibitemShut {NoStop}%
\bibitem [{\citenamefont {Aichelin}(1991)}]{Aichelin1991PR}%
  \BibitemOpen
  \bibfield  {author} {\bibinfo {author} {\bibfnamefont {J.}~\bibnamefont
  {Aichelin}},\ }\href {https://doi.org/10.1016/0370-1573(91)90094-3}
  {\bibfield  {journal} {\bibinfo  {journal} {Phys. Rep.}\ }\textbf {\bibinfo
  {volume} {202}},\ \bibinfo {pages} {233} (\bibinfo {year}
  {1991})}\BibitemShut {NoStop}%
\bibitem [{\citenamefont {Liu}\ \emph {et~al.}(2022)\citenamefont {Liu},
  \citenamefont {Deng},\ and\ \citenamefont {Ma}}]{Liu2022NST}%
  \BibitemOpen
  \bibfield  {author} {\bibinfo {author} {\bibfnamefont {C.}~\bibnamefont
  {Liu}}, \bibinfo {author} {\bibfnamefont {X.-G.}\ \bibnamefont {Deng}}, \
  and\ \bibinfo {author} {\bibfnamefont {Y.-G.}\ \bibnamefont {Ma}},\ }\href
  {\doibase 10.1007/s41365-022-01040-y} {\bibfield  {journal} {\bibinfo
  {journal} {Nucl. Sci. Tech.}\ }\textbf {\bibinfo {volume} {33}},\ \bibinfo
  {pages} {52} (\bibinfo {year} {2022})}\BibitemShut {NoStop}%
\bibitem [{\citenamefont {Li}\ \emph {et~al.}(2022)\citenamefont {Li},
  \citenamefont {Wang},\ and\ \citenamefont {Zhang}}]{Li2022NST}%
  \BibitemOpen
  \bibfield  {author} {\bibinfo {author} {\bibfnamefont {L.}~\bibnamefont
  {Li}}, \bibinfo {author} {\bibfnamefont {F.-Y.}\ \bibnamefont {Wang}}, \ and\
  \bibinfo {author} {\bibfnamefont {Y.-X.}\ \bibnamefont {Zhang}},\ }\href
  {\doibase 10.1007/s41365-022-01050-w} {\bibfield  {journal} {\bibinfo
  {journal} {Nucl. Sci. Tech.}\ }\textbf {\bibinfo {volume} {33}},\ \bibinfo
  {pages} {58} (\bibinfo {year} {2022})}\BibitemShut {NoStop}%
\bibitem [{\citenamefont {Wang}\ \emph {et~al.}(2022)\citenamefont {Wang},
  \citenamefont {Ou},\ and\ \citenamefont {Xiao}}]{Wang2022NST}%
  \BibitemOpen
  \bibfield  {author} {\bibinfo {author} {\bibfnamefont {R.-S.}\ \bibnamefont
  {Wang}}, \bibinfo {author} {\bibfnamefont {L.}~\bibnamefont {Ou}}, \ and\
  \bibinfo {author} {\bibfnamefont {Z.-G.}\ \bibnamefont {Xiao}},\ }\href
  {\doibase 10.1007/s41365-022-01075-1} {\bibfield  {journal} {\bibinfo
  {journal} {Nucl. Sci. Tech.}\ }\textbf {\bibinfo {volume} {33}},\ \bibinfo
  {pages} {92} (\bibinfo {year} {2022})}\BibitemShut {NoStop}%
\bibitem [{\citenamefont {Wang}\ \emph
  {et~al.}(2023{\natexlab{b}})\citenamefont {Wang}, \citenamefont {Yang},
  \citenamefont {Chen},  \emph {et~al.}}]{Wang2023NST}%
  \BibitemOpen
  \bibfield  {author} {\bibinfo {author} {\bibfnamefont {F.-Y.}\ \bibnamefont
  {Wang}}, \bibinfo {author} {\bibfnamefont {J.-P.}\ \bibnamefont {Yang}},
  \bibinfo {author} {\bibfnamefont {X.}~\bibnamefont {Chen}}, ,  \emph
  {et~al.},\ }\href {\doibase 10.1007/s41365-023-01241-z} {\bibfield  {journal}
  {\bibinfo  {journal} {Nucl. Sci. Tech.}\ }\textbf {\bibinfo {volume} {34}},\
  \bibinfo {pages} {94} (\bibinfo {year} {2023}{\natexlab{b}})}\BibitemShut
  {NoStop}%
\bibitem [{\citenamefont {Xiao}\ \emph {et~al.}(2023)\citenamefont {Xiao},
  \citenamefont {Li}, \citenamefont {Wang}, \citenamefont {Liu},\ and\
  \citenamefont {Li}}]{Xiao2023NST}%
  \BibitemOpen
  \bibfield  {author} {\bibinfo {author} {\bibfnamefont {K.}~\bibnamefont
  {Xiao}}, \bibinfo {author} {\bibfnamefont {P.-C.}\ \bibnamefont {Li}},
  \bibinfo {author} {\bibfnamefont {Y.-J.}\ \bibnamefont {Wang}}, \bibinfo
  {author} {\bibfnamefont {F.-H.}\ \bibnamefont {Liu}}, \ and\ \bibinfo
  {author} {\bibfnamefont {Q.-F.}\ \bibnamefont {Li}},\ }\href {\doibase
  10.1007/s41365-023-01205-3} {\bibfield  {journal} {\bibinfo  {journal} {Nucl.
  Sci. Tech.}\ }\textbf {\bibinfo {volume} {34}},\ \bibinfo {pages} {62}
  (\bibinfo {year} {2023})}\BibitemShut {NoStop}%
\bibitem [{\citenamefont {Wei}\ and\ \citenamefont {Feng}(2024)}]{Wei2024NST}%
  \BibitemOpen
  \bibfield  {author} {\bibinfo {author} {\bibfnamefont {S.-N.}\ \bibnamefont
  {Wei}}\ and\ \bibinfo {author} {\bibfnamefont {Z.-Q.}\ \bibnamefont {Feng}},\
  }\href {\doibase 10.1007/s41365-024-01380-x} {\bibfield  {journal} {\bibinfo
  {journal} {Nucl. Sci. Tech.}\ }\textbf {\bibinfo {volume} {35}},\ \bibinfo
  {pages} {15} (\bibinfo {year} {2024})}\BibitemShut {NoStop}%
\bibitem [{\citenamefont {Maruyama}\ \emph {et~al.}(1996)\citenamefont
  {Maruyama}, \citenamefont {Niita},\ and\ \citenamefont
  {Iwamoto}}]{Maruyama1996PRC}%
  \BibitemOpen
  \bibfield  {author} {\bibinfo {author} {\bibfnamefont {T.}~\bibnamefont
  {Maruyama}}, \bibinfo {author} {\bibfnamefont {K.}~\bibnamefont {Niita}}, \
  and\ \bibinfo {author} {\bibfnamefont {A.}~\bibnamefont {Iwamoto}},\ }\href
  {\doibase 10.1103/PhysRevC.53.297} {\bibfield  {journal} {\bibinfo  {journal}
  {Phys. Rev. C}\ }\textbf {\bibinfo {volume} {53}},\ \bibinfo {pages} {297}
  (\bibinfo {year} {1996})}\BibitemShut {NoStop}%
\bibitem [{\citenamefont {Feldmeier}(1990)}]{Feldmeier1990NPA}%
  \BibitemOpen
  \bibfield  {author} {\bibinfo {author} {\bibfnamefont {H.}~\bibnamefont
  {Feldmeier}},\ }\href {https://doi.org/10.1016/0375-9474(90)90328-J}
  {\bibfield  {journal} {\bibinfo  {journal} {Nucl. Phys. A}\ }\textbf
  {\bibinfo {volume} {515}},\ \bibinfo {pages} {147} (\bibinfo {year}
  {1990})}\BibitemShut {NoStop}%
\bibitem [{\citenamefont {Kiderlen}\ and\ \citenamefont
  {Danielewicz}(1997)}]{Kiderlen1997NPA}%
  \BibitemOpen
  \bibfield  {author} {\bibinfo {author} {\bibfnamefont {D.}~\bibnamefont
  {Kiderlen}}\ and\ \bibinfo {author} {\bibfnamefont {P.}~\bibnamefont
  {Danielewicz}},\ }\href {https://doi.org/10.1016/S0375-9474(97)00181-4}
  {\bibfield  {journal} {\bibinfo  {journal} {Nucl. Phys. A}\ }\textbf
  {\bibinfo {volume} {620}},\ \bibinfo {pages} {346} (\bibinfo {year}
  {1997})}\BibitemShut {NoStop}%
\bibitem [{\citenamefont {Colonna}\ and\ \citenamefont
  {Chomaz}(1998)}]{Colonna1998PLB}%
  \BibitemOpen
  \bibfield  {author} {\bibinfo {author} {\bibfnamefont {M.}~\bibnamefont
  {Colonna}}\ and\ \bibinfo {author} {\bibfnamefont {P.}~\bibnamefont
  {Chomaz}},\ }\href {https://doi.org/10.1016/S0370-2693(98)00882-X} {\bibfield
   {journal} {\bibinfo  {journal} {Phys. Lett. B}\ }\textbf {\bibinfo {volume}
  {436}},\ \bibinfo {pages} {1} (\bibinfo {year} {1998})}\BibitemShut {NoStop}%
\bibitem [{\citenamefont {He}\ \emph {et~al.}(2014)\citenamefont {He},
  \citenamefont {Ma}, \citenamefont {Cao}, \citenamefont {Cai},\ and\
  \citenamefont {Zhang}}]{He2014PRL}%
  \BibitemOpen
  \bibfield  {author} {\bibinfo {author} {\bibfnamefont {W.~B.}\ \bibnamefont
  {He}}, \bibinfo {author} {\bibfnamefont {Y.~G.}\ \bibnamefont {Ma}}, \bibinfo
  {author} {\bibfnamefont {X.~G.}\ \bibnamefont {Cao}}, \bibinfo {author}
  {\bibfnamefont {X.~Z.}\ \bibnamefont {Cai}}, \ and\ \bibinfo {author}
  {\bibfnamefont {G.~Q.}\ \bibnamefont {Zhang}},\ }\href {\doibase
  10.1103/PhysRevLett.113.032506} {\bibfield  {journal} {\bibinfo  {journal}
  {Phys. Rev. Lett.}\ }\textbf {\bibinfo {volume} {113}},\ \bibinfo {pages}
  {032506} (\bibinfo {year} {2014})}\BibitemShut {NoStop}%
\bibitem [{\citenamefont {Huang}\ \emph {et~al.}(2017)\citenamefont {Huang},
  \citenamefont {Ma},\ and\ \citenamefont {He}}]{Huang2017PRC}%
  \BibitemOpen
  \bibfield  {author} {\bibinfo {author} {\bibfnamefont {B.~S.}\ \bibnamefont
  {Huang}}, \bibinfo {author} {\bibfnamefont {Y.~G.}\ \bibnamefont {Ma}}, \
  and\ \bibinfo {author} {\bibfnamefont {W.~B.}\ \bibnamefont {He}},\ }\href
  {\doibase 10.1103/PhysRevC.95.034606} {\bibfield  {journal} {\bibinfo
  {journal} {Phys. Rev. C}\ }\textbf {\bibinfo {volume} {95}},\ \bibinfo
  {pages} {034606} (\bibinfo {year} {2017})}\BibitemShut {NoStop}%
\bibitem [{\citenamefont {Wang}\ \emph
  {et~al.}(2023{\natexlab{c}})\citenamefont {Wang}, \citenamefont {Ma},
  \citenamefont {He}, \citenamefont {Fang},\ and\ \citenamefont
  {Cao}}]{Wang2022PRC}%
  \BibitemOpen
  \bibfield  {author} {\bibinfo {author} {\bibfnamefont {S.~S.}\ \bibnamefont
  {Wang}}, \bibinfo {author} {\bibfnamefont {Y.~G.}\ \bibnamefont {Ma}},
  \bibinfo {author} {\bibfnamefont {W.~B.}\ \bibnamefont {He}}, \bibinfo
  {author} {\bibfnamefont {D.~Q.}\ \bibnamefont {Fang}}, \ and\ \bibinfo
  {author} {\bibfnamefont {X.~G.}\ \bibnamefont {Cao}},\ }\href {\doibase
  10.1103/PhysRevC.108.014609} {\bibfield  {journal} {\bibinfo  {journal}
  {Phys. Rev. C}\ }\textbf {\bibinfo {volume} {108}},\ \bibinfo {pages}
  {014609} (\bibinfo {year} {2023}{\natexlab{c}})}\BibitemShut {NoStop}%
\bibitem [{\citenamefont {Cao}\ \emph {et~al.}(2022)\citenamefont {Cao},
  \citenamefont {Deng},\ and\ \citenamefont {Ma}}]{Cao2022PRC}%
  \BibitemOpen
  \bibfield  {author} {\bibinfo {author} {\bibfnamefont {Y.~T.}\ \bibnamefont
  {Cao}}, \bibinfo {author} {\bibfnamefont {X.~G.}\ \bibnamefont {Deng}}, \
  and\ \bibinfo {author} {\bibfnamefont {Y.~G.}\ \bibnamefont {Ma}},\ }\href
  {\doibase 10.1103/PhysRevC.106.014611} {\bibfield  {journal} {\bibinfo
  {journal} {Phys. Rev. C}\ }\textbf {\bibinfo {volume} {106}},\ \bibinfo
  {pages} {014611} (\bibinfo {year} {2022})}\BibitemShut {NoStop}%
\bibitem [{\citenamefont {Cao}\ \emph {et~al.}(2023)\citenamefont {Cao},
  \citenamefont {Deng},\ and\ \citenamefont {Ma}}]{Cao2023PRC}%
  \BibitemOpen
  \bibfield  {author} {\bibinfo {author} {\bibfnamefont {Y.~T.}\ \bibnamefont
  {Cao}}, \bibinfo {author} {\bibfnamefont {X.~G.}\ \bibnamefont {Deng}}, \
  and\ \bibinfo {author} {\bibfnamefont {Y.~G.}\ \bibnamefont {Ma}},\ }\href
  {\doibase 10.1103/PhysRevC.108.024610} {\bibfield  {journal} {\bibinfo
  {journal} {Phys. Rev. C}\ }\textbf {\bibinfo {volume} {108}},\ \bibinfo
  {pages} {024610} (\bibinfo {year} {2023})}\BibitemShut {NoStop}%
\bibitem [{\citenamefont {Ma}(2023)}]{Ma2023NT}%
  \BibitemOpen
  \bibfield  {author} {\bibinfo {author} {\bibfnamefont {Y.~G.}\ \bibnamefont
  {Ma}},\ }\href {\doibase 10.11889/j.0253-3219.2023.hjs.46.080001} {\bibfield
  {journal} {\bibinfo  {journal} {Nucl. Tech.}\ }\textbf {\bibinfo {volume}
  {46}},\ \bibinfo {pages} {080001} (\bibinfo {year} {2023})}\BibitemShut
  {NoStop}%
\bibitem [{\citenamefont {Ono}\ \emph {et~al.}(1992{\natexlab{a}})\citenamefont
  {Ono}, \citenamefont {Horiuchi}, \citenamefont {Maruyama},\ and\
  \citenamefont {Ohnishi}}]{Ono1992PTP}%
  \BibitemOpen
  \bibfield  {author} {\bibinfo {author} {\bibfnamefont {A.}~\bibnamefont
  {Ono}}, \bibinfo {author} {\bibfnamefont {H.}~\bibnamefont {Horiuchi}},
  \bibinfo {author} {\bibfnamefont {T.}~\bibnamefont {Maruyama}}, \ and\
  \bibinfo {author} {\bibfnamefont {A.}~\bibnamefont {Ohnishi}},\ }\href
  {\doibase 10.1143/ptp/87.5.1185} {\bibfield  {journal} {\bibinfo  {journal}
  {Prog. Theo. Phys.}\ }\textbf {\bibinfo {volume} {87}},\ \bibinfo {pages}
  {1185} (\bibinfo {year} {1992}{\natexlab{a}})}\BibitemShut {NoStop}%
\bibitem [{\citenamefont {Ono}\ \emph {et~al.}(1992{\natexlab{b}})\citenamefont
  {Ono}, \citenamefont {Horiuchi}, \citenamefont {Maruyama},\ and\
  \citenamefont {Ohnishi}}]{Ono1992PRL}%
  \BibitemOpen
  \bibfield  {author} {\bibinfo {author} {\bibfnamefont {A.}~\bibnamefont
  {Ono}}, \bibinfo {author} {\bibfnamefont {H.}~\bibnamefont {Horiuchi}},
  \bibinfo {author} {\bibfnamefont {T.}~\bibnamefont {Maruyama}}, \ and\
  \bibinfo {author} {\bibfnamefont {A.}~\bibnamefont {Ohnishi}},\ }\href
  {\doibase 10.1103/PhysRevLett.68.2898} {\bibfield  {journal} {\bibinfo
  {journal} {Phys. Rev. Lett.}\ }\textbf {\bibinfo {volume} {68}},\ \bibinfo
  {pages} {2898} (\bibinfo {year} {1992}{\natexlab{b}})}\BibitemShut {NoStop}%
\bibitem [{\citenamefont {Ono}\ and\ \citenamefont
  {Horiuchi}(2004)}]{Ono2004PPNP}%
  \BibitemOpen
  \bibfield  {author} {\bibinfo {author} {\bibfnamefont {A.}~\bibnamefont
  {Ono}}\ and\ \bibinfo {author} {\bibfnamefont {H.}~\bibnamefont {Horiuchi}},\
  }\href {\doibase https://doi.org/10.1016/j.ppnp.2004.05.002} {\bibfield
  {journal} {\bibinfo  {journal} {Prog. Part.Nucl. Phys.}\ }\textbf {\bibinfo
  {volume} {53}},\ \bibinfo {pages} {501} (\bibinfo {year} {2004})}\BibitemShut
  {NoStop}%
\bibitem [{\citenamefont {Ikeno}\ \emph {et~al.}(2016)\citenamefont {Ikeno},
  \citenamefont {Ono}, \citenamefont {Nara},\ and\ \citenamefont
  {Ohnishi}}]{Ikeno2016PRC}%
  \BibitemOpen
  \bibfield  {author} {\bibinfo {author} {\bibfnamefont {N.}~\bibnamefont
  {Ikeno}}, \bibinfo {author} {\bibfnamefont {A.}~\bibnamefont {Ono}}, \bibinfo
  {author} {\bibfnamefont {Y.}~\bibnamefont {Nara}}, \ and\ \bibinfo {author}
  {\bibfnamefont {A.}~\bibnamefont {Ohnishi}},\ }\href {\doibase
  10.1103/PhysRevC.93.044612} {\bibfield  {journal} {\bibinfo  {journal} {Phys.
  Rev. C}\ }\textbf {\bibinfo {volume} {93}},\ \bibinfo {pages} {044612}
  (\bibinfo {year} {2016})}\BibitemShut {NoStop}%
\bibitem [{\citenamefont {Ono}\ \emph {et~al.}(1993{\natexlab{a}})\citenamefont
  {Ono}, \citenamefont {Horiuchi}, \citenamefont {Maruyama},\ and\
  \citenamefont {Ohnishi}}]{Ono1993PRC}%
  \BibitemOpen
  \bibfield  {author} {\bibinfo {author} {\bibfnamefont {A.}~\bibnamefont
  {Ono}}, \bibinfo {author} {\bibfnamefont {H.}~\bibnamefont {Horiuchi}},
  \bibinfo {author} {\bibfnamefont {T.}~\bibnamefont {Maruyama}}, \ and\
  \bibinfo {author} {\bibfnamefont {A.}~\bibnamefont {Ohnishi}},\ }\href
  {\doibase 10.1103/PhysRevC.47.2652} {\bibfield  {journal} {\bibinfo
  {journal} {Phys. Rev. C}\ }\textbf {\bibinfo {volume} {47}},\ \bibinfo
  {pages} {2652} (\bibinfo {year} {1993}{\natexlab{a}})}\BibitemShut {NoStop}%
\bibitem [{\citenamefont {Ono}\ \emph {et~al.}(1993{\natexlab{b}})\citenamefont
  {Ono}, \citenamefont {Horiuchi},\ and\ \citenamefont
  {Maruyama}}]{Ono1993PRC2}%
  \BibitemOpen
  \bibfield  {author} {\bibinfo {author} {\bibfnamefont {A.}~\bibnamefont
  {Ono}}, \bibinfo {author} {\bibfnamefont {H.}~\bibnamefont {Horiuchi}}, \
  and\ \bibinfo {author} {\bibfnamefont {T.}~\bibnamefont {Maruyama}},\ }\href
  {\doibase 10.1103/PhysRevC.48.2946} {\bibfield  {journal} {\bibinfo
  {journal} {Phys. Rev. C}\ }\textbf {\bibinfo {volume} {48}},\ \bibinfo
  {pages} {2946} (\bibinfo {year} {1993}{\natexlab{b}})}\BibitemShut {NoStop}%
\bibitem [{\citenamefont {Ono}\ and\ \citenamefont
  {Horiuchi}(1996{\natexlab{a}})}]{Ono1996PRC}%
  \BibitemOpen
  \bibfield  {author} {\bibinfo {author} {\bibfnamefont {A.}~\bibnamefont
  {Ono}}\ and\ \bibinfo {author} {\bibfnamefont {H.}~\bibnamefont {Horiuchi}},\
  }\href {\doibase 10.1103/PhysRevC.53.845} {\bibfield  {journal} {\bibinfo
  {journal} {Phys. Rev. C}\ }\textbf {\bibinfo {volume} {53}},\ \bibinfo
  {pages} {845} (\bibinfo {year} {1996}{\natexlab{a}})}\BibitemShut {NoStop}%
\bibitem [{\citenamefont {Ono}\ and\ \citenamefont
  {Horiuchi}(1996{\natexlab{b}})}]{Ono1996PRC2}%
  \BibitemOpen
  \bibfield  {author} {\bibinfo {author} {\bibfnamefont {A.}~\bibnamefont
  {Ono}}\ and\ \bibinfo {author} {\bibfnamefont {H.}~\bibnamefont {Horiuchi}},\
  }\href {\doibase 10.1103/PhysRevC.53.2958} {\bibfield  {journal} {\bibinfo
  {journal} {Phys. Rev. C}\ }\textbf {\bibinfo {volume} {53}},\ \bibinfo
  {pages} {2958} (\bibinfo {year} {1996}{\natexlab{b}})}\BibitemShut {NoStop}%
\bibitem [{\citenamefont {Ono}(1999)}]{Ono1999PRC}%
  \BibitemOpen
  \bibfield  {author} {\bibinfo {author} {\bibfnamefont {A.}~\bibnamefont
  {Ono}},\ }\href {\doibase 10.1103/PhysRevC.59.853} {\bibfield  {journal}
  {\bibinfo  {journal} {Phys. Rev. C}\ }\textbf {\bibinfo {volume} {59}},\
  \bibinfo {pages} {853} (\bibinfo {year} {1999})}\BibitemShut {NoStop}%
\bibitem [{\citenamefont {Ono}\ \emph {et~al.}(2002)\citenamefont {Ono},
  \citenamefont {Hudan}, \citenamefont {Chbihi},\ and\ \citenamefont
  {Frankland}}]{Ono2002PRC}%
  \BibitemOpen
  \bibfield  {author} {\bibinfo {author} {\bibfnamefont {A.}~\bibnamefont
  {Ono}}, \bibinfo {author} {\bibfnamefont {S.}~\bibnamefont {Hudan}}, \bibinfo
  {author} {\bibfnamefont {A.}~\bibnamefont {Chbihi}}, \ and\ \bibinfo {author}
  {\bibfnamefont {J.~D.}\ \bibnamefont {Frankland}},\ }\href {\doibase
  10.1103/PhysRevC.66.014603} {\bibfield  {journal} {\bibinfo  {journal} {Phys.
  Rev. C}\ }\textbf {\bibinfo {volume} {66}},\ \bibinfo {pages} {014603}
  (\bibinfo {year} {2002})}\BibitemShut {NoStop}%
\bibitem [{\citenamefont {Lin}\ \emph {et~al.}(2016)\citenamefont {Lin},
  \citenamefont {Liu}, \citenamefont {Wada}, \citenamefont {Huang},
  \citenamefont {Ren}, \citenamefont {Tian}, \citenamefont {Luo}, \citenamefont
  {Sun}, \citenamefont {Chen}, \citenamefont {Xiao}, \citenamefont {Han},
  \citenamefont {Shi}, \citenamefont {Liu},\ and\ \citenamefont
  {Gou}}]{Lin2016PRC}%
  \BibitemOpen
  \bibfield  {author} {\bibinfo {author} {\bibfnamefont {W.}~\bibnamefont
  {Lin}}, \bibinfo {author} {\bibfnamefont {X.}~\bibnamefont {Liu}}, \bibinfo
  {author} {\bibfnamefont {R.}~\bibnamefont {Wada}}, \bibinfo {author}
  {\bibfnamefont {M.}~\bibnamefont {Huang}}, \bibinfo {author} {\bibfnamefont
  {P.}~\bibnamefont {Ren}}, \bibinfo {author} {\bibfnamefont {G.}~\bibnamefont
  {Tian}}, \bibinfo {author} {\bibfnamefont {F.}~\bibnamefont {Luo}}, \bibinfo
  {author} {\bibfnamefont {Q.}~\bibnamefont {Sun}}, \bibinfo {author}
  {\bibfnamefont {Z.}~\bibnamefont {Chen}}, \bibinfo {author} {\bibfnamefont
  {G.~Q.}\ \bibnamefont {Xiao}}, \bibinfo {author} {\bibfnamefont
  {R.}~\bibnamefont {Han}}, \bibinfo {author} {\bibfnamefont {F.}~\bibnamefont
  {Shi}}, \bibinfo {author} {\bibfnamefont {J.}~\bibnamefont {Liu}}, \ and\
  \bibinfo {author} {\bibfnamefont {B.}~\bibnamefont {Gou}},\ }\href {\doibase
  10.1103/PhysRevC.94.064609} {\bibfield  {journal} {\bibinfo  {journal} {Phys.
  Rev. C}\ }\textbf {\bibinfo {volume} {94}},\ \bibinfo {pages} {064609}
  (\bibinfo {year} {2016})}\BibitemShut {NoStop}%
\bibitem [{\citenamefont {Chabanat}\ \emph {et~al.}(1998)\citenamefont
  {Chabanat}, \citenamefont {Bonche}, \citenamefont {Haensel}, \citenamefont
  {Meyer},\ and\ \citenamefont {Schaeffer}}]{Chabanat1998NPA}%
  \BibitemOpen
  \bibfield  {author} {\bibinfo {author} {\bibfnamefont {E.}~\bibnamefont
  {Chabanat}}, \bibinfo {author} {\bibfnamefont {P.}~\bibnamefont {Bonche}},
  \bibinfo {author} {\bibfnamefont {P.}~\bibnamefont {Haensel}}, \bibinfo
  {author} {\bibfnamefont {J.}~\bibnamefont {Meyer}}, \ and\ \bibinfo {author}
  {\bibfnamefont {R.}~\bibnamefont {Schaeffer}},\ }\href {\doibase
  https://doi.org/10.1016/S0375-9474(98)00180-8} {\bibfield  {journal}
  {\bibinfo  {journal} {Nucl. Phys. A}\ }\textbf {\bibinfo {volume} {635}},\
  \bibinfo {pages} {231} (\bibinfo {year} {1998})}\BibitemShut {NoStop}%
\bibitem [{\citenamefont {Ikeno}\ and\ \citenamefont
  {Ono}(2023)}]{Ikeno2023PRC}%
  \BibitemOpen
  \bibfield  {author} {\bibinfo {author} {\bibfnamefont {N.}~\bibnamefont
  {Ikeno}}\ and\ \bibinfo {author} {\bibfnamefont {A.}~\bibnamefont {Ono}},\
  }\href {\doibase 10.1103/PhysRevC.108.044601} {\bibfield  {journal} {\bibinfo
   {journal} {Phys. Rev. C}\ }\textbf {\bibinfo {volume} {108}},\ \bibinfo
  {pages} {044601} (\bibinfo {year} {2023})}\BibitemShut {NoStop}%
\bibitem [{\citenamefont {Ono}(2013)}]{Ono2013JPCS}%
  \BibitemOpen
  \bibfield  {author} {\bibinfo {author} {\bibfnamefont {A.}~\bibnamefont
  {Ono}},\ }\href {\doibase 10.1088/1742-6596/420/1/012103} {\bibfield
  {journal} {\bibinfo  {journal} {Jour. Phys. Conf. Seri.}\ }\textbf {\bibinfo
  {volume} {420}},\ \bibinfo {pages} {012103} (\bibinfo {year}
  {2013})}\BibitemShut {NoStop}%
\bibitem [{\citenamefont {Cugnon}\ \emph {et~al.}(1996)\citenamefont {Cugnon},
  \citenamefont {L'H\^{o}te},\ and\ \citenamefont
  {Vandermeulen}}]{Cugnon1996NIM}%
  \BibitemOpen
  \bibfield  {author} {\bibinfo {author} {\bibfnamefont {J.}~\bibnamefont
  {Cugnon}}, \bibinfo {author} {\bibfnamefont {D.}~\bibnamefont {L'H\^{o}te}},
  \ and\ \bibinfo {author} {\bibfnamefont {J.}~\bibnamefont {Vandermeulen}},\
  }\href {\doibase https://doi.org/10.1016/0168-583X(95)01384-9} {\bibfield
  {journal} {\bibinfo  {journal} {Nucl. Instrum. Methods Phys. Res. B}\
  }\textbf {\bibinfo {volume} {111}},\ \bibinfo {pages} {215} (\bibinfo {year}
  {1996})}\BibitemShut {NoStop}%
\bibitem [{\citenamefont {Maruyama}\ \emph {et~al.}(1992)\citenamefont
  {Maruyama}, \citenamefont {Ono}, \citenamefont {Ohnishi},\ and\ \citenamefont
  {Horiuchi}}]{Maruyama1992PTP}%
  \BibitemOpen
  \bibfield  {author} {\bibinfo {author} {\bibfnamefont {T.}~\bibnamefont
  {Maruyama}}, \bibinfo {author} {\bibfnamefont {A.}~\bibnamefont {Ono}},
  \bibinfo {author} {\bibfnamefont {A.}~\bibnamefont {Ohnishi}}, \ and\
  \bibinfo {author} {\bibfnamefont {H.}~\bibnamefont {Horiuchi}},\ }\href
  {\doibase https://doi.org/10.1143/ptp/87.6.1367} {\bibfield  {journal}
  {\bibinfo  {journal} {Prog. Theo. Phys.}\ }\textbf {\bibinfo {volume} {87}},\
  \bibinfo {pages} {1367} (\bibinfo {year} {1992})}\BibitemShut {NoStop}%
\bibitem [{\citenamefont {P\"{u}hlhofer}(1977)}]{puhlhofer1977}%
  \BibitemOpen
  \bibfield  {author} {\bibinfo {author} {\bibfnamefont {F.}~\bibnamefont
  {P\"{u}hlhofer}},\ }\href {\doibase 10.1016/0375-9474(77)90308-6} {\bibfield
  {journal} {\bibinfo  {journal} {Nuclear Physics A}\ }\textbf {\bibinfo
  {volume} {280}},\ \bibinfo {pages} {267} (\bibinfo {year}
  {1977})}\BibitemShut {NoStop}%
\bibitem [{\citenamefont {Tian}\ \emph {et~al.}(2018)\citenamefont {Tian},
  \citenamefont {Chen}, \citenamefont {Han}, \citenamefont {Shi}, \citenamefont
  {Luo}, \citenamefont {Sun}, \citenamefont {Song}, \citenamefont {Zhang},
  \citenamefont {Xiao}, \citenamefont {Wada} \emph {et~al.}}]{tian2018}%
  \BibitemOpen
  \bibfield  {author} {\bibinfo {author} {\bibfnamefont {G.}~\bibnamefont
  {Tian}}, \bibinfo {author} {\bibfnamefont {Z.}~\bibnamefont {Chen}}, \bibinfo
  {author} {\bibfnamefont {R.}~\bibnamefont {Han}}, \bibinfo {author}
  {\bibfnamefont {F.}~\bibnamefont {Shi}}, \bibinfo {author} {\bibfnamefont
  {F.}~\bibnamefont {Luo}}, \bibinfo {author} {\bibfnamefont {Q.}~\bibnamefont
  {Sun}}, \bibinfo {author} {\bibfnamefont {L.}~\bibnamefont {Song}}, \bibinfo
  {author} {\bibfnamefont {X.}~\bibnamefont {Zhang}}, \bibinfo {author}
  {\bibfnamefont {G.}~\bibnamefont {Xiao}}, \bibinfo {author} {\bibfnamefont
  {R.}~\bibnamefont {Wada}},  \emph {et~al.},\ }\href {\doibase
  10.1103/physrevc.97.034610} {\bibfield  {journal} {\bibinfo  {journal} {Phys.
  Rev. C}\ }\textbf {\bibinfo {volume} {97}},\ \bibinfo {pages} {034610}
  (\bibinfo {year} {2018})}\BibitemShut {NoStop}%
\bibitem [{\citenamefont {Ono}(2022)}]{Ono2022PLB}%
  \BibitemOpen
  \bibfield  {author} {\bibinfo {author} {\bibfnamefont {A.}~\bibnamefont
  {Ono}},\ }\href {\doibase https://doi.org/10.1016/j.physletb.2022.136931}
  {\bibfield  {journal} {\bibinfo  {journal} {Physics Letters B}\ }\textbf
  {\bibinfo {volume} {826}},\ \bibinfo {pages} {136931} (\bibinfo {year}
  {2022})}\BibitemShut {NoStop}%
\bibitem [{\citenamefont {{LPC hadrontherapy web site}}()}]{GANIL}%
  \BibitemOpen
  \bibfield  {author} {\bibinfo {author} {\bibnamefont {{LPC hadrontherapy web
  site}}},\ }\href {http://hadrontherapy-data.in2p3.fr/} {\bibinfo  {journal}
  {http://hadrontherapy-data.in2p3.fr/}\ }\BibitemShut {NoStop}%
\bibitem [{\citenamefont {Dudouet}\ \emph {et~al.}(2014)\citenamefont
  {Dudouet}, \citenamefont {Labalme}, \citenamefont {Cussol}, \citenamefont
  {Finck}, \citenamefont {Rescigno}, \citenamefont {Rousseau}, \citenamefont
  {Salvador},\ and\ \citenamefont {Vanstalle}}]{Dudouet2014PRC}%
  \BibitemOpen
\bibfield  {journal} {  }\bibfield  {author} {\bibinfo {author} {\bibfnamefont
  {J.}~\bibnamefont {Dudouet}}, \bibinfo {author} {\bibfnamefont
  {M.}~\bibnamefont {Labalme}}, \bibinfo {author} {\bibfnamefont
  {D.}~\bibnamefont {Cussol}}, \bibinfo {author} {\bibfnamefont
  {C.}~\bibnamefont {Finck}}, \bibinfo {author} {\bibfnamefont
  {R.}~\bibnamefont {Rescigno}}, \bibinfo {author} {\bibfnamefont
  {M.}~\bibnamefont {Rousseau}}, \bibinfo {author} {\bibfnamefont
  {S.}~\bibnamefont {Salvador}}, \ and\ \bibinfo {author} {\bibfnamefont
  {M.}~\bibnamefont {Vanstalle}},\ }\href {\doibase 10.1103/PhysRevC.89.064615}
  {\bibfield  {journal} {\bibinfo  {journal} {Phys. Rev. C}\ }\textbf {\bibinfo
  {volume} {89}},\ \bibinfo {pages} {064615} (\bibinfo {year}
  {2014})}\BibitemShut {NoStop}%
\bibitem [{\citenamefont {Souliotis}\ \emph {et~al.}(1992)\citenamefont
  {Souliotis}, \citenamefont {Morrissey}, \citenamefont {Orr}, \citenamefont
  {Sherrill},\ and\ \citenamefont {Winger}}]{Souliotis1992}%
  \BibitemOpen
  \bibfield  {author} {\bibinfo {author} {\bibfnamefont {G.~A.}\ \bibnamefont
  {Souliotis}}, \bibinfo {author} {\bibfnamefont {D.~J.}\ \bibnamefont
  {Morrissey}}, \bibinfo {author} {\bibfnamefont {N.~A.}\ \bibnamefont {Orr}},
  \bibinfo {author} {\bibfnamefont {B.~M.}\ \bibnamefont {Sherrill}}, \ and\
  \bibinfo {author} {\bibfnamefont {J.~A.}\ \bibnamefont {Winger}},\ }\href
  {\doibase 10.1103/physrevc.46.1383} {\bibfield  {journal} {\bibinfo
  {journal} {Physical Review C}\ }\textbf {\bibinfo {volume} {46}},\ \bibinfo
  {pages} {1383} (\bibinfo {year} {1992})}\BibitemShut {NoStop}%
\bibitem [{\citenamefont {Momota}\ \emph {et~al.}(2018)\citenamefont {Momota},
  \citenamefont {Kanazawa}, \citenamefont {Kitagawa},\ and\ \citenamefont
  {Sato}}]{Momota2018PRC}%
  \BibitemOpen
  \bibfield  {author} {\bibinfo {author} {\bibfnamefont {S.}~\bibnamefont
  {Momota}}, \bibinfo {author} {\bibfnamefont {M.}~\bibnamefont {Kanazawa}},
  \bibinfo {author} {\bibfnamefont {A.}~\bibnamefont {Kitagawa}}, \ and\
  \bibinfo {author} {\bibfnamefont {S.}~\bibnamefont {Sato}},\ }\href {\doibase
  10.1103/PhysRevC.97.044604} {\bibfield  {journal} {\bibinfo  {journal} {Phys.
  Rev. C}\ }\textbf {\bibinfo {volume} {97}},\ \bibinfo {pages} {044604}
  (\bibinfo {year} {2018})}\BibitemShut {NoStop}%
\end{thebibliography}%
\end{document}